\documentclass[twocolumn]{aastex6}
\usepackage{graphicx}
\usepackage{amssymb,amsfonts,amsmath,amstext,amsgen,amsopn,amsxtra,indentfirst,times}
\usepackage{xcolor}

\begin{document}

\title{How Much Information Does the Sodium Doublet Encode? \\ Retrieval Analysis of Non-LTE Sodium Lines at Low and High Spectral Resolutions}

\author{Chloe Fisher\altaffilmark{1,2}}
\author{Kevin Heng\altaffilmark{1}}
\altaffiltext{1}{University of Bern, Center for Space and Habitability, Gesellschaftsstrasse 6, CH-3012, Bern, Switzerland.  Emails: chloe.fisher@csh.unibe.ch, kevin.heng@csh.unibe.ch}
\altaffiltext{2}{University of Bern International 2021 Ph.D Fellowship}

\begin{abstract}
Motivated by both ground- and space-based detections of the sodium doublet in the transmission spectra of exoplanetary atmospheres, we revisit the theory and interpretation of sodium lines in non-local thermodynamic equilibrium (NLTE), where collisions are not efficient enough to maintain a Boltzmann distribution for the excited and ground states of the sodium atom.  We consider non-Boltzmann distributions that account for the ineffectiveness of collisions.  We analyze the sodium doublet in transmission spectra measured at low (HAT-P-1b, HAT-P-12b, HD 189733b, WASP-6b, WASP-17b and WASP-39b) and high (WASP-49b) spectral resolutions.  Nested-sampling retrievals performed on low-resolution optical/visible transmission spectra are unable to break the normalization degeneracy if the spectral continuum is associated with Rayleigh scattering by small cloud particles.  Using mock retrievals, we demonstrate that un-normalized ground-based, high-resolution spectra centered on the sodium doublet alone are unable to precisely inform us about the pressure levels probed by the transit chord and hence to identify the region (i.e., thermosphere, exosphere) of the atmosphere being probed.  Retrievals performed on the HARPS transmission spectrum of WASP-49b support this conclusion.  Generally, we are unable to distinguish between LTE versus NLTE interpretations of the sodium doublet based on the computed Bayesian evidence with the implication that LTE interpretations tend to under-estimate the temperature probed by the transit chord.  With the current low-resolution data, the sodium line shapes are consistent with Voigt profiles without the need for sub-Lorentzian wings.  The retrieved sodium abundances are consistent with being sub-solar to solar.
\end{abstract}

\keywords{planets and satellites: atmospheres}

\section{Introduction}
\label{sect:intro}

The use of the resonant lines of the sodium atom to probe the atmospheres of exoplanets has a rich history.  The detectability of the resonant doublets of sodium and potassium in transmission spectra was first predicted by \cite{ss00}.  \textbf{In a contemporaneous study, \cite{s00} reached the same conclusion.}  Later, sodium was detected for the first time in an exoplanet by \cite{char02} for the hot Jupiter HD 209458b.  Since the predictions and discovery, the detection of sodium has become routine from space at low spectral resolution (e.g., \citealt{sing16}) and from the ground both at low (e.g., \citealt{nikolov18}) and high spectral resolution (e.g., \citealt{redfield08,snellen08,jensen12,w15,w17,k17,c18}).  \textbf{(But see \citealt{gibson19} for a retraction of the detection of potassium.)}  These detections motivate a series of theoretical and phenomenological questions concerning the sodium lines, which we address in the current study.

\subsection{Are sodium lines in non-local thermodynamic equilibrium?}

Local thermodynamic equilibrium (LTE) involves a set of four assumptions: the radiation field is described by a Planck or blackbody function, the velocity field follows a Maxwellian distribution, the distribution of neutrals versus electrons and ions is a solution of the Saha equation, and the energy levels of the ground versus excited states of the atom follow a Boltzmann distribution.  One of the few studies to examine non-LTE (NLTE) effects associated with the sodium lines in exoplanetary atmospheres is \cite{fortney03}, who considered non-Saha distributions of the electron density caused by photoionization.  

Here, we take a different approach: we study the NLTE effect associated with non-Boltzmann distributions of the ground and excited states of the sodium atom.  Physically, this departure from LTE arises from the diminished role of collisions.  To maintain a Boltzmann distribution, collisions between the sodium atoms and the atoms or molecules of the bulk gas they are embedded in (e.g., atomic or molecular hydrogen) need to be efficient.  There are two aspects to this NLTE effect.  The first is that the sodium doublet, also known as the Fraunhofer D$_1$ and D$_2$ lines, is a pair of resonant lines, where the difference in transit radii between line center and wing spans about 20 scale heights (e.g., \citealt{heng16}), corresponding to about 9 orders of magnitude in pressure.  It implies that the importance of collisions varies between the peak and wings of each sodium line---they are more important in the line wings, which probe deeper into the atmosphere, and less influential at the line peak.

\begin{figure}%[!t]
\begin{center}
%\vspace{-0.2in}
\includegraphics[width=\columnwidth]{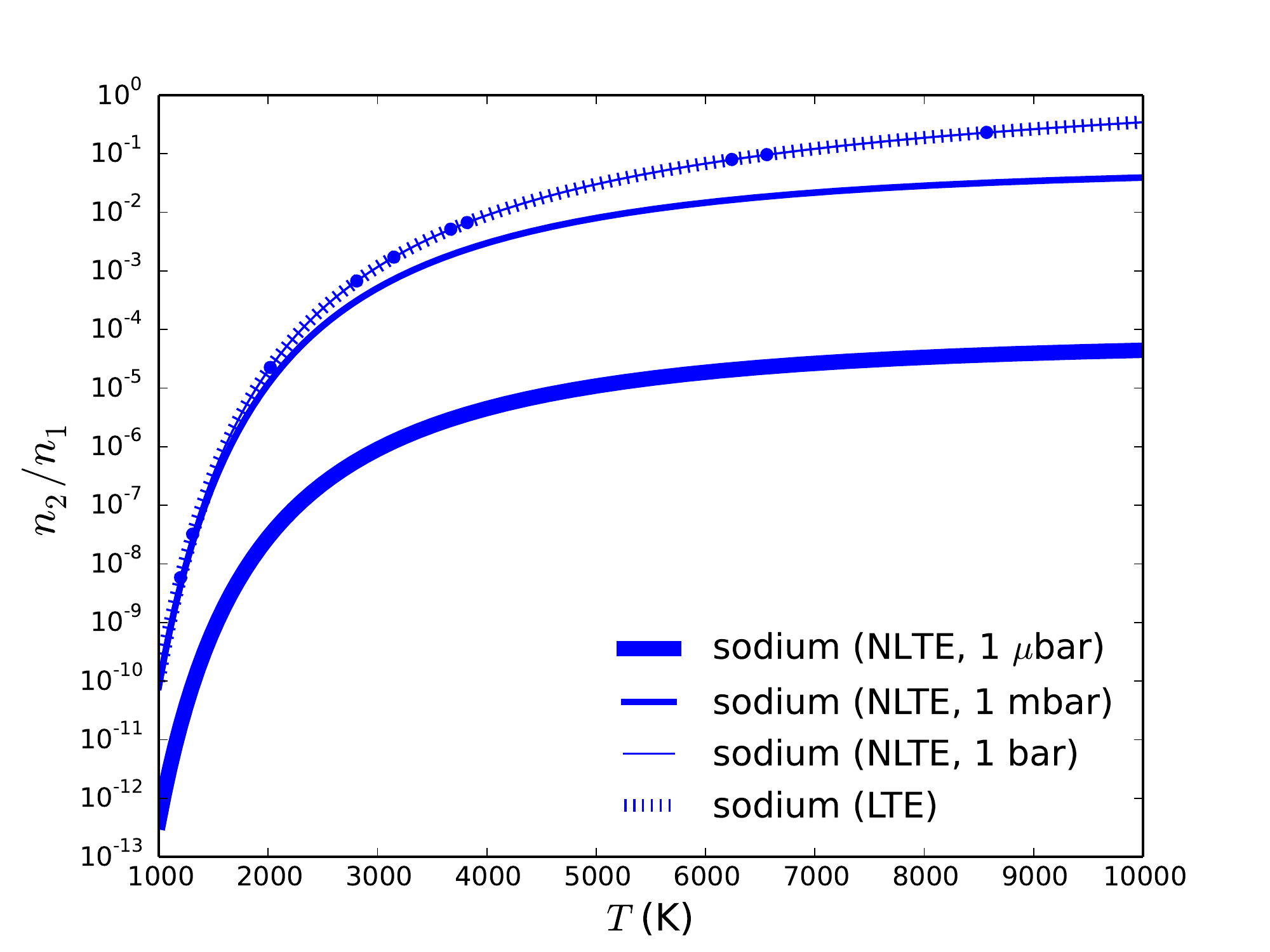}
\includegraphics[width=\columnwidth]{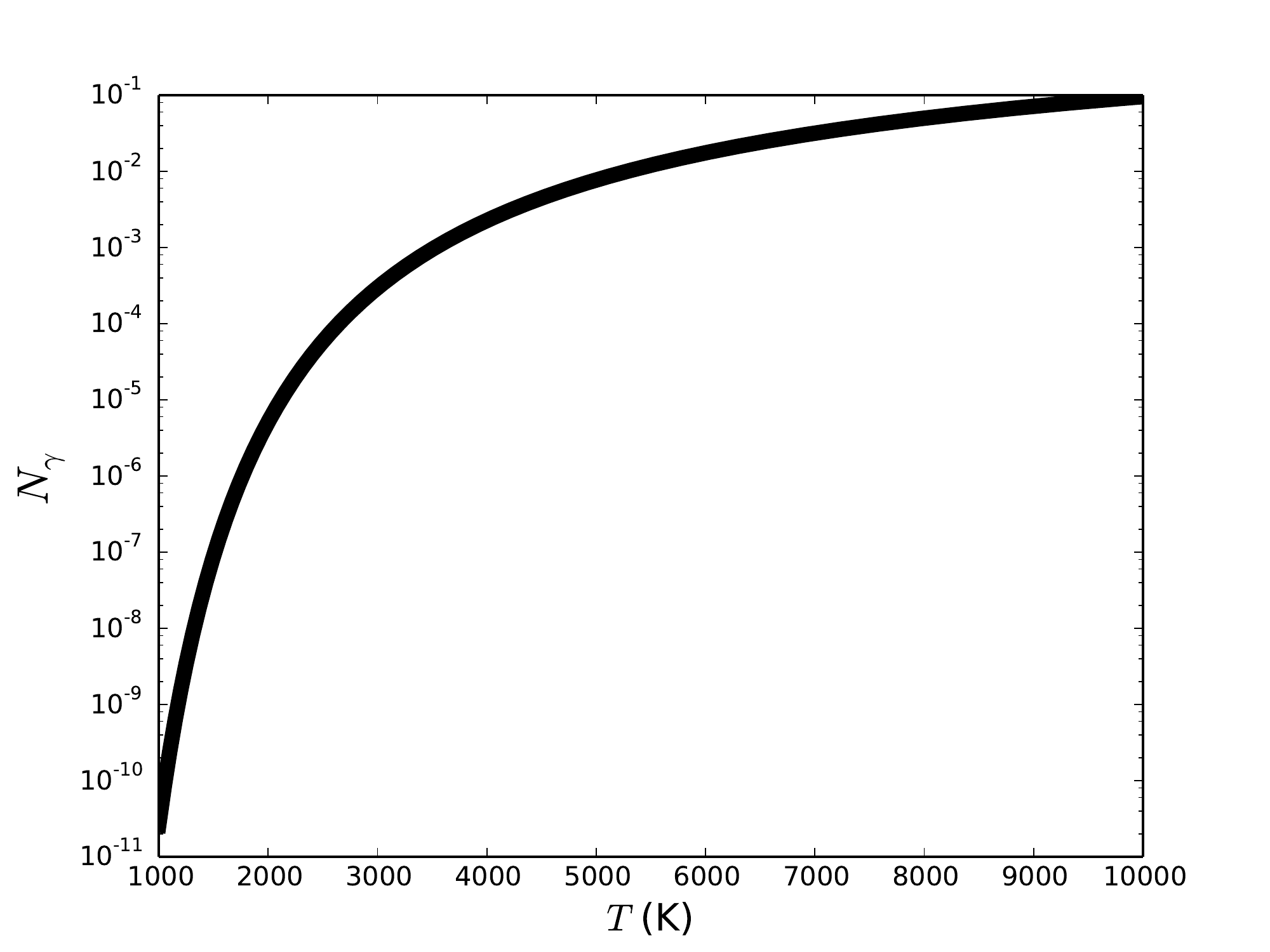}
\end{center}
\vspace{-0.1in}
\caption{Ratio of number densities of first excited level to ground state (top panel) and photon occupation number ($N_\gamma$; bottom panel) associated with the sodium lines.  As the Einstein A-coefficients and energy levels of the sodium doublet are very similar (see Table \ref{tab:sodium}), we take their average for illustration.  Also for illustration, we set $X_{\rm Na} = 10^{-6}$ (roughly the solar value) and chose three values of the pressure: 1 $\mu$bar, 1 mbar and 1 bar.  Only for a pressure of 1 bar is the ratio of number densities described by the LTE limit.  Since $N_\gamma \ll 1$, we can ignore both photoabsorption and stimulated emission (see text).}
%\vspace{-0.1in}
\label{fig:intro}
\end{figure}

Figure \ref{fig:intro} shows that the ratio of number densities of the first excited state of the sodium atom to its ground state may depart from the LTE value at the order-of-magnitude level, depending on the temperature and pressure.  The departure from LTE is non-negligible for pressures of 1 $\mu$bar and 1 mbar.  The LTE and NLTE ratios of number densities agree only at a pressure of 1 bar, as shown in \textbf{the top panel of} Figure \ref{fig:intro}.  Since transmission spectra are expected to probe pressures much less than 1 bar, this NLTE effect is worthy of investigation.  The departure from NLTE produces a correction factor, which reduces to unity in the limit of LTE, to the integrated strength of the sodium lines (as we will elucidate in \S\ref{sect:methods}).  When the sodium lines are in NLTE, the correction factor exceeds unity and produces weaker opacities.  Figure \ref{fig:trends} shows that the LTE model tends to over-estimate the overall strength of the sodium doublet in the transmission spectrum, regardless of whether clouds are present.

When applied to low-resolution Hubble Space Telescope (HST) transmission spectra \citep{sing16}, we find this NLTE effect produces minor corrections to the retrieved atmospheric properties, except for HD 189733b.  However, the interpretation of a high-resolution HARPS transmission spectrum of the hot Jupiter WASP-49b \citep{w17} is biased towards lower temperatures if LTE is assumed, as we will demonstrate in \S\ref{sect:results}.

\subsection{Are sub-Lorentzian line wings of sodium needed to interpret spectra?}

A classical approach to modeling the spectral lines of atoms or molecules is to assume that the shapes of these lines are described by the Voigt profile, which is a mathematical convolution of the Lorentz and Doppler profiles (e.g., Chapter 6.5 of \citealt{draine11}, Chapter 5.2 of \citealt{heng17}).  For the sodium line observed at $\sim 1000$ K, the damping parameter is low enough ($\sim 10^{-3}$) that the Voigt profile consists of a Doppler core and Lorentzian wings, if pressure broadening is ignored \citep{heng15}.  However, the studies of \cite{burrows00} and \cite{allard12} suggest that a more realistic description of the sodium line shape consists of sub-Lorentzian wings, even though these studies do not agree on the quantitative details of how to compute these line wings.  

\begin{figure*}
\vspace{-0.1in}
\begin{center}
\includegraphics[width=\columnwidth]{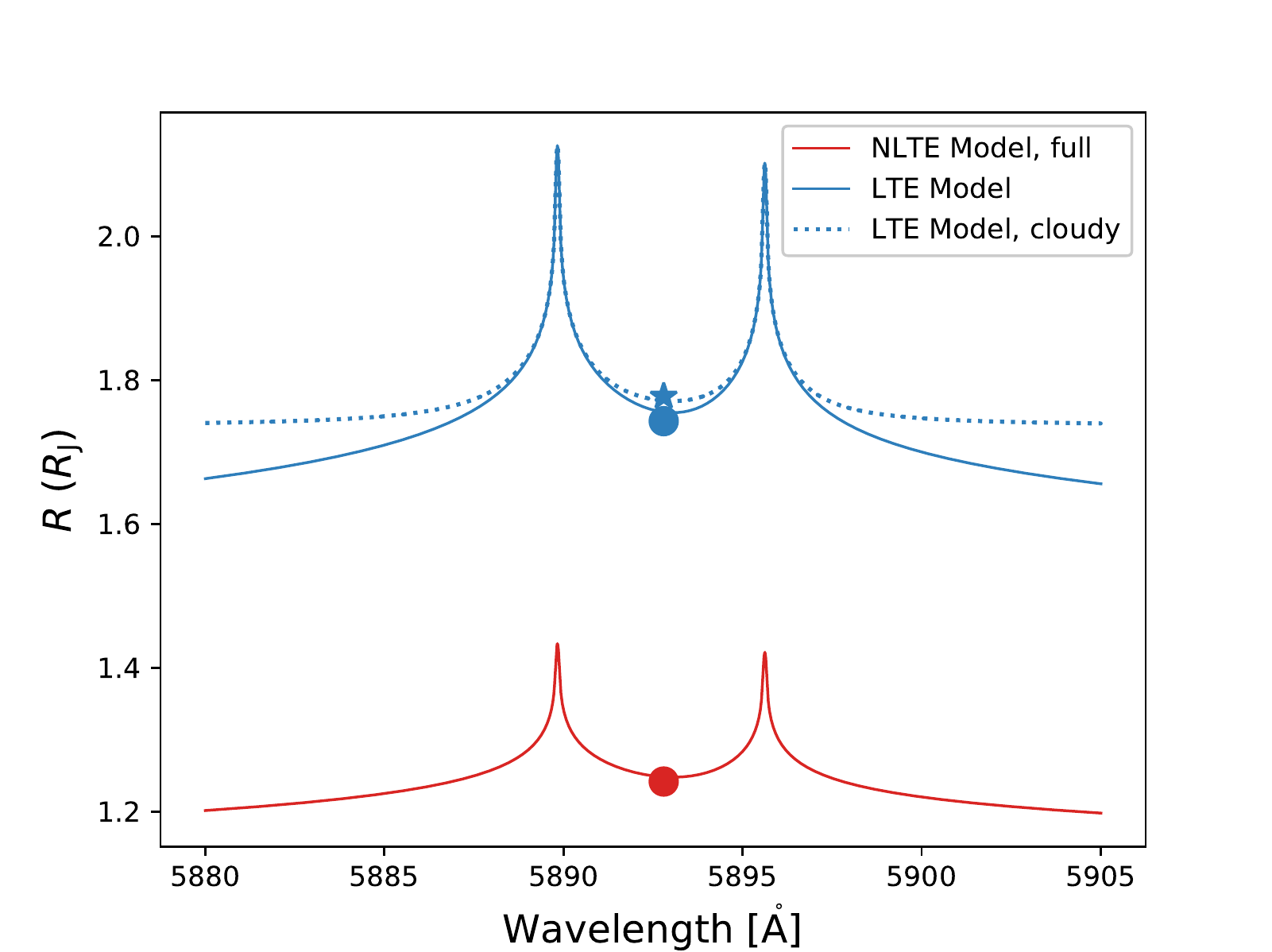}
\includegraphics[width=\columnwidth]{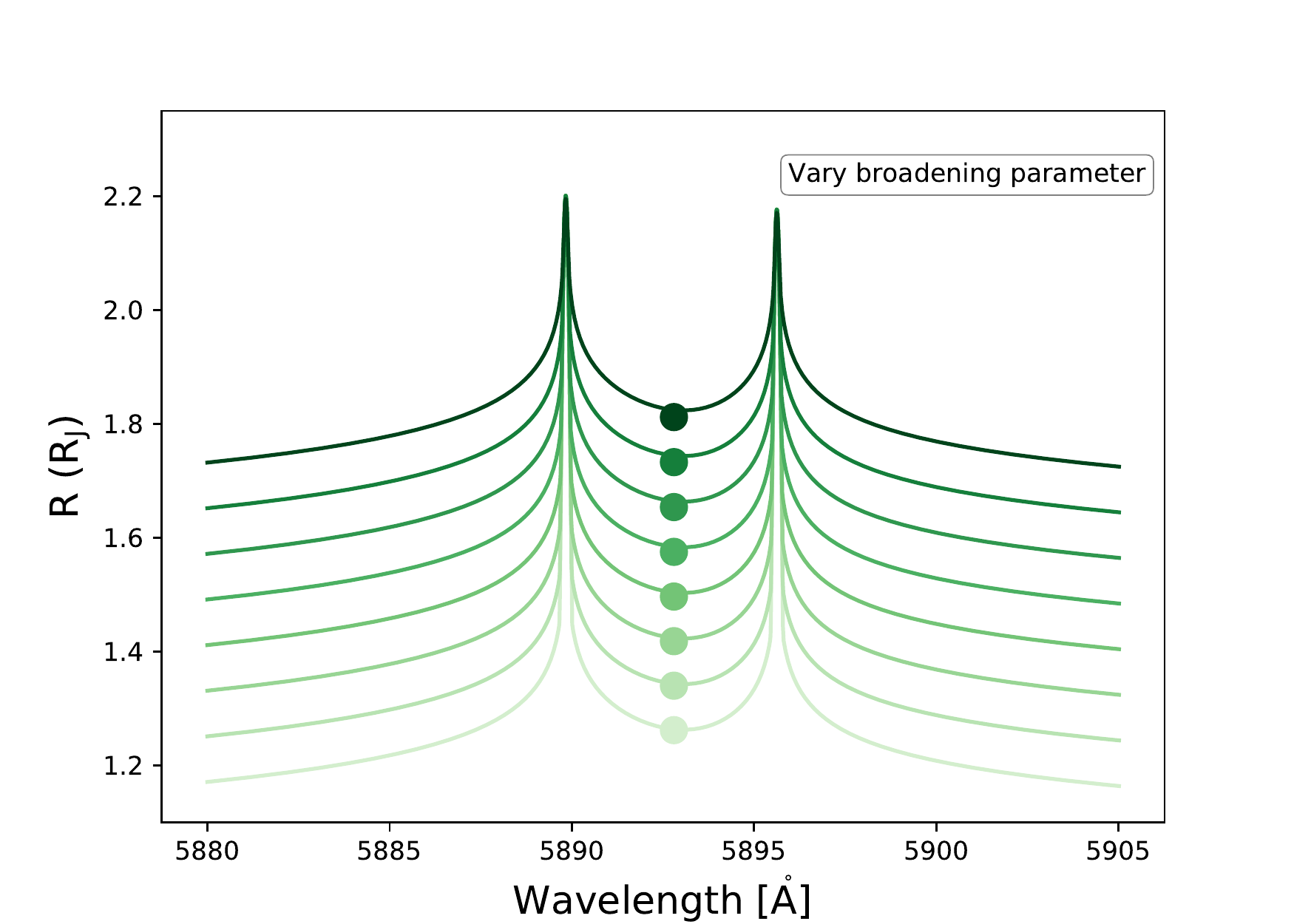}
\includegraphics[width=\columnwidth]{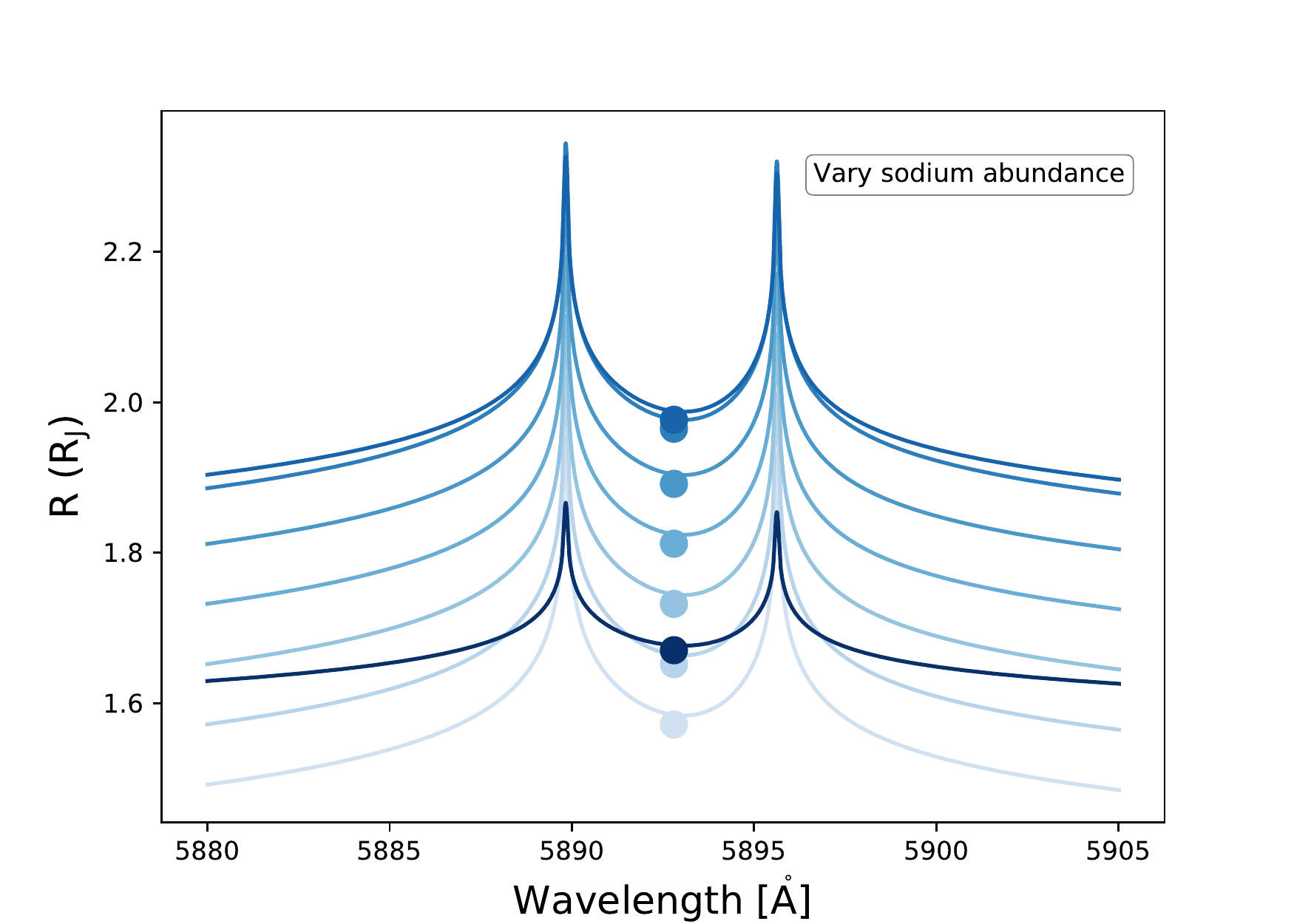}
\includegraphics[width=\columnwidth]{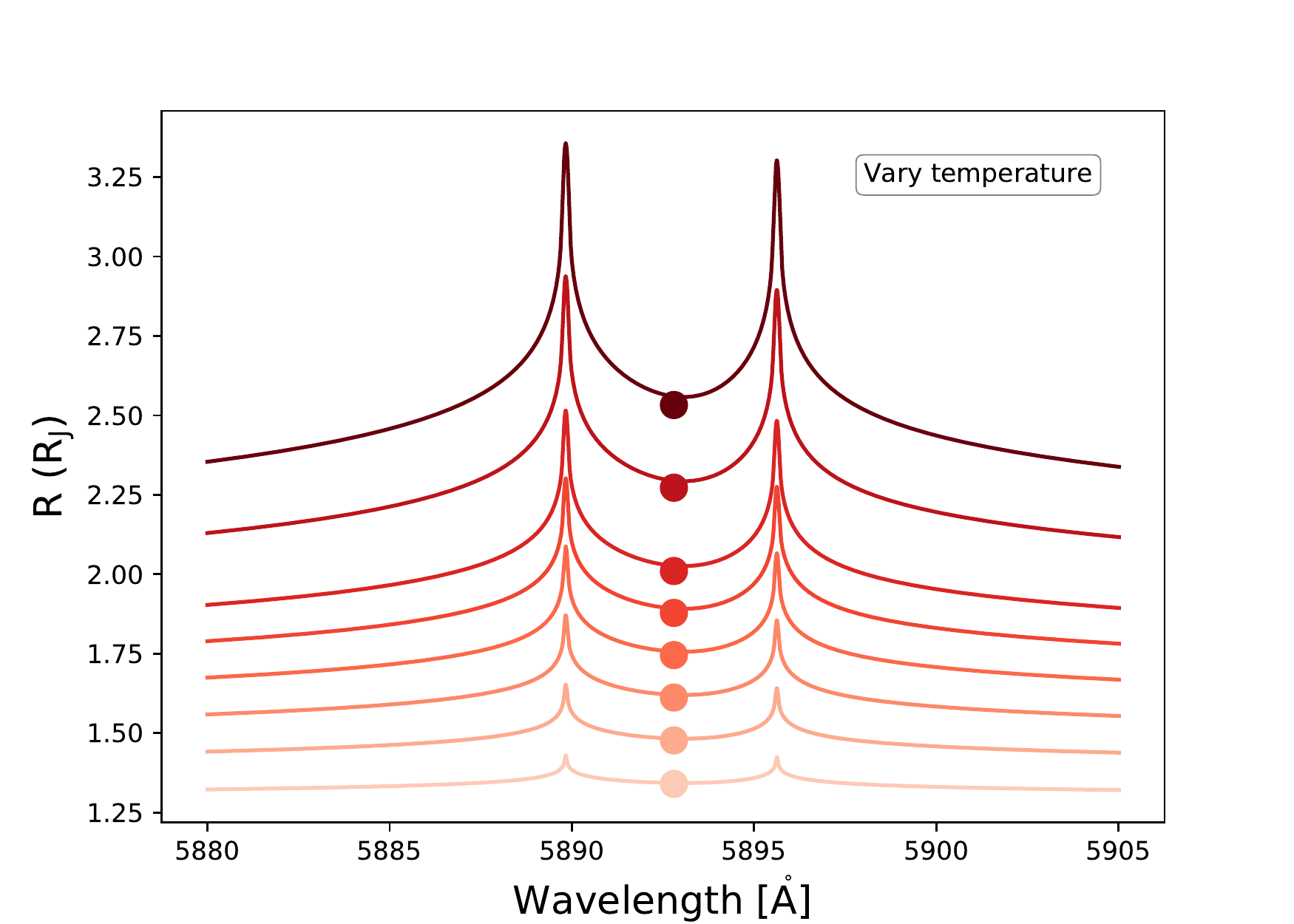}
\end{center}
\caption{Models of the transmission spectrum of the sodium doublet, assuming WASP-49b-like parameter values ($g=689$ cm$^2$ s$^{-1}$, $T=4500$ K).  For illustration, we assume $R_0=1.198 R_{\rm J}$ and $P_0=10$ bar for the theoretical normalization.  The fiducial model assumes $X_{\rm Na}=10^{-5}$ and $f_{\rm broad}=100$.  Top left panel: cloudfree NLTE versus cloudfree and cloudy ($\kappa_{\rm cloud}=100$ cm$^2$ g$^{-1}$) LTE models.  Top right panel: varying $f_{\rm broad}$ from $10^{-5}$ to $10^2$ with lighter colors corresponding to lower values of the broadening parameter.  Lower left panel: varying $X_{\rm Na}$ from $10^{-8}$ to $10^{-1}$ with lighter colors corresponding to lower sodium mixing ratios.  When $X_{\rm Na}=10^{-1}$ (darkest color), \textbf{the mean molecular mass increases beyond its hydrogen-dominated value ($\approx 2.4 m_{\rm amu}$) and the pressure scale height becomes small, leading to smaller spectral features.} Lower right panel: varying temperature from 1000 to 10,000 K, with lighter colors corresponding to lower temperatures. In all of the panels, the filled circle corresponds to the bandpass-averaged transit radius, except for the filled star which corresponds to that for the cloudy LTE model.}
\label{fig:trends}
\end{figure*}

From a phenomenological viewpoint, we wish to ask if sub-Lorentzian wings are a necessary ingredient for fitting transmission spectra of sodium lines, both at low and high resolutions? In the current study, we adopt an agnostic and data-driven approach, which is to describe the line wings of sodium by a dimensionless broadening parameter ($f_{\rm broad}$).  A classical Voigt profile with no pressure broadening present has $f_{\rm broad}=1$.  If pressure broadening is present, we have $f_{\rm broad}>1$. \textbf{ In high-resolution transmission spectra, a resolved sodium line with $f_{\rm broad}>1$ may also encode information associated with the Rossiter-McLaughlin effect and large-scale atmospheric winds (e.g., \citealt{lw15})}.  Any value of $f_{\rm broad}<1$ may be interpreted as the presence of sub-Lorentzian wings.  Figure \ref{fig:trends} shows that as the broadening parameter increases, the sodium lines become stronger in the wings (while the line-center strengths remain invariant).  One expects a degeneracy between the broadening parameter and the mixing ratio of sodium, because the strength of the line wings may be negated by assuming lower abundances of sodium (Figure \ref{fig:trends}).

From performing retrievals on low-resolution transmission spectra, we find that sub-Lorentzian wings are not necessary to explain the current state of the data (\S\ref{sect:results}).  This is a phenomenological, rather than theoretical, statement.

\subsection{What are the limitations of interpreting low- versus high-resolution spectra?}

There is a debate in the exoplanet literature concerning how to leverage the value of low- versus high-resolution data off each other to maximise the outcome of retrievals.  For example, \cite{brogi17} studied a joint emission spectrum of HD 209458b with data from the HST Wide Field Camera 3 (WFC3), the Spitzer Space Telescope Infrared Array Camera (IRAC) and the VLT CRIRES spectrograph.  These authors demonstrated that the posterior distributions of molecular abundances become narrower at the order-of-magnitude level when the high-resolution data are considered in the retrieval analysis of the low-resolution spectrum.  However, as the computations were expensive they were unable to explore the entire parameter space and instead used low-resolution retrievals to guide the exploration.  The conclusions of \cite{brogi17} on whether the dayside of HD 209458b is isothermal, as well as the retrieved range of metallicity values for the atmosphere, appear to be dependent on their choice of parametrization of the temperature-pressure profile.

As another example, \cite{pino18} combined low-resolution data from several HST instruments with high-resolution transmission spectra from HARPS to study the hot Jupiter HD 189733b, which includes the detection of the sodium doublet by \cite{w15}.  Instead of performing a retrieval, they fixed the temperature-pressure profile to that reported by \cite{w15}, which was itself derived from an isothermal treatment.  By paying attention to the line spread function of the different instruments, \cite{pino18} were able to fit the combined low- and high-resolution spectrum with a solar-composition model, i.e., the water, sodium and potassium abundances were held fixed.  \cite{pino18} explicitly acknowledged that they did not explore the normalization degeneracy \citep{bs12,g14,hk17}.

In the current study, we ask a different question: what are the limitations encountered when \textit{separately} interpreting low- versus high-resolution transmission spectra of the sodium doublet?  This question is non-trivial, because high-resolution, ground-based spectroscopy is a ``double differential" technique, where the absolute \textit{empirical} normalization of the transmission spectrum is lost during data calibration and reduction \citep{w15,w17,pino18}.  In other words, only the relative, and not the absolute, transit depths are measured.  Quantitatively, it is unclear how the enhanced spectral resolution (relative to, e.g., HST transmission spectra), diminished wavelength range and lack of an empirical normalization play off or negate one another.  Space-based HST transmission spectra cover a wider wavelength range and do possess an absolute empirical normalization, but may not encode enough information (due to the low spectral resolution) to completely break the normalization degeneracy \citep{fh18}, which states that the \textit{theoretical} normalization of the transmission spectrum is uncertain and leads to order-of-magnitude uncertainties in the chemical abundances \citep{bs12,g14,hk17}.

To this end, we apply our retrieval technique to the HST Space Telescope Imaging Spectrograph (STIS) transmission spectra curated by \cite{sing16}, which includes HAT-P-1b, HAT-P-12b, HD 189733b, WASP-6b, WASP-17b and WASP-39b. The sodium doublet is not resolved in these STIS spectra, but is rather detected as a single, unresolved spike of two blended lines in absorption.  As a single case study of high-resolution, ground-based spectra, we analyze the HARPS transmission spectrum of the hot Jupiter WASP-49b measured by \cite{w17}.  The sodium doublet is resolved in this HARPS spectrum.  Besides the deep transits associated with the sodium doublet (about 2\% relative to the continuum), \textbf{the spectral signature of the Rossiter-McLaughlin effect was not detected} \citep{w17}, which makes WASP-49b an ideal high-resolution case study.

Previously, \cite{fh18} demonstrated that HST WFC3 near-infrared transmission spectra encode enough information to partially break the normalization degeneracy.  Specifically, the caveat is that the lower limits to the posterior distributions correspond to the maximum value assumed for the prior distribution of the reference pressure.  The HST STIS transmission spectra of \cite{sing16} cover the optical/visible range of wavelengths, where it has previously been established (e.g., \citealt{lec08,fh18}) that if the optical/visible spectral slope may be uniquely attributed to molecular hydrogen alone then the normalization degeneracy is completely broken.  However, the expectation is that if the optical/visible spectral slope is due to clouds or hazes, then the normalizaton degeneracy cannot be broken \citep{heng16}.  In the current study, we interpret the STIS transmission spectra by assuming that absorption by sodium atoms, Rayleigh scattering associated with small cloud particles and gray absorption by large cloud particles are present.  We find that the normalization degeneracy cannot be broken in the context of such a model interpretation using optical/visible data alone.

For the high-resolution spectra, we find that the theoretical normalization cannot be uniquely determined, but given the lack of an empirical normalization this has little to no effect on the posterior distributions of the retrieved atmospheric properties.

\subsection{Is the degree of cloudiness of an exoplanetary atmosphere correlated with any exoplanet property?}

Previously, \cite{s16} and \cite{heng16} used WFC3 transmission spectra dominated by water and STIS observations of the sodium doublet, respectively, to study the degree of cloudiness of exoplanetary atmospheres and if it is correlated with any exoplanet property.  Collectively, these studies find that hotter objects \citep{heng16,s16} with higher gravity \citep{s16} are more likely to have cloudfree atmospheres.

One of the objectives of the present study is to improve upon and revisit the study of \cite{heng16}, who designed a diagnostic to interpret the sodium and potassium lines observed in transmission spectra at low resolution, building on the theory developed in \cite{heng15}.  It is based on the reasoning that, since these are strong resonant lines, the line peaks are unaffected by the opacities of clouds or hazes, while the line wings are significantly affected, thus allowing for the peak-to-wing distance to be used as a diagnostic for the degree of cloudiness.  The diagnostic was constructed to apply at two points: one at the line peak, one in the line wing.  This two-point approach was previously applied to the high-resolution HARPS spectrum of WASP-49b \citep{w17}.  

A more accurate approach is to perform a retrieval that fits for the partially resolved shape of the sodium lines.  From the fit, one can compute the distance between the line peak and wing, $\Delta R_{\rm fit}$.  One can then remove cloud opacity from the model and compute the corresponding distance in a cloudfree atmosphere, $\Delta R_{\rm fit, CF}$.  By taking the ratio of these quantities, we obtain the cloudiness index \citep{heng16},
\begin{equation}
C = \frac{\Delta R_{\rm fit, CF}}{\Delta R_{\rm fit}},
\end{equation}
such that cloudfree and cloudy atmospheres have $C=1$ and $C>1$, respectively.  The reason to use the best-fit model, rather than the actual data points, in the denominator of $C$ is to avoid situations where an imperfect fit to the measured transmission spectrum produces a best-fit model line peak that sits below the data point, which produces spurious values of $C<1$.  We further investigate if estimates of $C$ are affected by the assumption of LTE.

\subsection{Layout of paper}

\begin{table}
         \caption{Relevant physical properties of Na D lines}
         \label{tab:sodium}
     $$ 
         \begin{array}{ll}
            \hline
            \noalign{\smallskip}
            \mathrm{Quantity} & \mathrm{Value} \\
            \noalign{\smallskip}
            \hline
            \noalign{\smallskip}
		\lambda_{\rm D_1} ~(\AA) & 5895.92424 \\
		\nu_{\rm D_1} ~(\mathrm{Hz}) & 5.085 \times 10^{14} \\
		\lambda_{\rm D_2} ~(\AA) & 5889.95095 \\
		\nu_{\rm D_2} ~(\mathrm{Hz}) & 5.090 \times 10^{14} \\
		A_{\rm D_1} ~(\mathrm{s}^{-1}) & 6.14 \times 10^7 \\
		A_{\rm D_2} ~(\mathrm{s}^{-1}) & 6.16 \times 10^7 \\
		E_{\rm D_1} ~(\mathrm{erg}) & 3.369 \times 10^{-12} \\
		E_{\rm D_1} ~(\mathrm{eV}) & 2.1 \\
		E_{\rm D_2} ~(\mathrm{erg}) & 3.373 \times 10^{-12} \\
		E_{\rm D_2} ~(\mathrm{eV}) & 2.1 \\
		f_{\rm D_1} & 0.320 \\
		f_{\rm D_2} & 0.641 \\
		\hline
		
            \noalign{\smallskip}
            \hline
         \end{array}
     $$ 
     All data extracted from NIST database except for oscillator strengths ($f_{\rm D_i}$), which are taken from Table 9.3 of \cite{draine11}.  $A_{\rm D_1}$ and $A_{\rm D_2}$ are the $A_{21}$ values for the D$_1$ and D$_2$ lines, respectively.
     \vspace{0.1in}
   \end{table}

\begin{table*}
\begin{center}
\caption{Retrieved parameters and their prior ranges}
\label{tab:priors}
\begin{tabular}{lccccccccc}
\hline
Quantity & Symbol & Units & High-res Range & Low-res Range & Prior Type \\
\hline
Temperature & $T$ & K & 0--10,000 & 0--6000 & uniform \\
Sodium volume mixing ratio & $X_{\rm Na}$ & -- & $10^{-13}$--1 & $10^{-13}$--1 & log-uniform \\
Broadening parameter & $f_{\rm broad}$ & -- & $10^{-1}$--$10^4$ & $10^{-5}$--$10^{10}$ & log-uniform \\
Gray cloud opacity & $\kappa_{\rm cloud}$ & cm$^2$ g$^{-1}$ & $10^{-4}$--$10^{5}$ & $10^{-10}$--$10^{5}$ & log-uniform \\
Rayleigh scattering scaling parameter (clouds) & $f_{\rm cloud}$ & -- & -- & 1--$10^7$ & log-uniform \\
Reference pressure & $P_0$ & bar & 1--100 & 1--100 & log-uniform \\
\hline
\hline
\end{tabular}\\
Note: $1 \mbox{ bar} = 10^6$ dyn cm$^{-2}$ (cgs units) = $10^5$ N m$^{-2}$ (mks units).
\end{center}
\end{table*}

In Section \ref{sect:methods}, we lay out our methodology, including the theory behind the two-level atom treatment, the cross section of the sodium lines, models of the transmission spectrum and treatments of the high-resolution data format.  In Section \ref{sect:results}, we study a suite of mock retrievals and perform retrieval analyses on the HST STIS and HARPS transmission spectra.  In Section \ref{sect:discussion}, we discuss the implications of our results, including a scrutiny of the information content of high-resolution spectra with respect to detecting NLTE effects.  Table \ref{tab:sodium} contains the physical constants associated with the sodium doublet.  Table \ref{tab:priors} states the physical units and prior ranges of the retrieved parameters.  Table \ref{tab:retrievals} summarizes the values of the retrieved atmospheric properties for the sample of 7 exoplanets considered in the current study.

\section{Theory \& Methodology}
\label{sect:methods}

\subsection{Two-level sodium atom}

The two-level atom is not a novel concept and has previously been described in monographs (e.g., Chapter 17.1 of \citealt{draine11}).  Here, we recast it in a form that is suitable for computation and application to sodium atoms.

The two-level atom has a ground state and an excited state.  The number density of atoms in the excited state ($n_2$) is described by the following evolution equation (e.g., equation [17.5] of \citealt{draine11}),
\begin{equation}
\begin{split}
\frac{\partial n_2}{\partial t} =& \left( n_e + n_{\rm total} \right) n_1 C_{12} + \frac{n_1 N_\gamma g_2 A_{21}}{g_1}\\
&- \left( n_e + n_{\rm total} \right) n_2 C_{21} - n_2 A_{21} \left(1 + N_\gamma \right),
\label{eq:twolevel}
\end{split}
\end{equation}
where $t$ denotes the time, $n_1$ is the number density of atoms in the ground state, $n_e$ is the number density of electrons, $n_{\rm total}$ is the total number density (which is the number density of hydrogen molecules in a hydrogen-dominated atmosphere), $A_{21}$ is the Einstein A-coefficient (with units of s$^{-1}$) and $N_\gamma$ is the photon occupation number.  The statistical weights are $g_j = 2j^2$, where $j=1,2$.  The rate coefficent for collisional de-excitation is given by equation (2.27) of \cite{draine11},
\begin{equation}
C_{21} = 8.629 \times 10^{-8} \mbox{ cm}^3 \mbox{ s}^{-1} ~T_4^{-1/2} ~\frac{\Omega_{21}}{g_2},
\end{equation}
where $T_4 \equiv T / 10^4$ K and $T$ is the temperature.  As explained by \cite{draine11}, the collision strength $\Omega_{21}$ is approximately independent of temperature for $T<10^4$ K and typically has values between 1 and 10.  To maximise the NLTE effect, we approximate $\Omega_{21}\approx 1$, which produces the highest possible value of the critical density.

Physically, besides the collisional excitation ($n_{\rm total} n_1 C_{12}$) and de-excitation ($n_{\rm total} n_2 C_{21}$) of the excited level as well as de-excitation by spontaneous emission ($n_2 A_{21}$), there is also photoabsorption ($n_1 N_\gamma g_2 A_{21}/g_1$) and stimulated emission ($n_2 N_\gamma A_{21}$), as described in Chapter 17.1 of \cite{draine11}.  However, we show in \textbf{the bottom panel of} Figure \ref{fig:intro} that the photon occupation number,
\begin{equation}
N_\gamma = \left( e^{h \nu / k_{\rm B} T} - 1 \right)^{-1},
\end{equation}
with $h$ being Planck's constant, $k_{\rm B}$ being Boltzmann's constant and $\nu$ being the frequency, is much less than unity, implying that the terms associated with $N_\gamma$ in equation (\ref{eq:twolevel}) may be dropped.  Note that the preceding expression for $N_\gamma$ assumes a Planckian radiation field.

Demanding a steady state ($\partial n_2 / \partial t = 0$) yields
\begin{equation}
\frac{n_2}{n_1} = \frac{C_{12}}{C_{21}} ~\left( 1 + f_e \right) ~\left( 1 + f_e +  \frac{n_{\rm crit}}{n_{\rm total}} \right)^{-1},
\end{equation}
where $E_{12}$ is the energy difference between the ground and excited levels and $f_e \equiv n_e/n_{\rm total}$ is the ionization fraction.  For $f_e \ll 1$, we obtain the standard NLTE expression for a two-level atom,
\begin{equation}
\frac{n_2}{n_1} = \frac{g_2}{g_1} ~e^{-E_{12} / k_{\rm B} T} ~\left( 1 +  \frac{n_{\rm crit}}{n_{\rm total}} \right)^{-1}.
\label{eq:twolevel2}
\end{equation}
To arrive at equation (\ref{eq:twolevel2}), we invoke the principle of detailed balance (e.g., Section 3.5 of \citealt{draine11}),
\begin{equation}
\frac{C_{12}}{C_{21}} = \frac{g_2}{g_1} ~e^{-E_{12} / k_{\rm B} T}.
\end{equation}
The critical density is defined as
\begin{equation}
n_{\rm crit} \equiv \frac{A_{21}}{C_{21}}.
\end{equation}
When $n_{\rm total} \gg n_{\rm crit}$ (i.e., collisions are important), we recover the Boltzmann distribution and the states obey LTE.  At a pressure of 1 bar, $n_2/n_1$ is well described by its LTE limit for temperatures between $10^3$--$10^4$ K (Figure \ref{fig:intro}).  However, for lower pressures, the departures from LTE may become significant at higher temperatures.

For completeness, Appendix \ref{append:saha} describes an analytical expression for the electron density if all of the electrons are being sourced by collisional ionization of the sodium atom.

\subsection{Cross section of sodium lines}

The absorption cross section of the sodium atom is the product of the integrated line strength ($S$) and the line shape ($\Phi$),
\begin{equation}
\sigma_{\rm Na} = S \Phi.
\label{eq:cross_section}
\end{equation}
There are three different ways to write $S$, depending on if the line shape is written in per wavenumber, per frequency or per wavelength units.  We use per frequency units, such that $S$ has units of cm$^2$ s$^{-1}$ and $\Phi$ has units of s.

\subsubsection{Integrated line strength}

The integrated line strength is \citep{penner52}
\begin{equation}
S = \frac{n_2 A_{21} h \nu}{\epsilon c n_{\rm total}}.
\label{eq:ss}
\end{equation}
Following Appendix 2 of \cite{gy89}, a factor of $1/n_{\rm total}$ has been inserted to give $S$ the correct physical units, such that $\sigma_{\rm Na}$ has units of cm$^2$.  The energy density per unit frequency is given by 
\begin{equation}
\epsilon = \frac{4\pi B_\nu}{c},
\end{equation}
where $c$ is the speed of light and
\begin{equation}
B_\nu = \frac{2 h \nu^3}{c^2} \left( e^{h \nu / k_{\rm B} T} - 1 \right)^{-1}
\end{equation}
is the Planck function in per frequency units.  The Einstein A-coefficient is related to the oscillator strength, $f_{12}$, by (e.g., equation [6.20] of \citealt{draine11})
\begin{equation}
A_{21} = \frac{8 \pi^2 e^2 \nu^2}{m_e c^3} \frac{g_1}{g_2} f_{12},
\end{equation}
where $e$ is the elementary charge and $m_e$ is the mass of the electron.

It follows that the integrated line strength becomes
\begin{equation}
S = \frac{\pi e^2 f_{12}}{m_e c} \frac{n_1}{n_{\rm total}} \left( 1 - e^{-h \nu / k_{\rm B} T} \right) \left( 1 + \frac{n_{\rm crit}}{n_{\rm total}} \right)^{-1}.
\end{equation}
There are several noteworthy aspects of the preceding expression.  First, since $N_\gamma \ll 1$ and $n_1 \approx n_{\rm total}$, we obtain
\begin{equation}
S = \frac{\pi e^2 f_{12}}{m_e c} \left( 1 + \frac{n_{\rm crit}}{n_{\rm total}} \right)^{-1},
\label{eq:ss2}
\end{equation}
which is our final NLTE expression for the two-level atom.  Second, in the LTE limit we obtain $S \approx \pi e^2 f_{12} / m_e c$, which is the same as equation (6.25) of \cite{draine11}.  Third, since equation (\ref{eq:ss}) is the same starting point for deriving the standard expression for the line strength in HITRAN / HITEMP, it implies that equation (\ref{eq:ss2}) is equivalent to equation (A5) of \cite{rothman98}.

\subsubsection{Line shape}

The Voigt profile is standard knowledge in monographs (e.g., Chapter 6.5 of \citealt{draine11}, Chapter 5.2 of \citealt{heng17}).  However, there is more than one way to define the Lorentz and Doppler widths.  Here, we concisely restate the formalism to be explicit about the conventions adopted in the current study.

The Lorentz profile is
\begin{equation}
\Phi_{\rm L} = \frac{\Gamma_{\rm L} / \pi}{\left( \nu - \nu_0 \right)^2 + \Gamma_{\rm L}^2},
\end{equation}
where $\nu_0$ is the frequency at line center and $\Gamma_{\rm L} = f_{\rm broad} A_{21}/4\pi$ is the half-width at half-maximum (HWHM) of the Lorentz profile.  This convention differs from, for example, \cite{draine11} who defines it as the full-width at half-maximum.  We use the dimensionless broadening parameter $f_{\rm broad}$, which expresses the strength of the Lorentzian wings in terms of the natural width, as an agnostic way of including all sources of broadening.  In particular, since the theory of pressure broadening is incomplete, the use of $f_{\rm broad}>1$ allows us to report retrieved broadening values in terms of the natural width, which will motivate future studies of the sodium line shape.  It also allows us to isolate the pressure dependence of the cross section due to the NLTE term associated with the total density.  The use of $f_{\rm broad}<1$ allows us to quantify the extent to which the line wings are sub-Lorentzian.  We do not truncate the sodium line wings.

The Doppler profile is
\begin{equation}
\Phi_{\rm D} = \Gamma^{-1}_{\rm D} \sqrt{\frac{\ln{2}}{\pi}} e^{-\left(\nu-\nu_0\right)^2 / \sigma_{\rm th}^2}.
\end{equation}
The thermal speed is $v_{\rm th} = \sqrt{2 k_{\rm B} T / m}$, while the Doppler shift associated with the thermal speed is $\sigma_{\rm th} = \nu_0 v_{\rm th}/c$.  $\Gamma_{\rm D} = \sigma_{\rm th} \sqrt{\ln{2}}$ is the HWHM of the Doppler profile.

By convolving the Lorentz and Doppler profiles, the Voigt profile obtains,
\begin{equation}
\Phi = \sqrt{\frac{\ln{2}}{\pi}} \frac{H_{\rm V}}{\Gamma_{\rm D}},
\end{equation}
where the Voigt H-function is
\begin{equation}
H_{\rm V} \equiv \frac{a_0}{\pi} \int^{+\infty}_{-\infty} \frac{e^{-y^2}}{\left(x-y\right)^2 + a^2_0} ~dy,
\end{equation}
and $x \equiv (\nu-\nu_0)/\sigma_{\rm th}$.  The damping parameter is defined as $a_0 \equiv \Gamma_{\rm L}/\sigma_{\rm th}$.

In theory, the Voigt H-function is evaluated using \citep{ro05,z07}
\begin{equation}
\begin{split}
H_{\rm V}  =& ~e^{a^2_0 - x^2} ~\texttt{erfc}\left(a_0\right) ~\cos\left(2a_0x\right) \\
&+ \frac{2}{\sqrt{\pi}} \int^x_0 e^{y^2-x^2} \sin{\left[ 2a_0 \left(x-y\right)\right]} ~dy,
\end{split}
\end{equation}
where $\texttt{erfc}(a_0)$ is the complementary error function (e.g., Chapter 10 of \citealt{aw}) with $a_0$ as its argument.

In practice, the Voigt profile is evaluated using \citep{schreier92}
\begin{equation}
H_{\rm V} = \Re \left[ w\left(z\right) \right], 
\end{equation}
where $w(z)$ is the Faddeeva function (equation [7.1.3]\footnote{The Faddeeva function appears without being specifically attributed to Faddeeva in \cite{abram}.} of \citealt{abram}) evaluated at \citep{schreier92}
\begin{equation}
z = x + i a_0.
\end{equation}
This representation of the Voigt function allows for faster and simpler computation of the line shape.

\subsection{Transmission spectrum}

\subsubsection{LTE}

At a given wavelength or frequency, the expression for the transit radius corresponding to an isothermal transit chord \citep{lec08,ds13,bs17,hk17,je18} is
\begin{equation}
R = R_0 + H \left[ \gamma + \ln{\left( \frac{P_0 \kappa}{g} \sqrt{\frac{2 \pi R_0}{H}} \right)} \right],
\label{eq:transit}
\end{equation}
where $H = k_{\rm B} T / m g$ is the isothermal pressure scale height, $m$ is the mean molecular mass, $g$ is the gravity, $\gamma = 0.57721$ is the Euler-Mascheroni constant and $\kappa$ is the opacity (cross section per unit mass).  The reference pressure, $P_0$, corresponds to a reference transit radius, $R_0$. 

For our model atmospheres, the opacity is
\begin{equation}
\kappa = \frac{X_{\rm Na} \pi e^2 f_{12}}{m m_e c} \Phi + \kappa_{\rm cloud},
\end{equation}
where $X_{\rm Na}$ is the volume mixing ratio of sodium and $\kappa_{\rm cloud}$ is a gray/constant cloud opacity.  The assumption is the wavelength range probed is narrow enough that it is insensitive to changes in cloud opacity caused by small particles, or that the particles are large (compared to the wavelength).

Denoting the atomic mass unit by $m_{\rm amu}$, the mean molecular mass is given by
\begin{equation}
m = 2.4 X_{\rm H_2} m_{\rm amu} +  X_{\rm Na} m_{\rm Na},
\label{eq:mmm}
\end{equation}
where $m_{\rm Na}= 23 m_{\rm amu}$ is the mass of the sodium atom.  The mixing ratio of molecular hydrogen is determined by demanding that all mixing ratios sum to unity,
\begin{equation}
1.1 X_{\rm H_2} + X_{\rm Na} = 1,
\end{equation}
where we have assumed that the helium mixing ratio follows cosmic abundance ($X_{\rm He} = 0.1 X_{\rm H_2}$).  When $X_{\rm Na} \ll 1$, we have $X_{\rm H_2} \approx 0.91$.  \textbf{A continuum of possibilities, from hydrogen-dominated to sodium-dominated atmospheres, is allowed.}

\subsubsection{NLTE}
\label{subsect:NLTE_R}

The NLTE expression describing the number densities of the excited and ground states of sodium explicitly contains the pressure $P$, because $n_{\rm total} = P/k_{\rm B} T$ if we assume the ideal gas law.  This means that we need to explicitly account for the variation of pressure across the sodium line profile.  To keep the problem tractable, we assume hydrostatic equilibrium, $P=P_0 \exp{[(R_0-R)/H]}$.  One of the assumptions involved in deriving equation (\ref{eq:transit}) is worth emphasizing.  Early in the derivation, one assumes that, in writing the optical depth as an integral involving the cross section and number density, one may bring the cross section out of the integral, i.e., assume that it is a constant with respect to the integration variable.  Since the integral is carried out over the transit chord, this is equivalent to assuming that the cross section is constant across the chord at a given wavelength.  It does not preclude the fact that the pressure probed by the transit chord may be rather different at different wavelengths.  In that sense, equation (\ref{eq:transit}) is only isobaric at a given wavelength and may be used to describe spectral lines where the pressure probed across the line (across different wavelengths) may vary markedly.

The opacity is given by
\begin{equation}
\kappa = \kappa_{\rm LTE}  \left( 1 + \frac{n_{\rm crit} k_{\rm B} T}{P} \right)^{-1},
\end{equation}
where we have defined
\begin{equation}
\kappa_{\rm LTE} \equiv \frac{X_{\rm Na} \pi e^2 f_{12}}{m m_e c} \Phi + \kappa_{\rm cloud}.
\end{equation}
We assume that the particles associated with the gray cloud opacity track the behavior of the sodium atoms with regards to departures from LTE.

Substituting the expression for the opacity into equation (\ref{eq:transit}) yields
\begin{equation}
R = R_{\rm LTE} - H \ln{ \left( 1 + \frac{n_{\rm crit} k_{\rm B} T}{P} \right) },
\label{eq:r_intermediate}
\end{equation}
where we have defined
\begin{equation}
R_{\rm LTE} \equiv R_0 + H \left[ \gamma + \ln{\left(\frac{P_0 \kappa_{\rm LTE}}{g} \sqrt{\frac{2\pi R_0}{H}} \right)} \right].
\end{equation}
This way of writing equation (\ref{eq:r_intermediate}) means that the two effects of $P_0$ have been separated: its regular appearance in $R_{\rm LTE}$ versus its appearance in the NLTE term, which is related to the varying importance of collisions across the sodium line.

By defining
\begin{equation}
x \equiv e^{(R-R_0)/H},
\end{equation}
equation (\ref{eq:r_intermediate}) may be written as a quadratic equation,
\begin{equation}
A x^2 + x - x_0 = 0,
\end{equation}
where $A \equiv n_{\rm crit} k_{\rm B} T/P_0$ and $x_0 \equiv \exp{[(R_{\rm LTE} - R_0)/H]}$.  By solving this quadratic equation, we obtain a closed-form solution for the transit radius,
\begin{equation}
R = R_0 + H \ln{\left( \frac{\sqrt{1 + 4 A x_0} - 1}{2A} \right)}.
\end{equation}
It is apparent that the dependence of the transit radius on $A$, and hence the $P_0$ term responsible for the varying importance of collisions across the sodium line, is weak.

\subsubsection{Values used for the reference transit radius and reference pressure}

Following \cite{fh18}, we estimate the reference transit radius using
\begin{equation}
R_0 = \bar{R} - 6.908 H.
\end{equation}
In \cite{fh18}, $\bar{R}$ is the average transit radius in the WFC3 bandpass.  Here, we take $\bar{R}$ to be the HST whitelight radius as stated in \cite{sing16}.

Based on WASP-17b, \cite{fh18} estimated that the WFC3 and STIS bandpasses probe $\sim 10$ mbar.  This implies that $P_0 \sim 10$ bar.  For the low-resolution HST STIS retrievals, we therefore set a tight, log-uniform prior of $1 \le P_0 \le 100$ bar.

For the high-resolution spectrum of WASP-49b, we match the spectral continuum to the HST whitelight radius of $1.198 ~R_{\rm J}$ \citep{w17}, where $R_{\rm J}$ is the radius of Jupiter. This requires setting $R_0 = 0.752 R_{\rm J}$ for $P_0 = 10$ bar.

\subsection{Nested-sampling retrievals and Bayesian evidence}

Our exploration of the parameter space of models is performed using a nested-sampling framework \citep{feroz08,feroz09,feroz13}.  We use the open-source \texttt{PyMultiNest} package \citep{buchner14}, which was previously implemented in \cite{fh18}.  Nested sampling allows us to compute the Bayesian evidence and Bayes factor, which allows us to identify the best model given the quality of the data \citep{trotta08}.  It also allows us to identify the family of models consistent with the data (e.g., \citealt{fh18}).

\subsection{Equivalence with \cite{heng16}}

Let the transit radius at line center and wing be $R_{\rm c}$ and $R_{\rm w}$, respectively.  Let also $\Delta R \equiv R_{\rm c} - R_{\rm w}$.  It follows from equation (\ref{eq:transit}) that
\begin{equation}
\Delta R = H \ln{\left( \frac{\Phi_{\rm c}}{\Phi_{\rm w}} \right)},
\end{equation}
where $\Phi_{\rm c}$ and $\Phi_{\rm w}$ are the Voigt profiles at line center and wing, respectively.  If we approximate the Voigt profile as consisting of a Doppler core and Lorentzian wings, then we recover equation (10) of \cite{heng16},
\begin{equation}
\Delta R = H \ln{\left[ \frac{4 \pi^2 c^2 \left( \lambda - \lambda_0 \right)^2 \lambda_0 }{ A_{21} \lambda^2 \lambda^2_0 \sqrt{2\pi H g} } \right]}.
\end{equation}
Thus, the pointwise approach of \cite{heng16}, as designed for analyzing low-resolution spectra, is a limiting case of the continuous approach given by equation (\ref{eq:transit}) that we are applying in the current study to high-resolution spectra.

\begin{figure*}
\centering
\includegraphics[width=0.9\columnwidth]{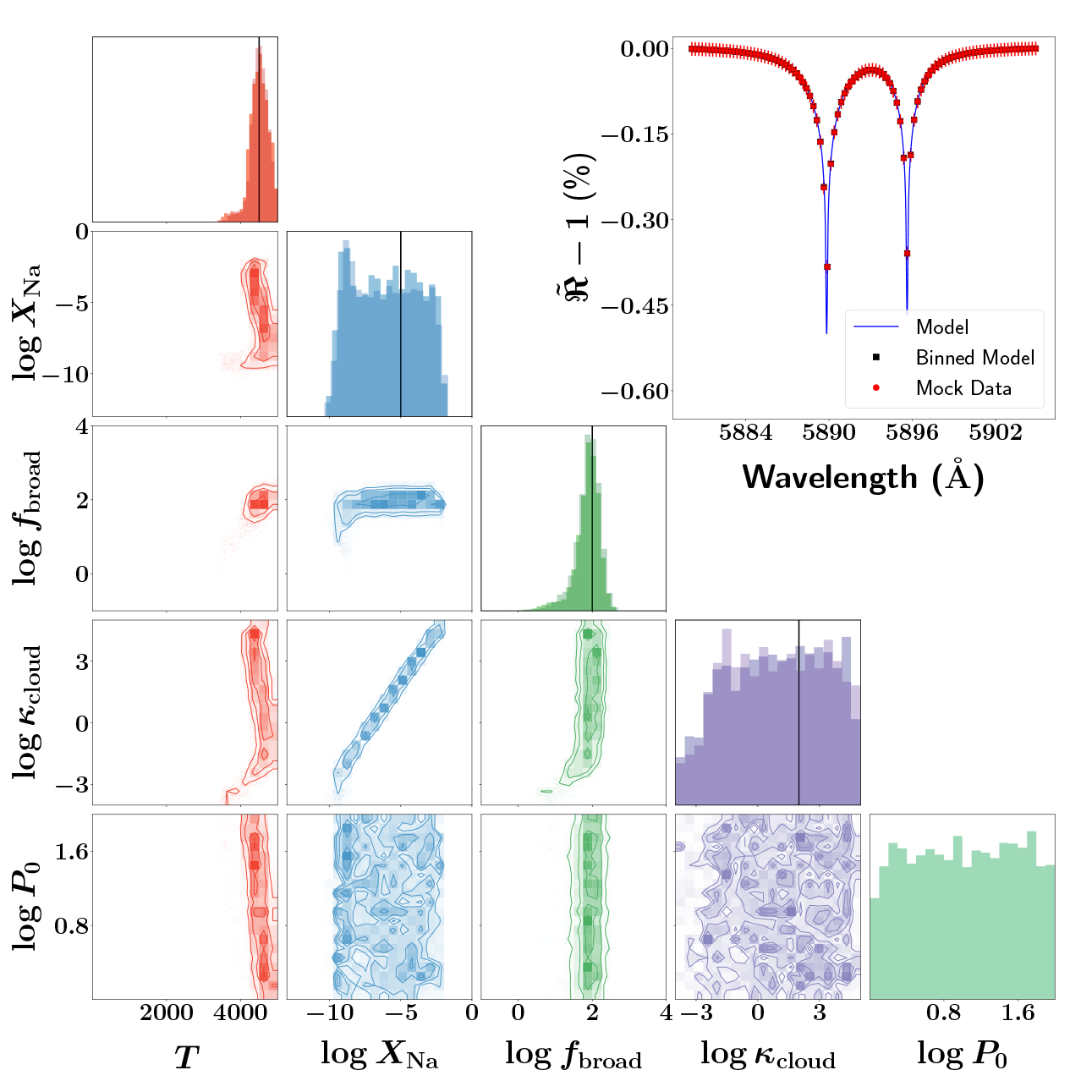}
\includegraphics[width=0.9\columnwidth]{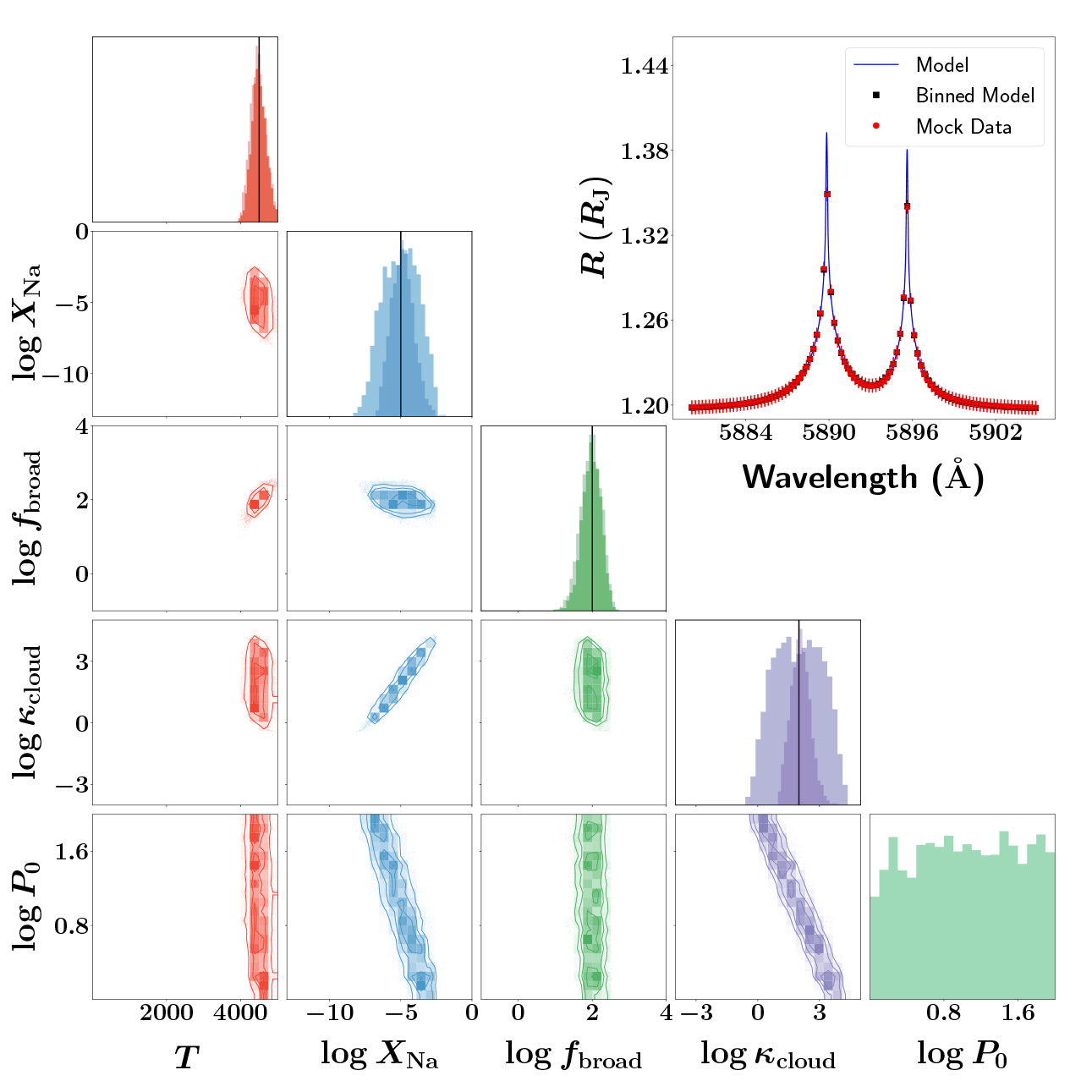}
\includegraphics[width=0.9\columnwidth]{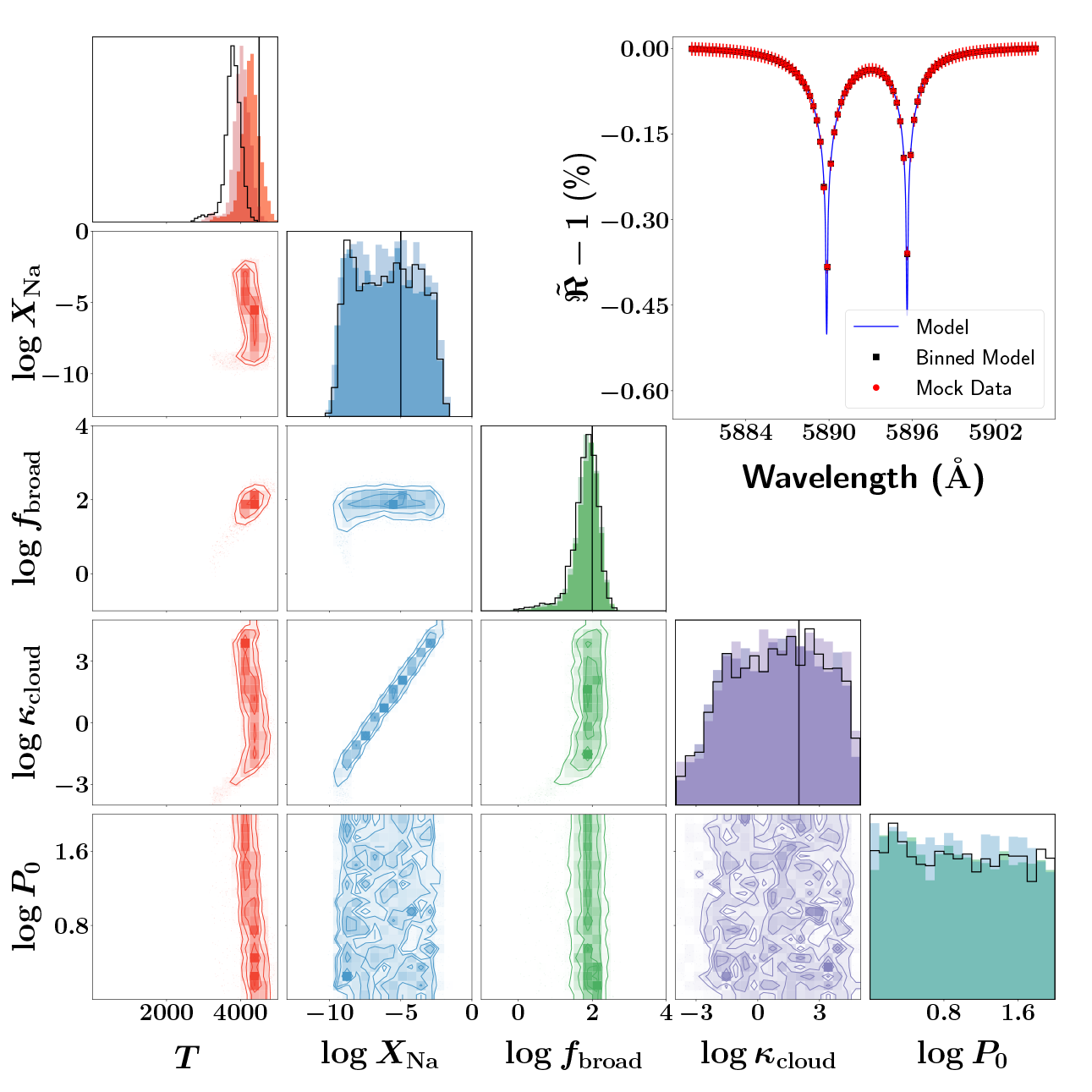}
\includegraphics[width=0.9\columnwidth]{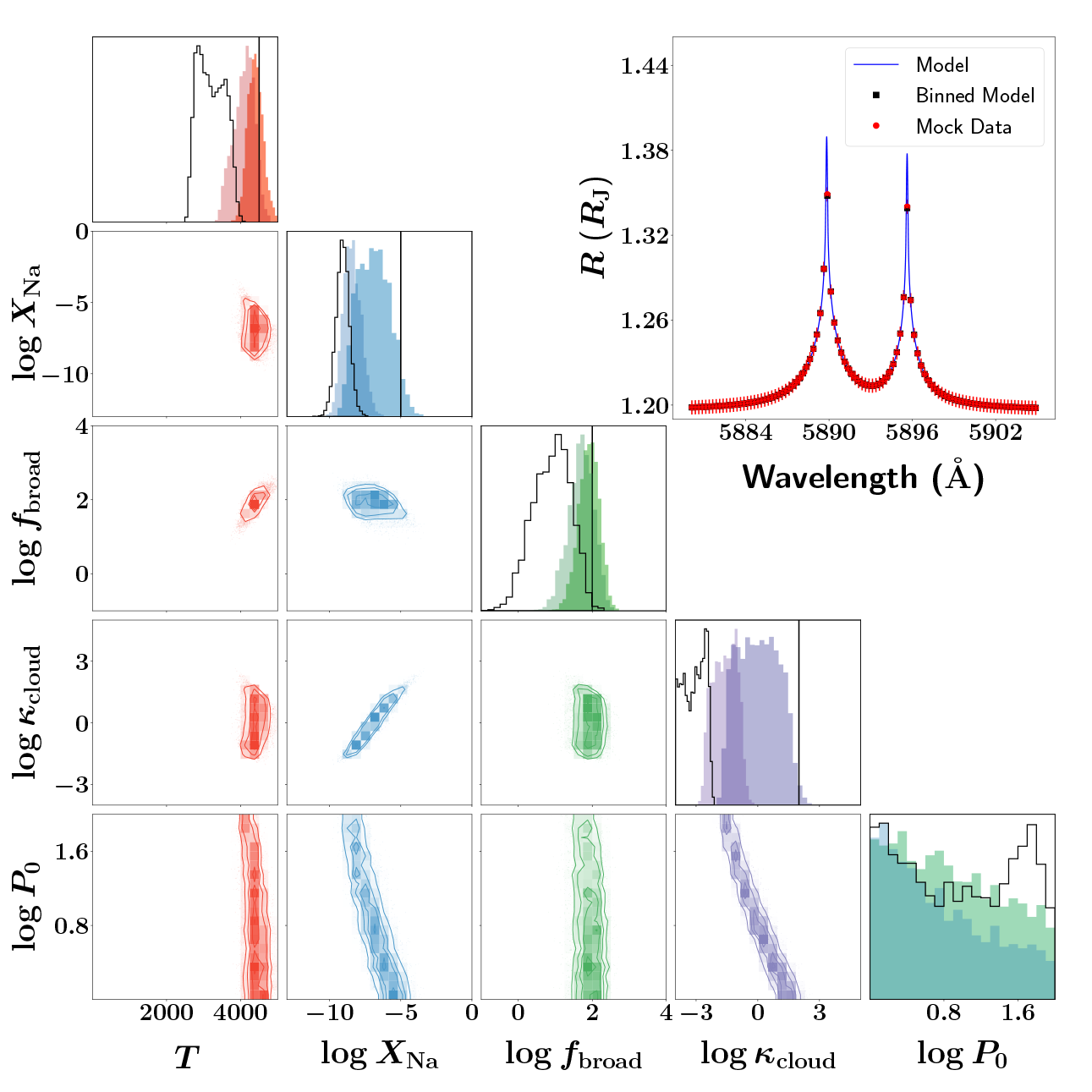}
\caption{Elucidating the influence of the normalization degeneracy on un-normalized (left column) versus normalized (right column) spectra, where the spectral continuum of the latter has been matched to the measured white-light HST radius of WASP-49b (see text).  This matching produces $R_0=0.752 ~R_{\rm J}$, while we have arbitrarily chosen $P_0=10$ bar for illustration.  We have set $T=4500$ K, $X_{\rm Na}=10^{-5}$, $f_{\rm broad}=100$ and $\kappa_{\rm cloud}=100$ cm$^2$ g$^{-1}$, again for illustration.  The top row shows mock retrievals where the values of $R_0$ and $P_0$ have been fixed to their input values (darker posteriors) versus those where $P_0$ is allowed to be a fitting parameter but $R_0=0.752 ~R_{\rm J}$ (lighter posteriors).  The bottom row shows mock retrievals where the value of $R_0$ is varied to mimic the real situation where we have no prior knowledge of what its value should be.  The light (shaded), dark (shaded) and solid-curve posteriors corresponds to good, not-so-good and bad guesses for $R_0$, respectively (see text).  Wherever relevant, the vertical lines indicate the true input value of a given parameter.}
\label{fig:degeneracies}
\end{figure*}

\begin{figure}
\centering
\includegraphics[width=\columnwidth]{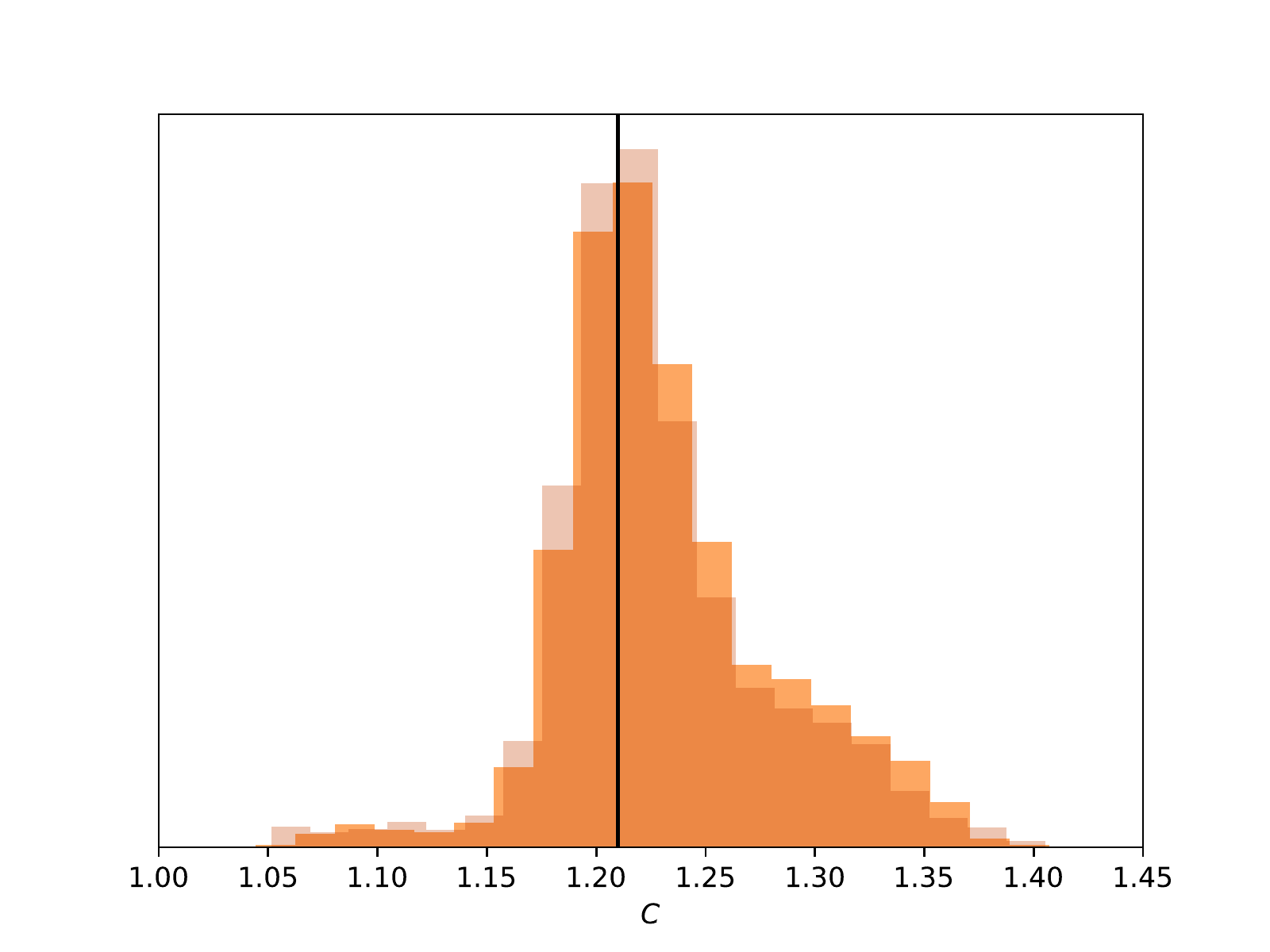}
\includegraphics[width=\columnwidth]{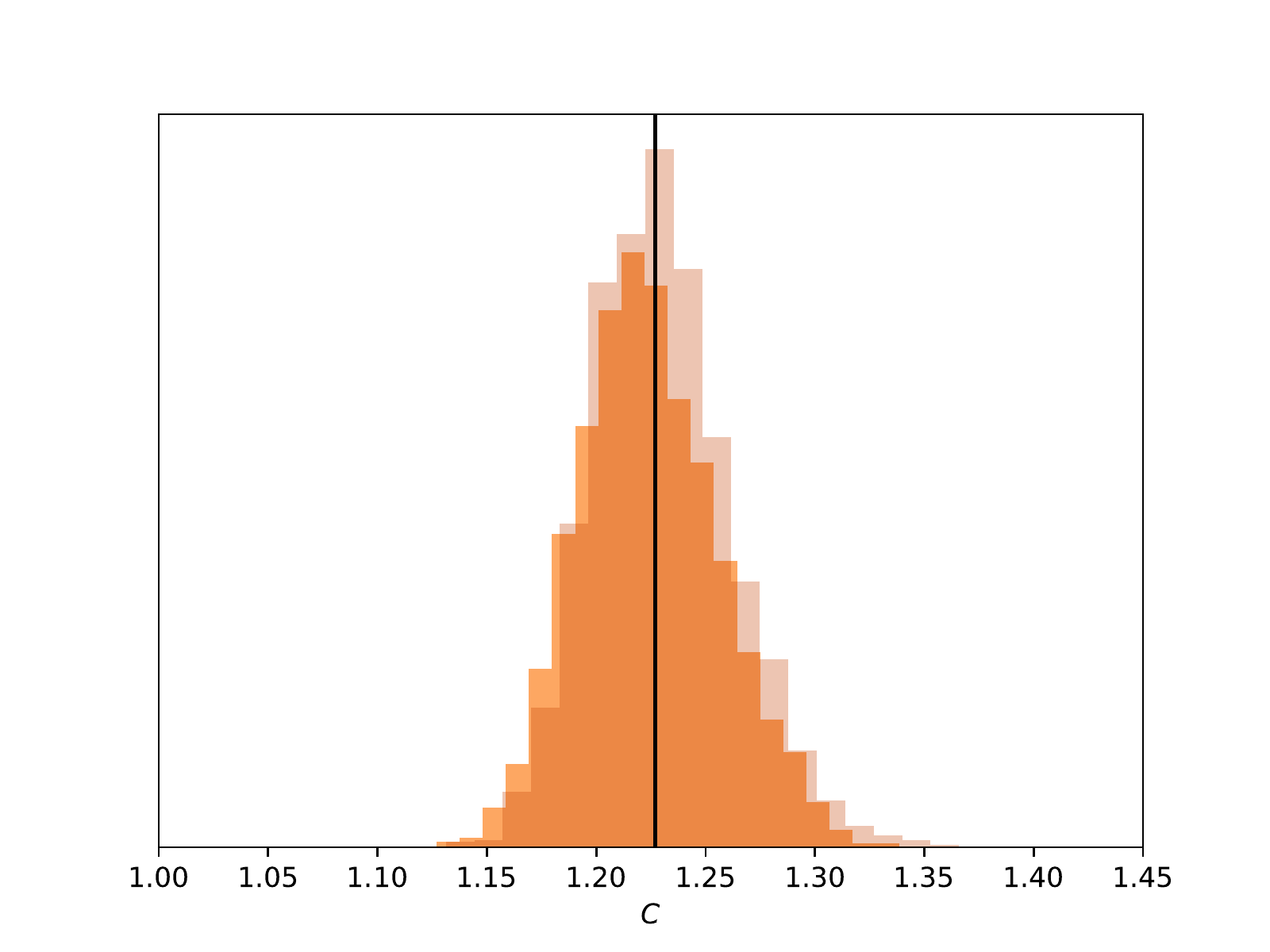}
\caption{Posterior distribution of the cloudiness index associated with the mock retrieval in Figure \ref{fig:degeneracies} for the un-normalized (top panel) and normalized (bottom panel) transmission spectra.  The darker posteriors correspond to mock retrievals where $P_0$ is treated as a fitting parameter.}
\label{fig:mock_cloudiness_index}
\end{figure}

\subsection{Data format for un-normalized high-resolution spectra}
\label{subsect:data_format}

Let the transit depth be $D = (R/R_\star)^2$, where $R_\star$ is the stellar radius.  Ground-based, high-resolution data do not measure the transit depth, but rather the differential transit depth, which is usually denoted by ${\tilde{\Re}}$ and lacks an absolute empirical normalization.  We wish to understand the relationship between $D$ and ${\tilde{\Re}}$.

Generally, we have $D \ll 1$.  Thus, $1-D$ is a number that is almost unity across wavelength.  All of the peaks of $D$ are converted into troughs for $1-D$.  Let the minimum value of $D$, which corresponds to the continuum, be $D_{\rm min}$.  The quantity ${\tilde{\Re}} = (1-D)/(1-D_{\rm min})$ \textbf{shifts the minimum value of the continuum to unity.}  The quantity that is typically used for measured spectra is ${\tilde{\Re}}-1$ \citep{w15,w17}, which shifts the continuum to zero.  

\section{Results}
\label{sect:results}

\begin{figure*}
\centering
\includegraphics[width=0.9\columnwidth]{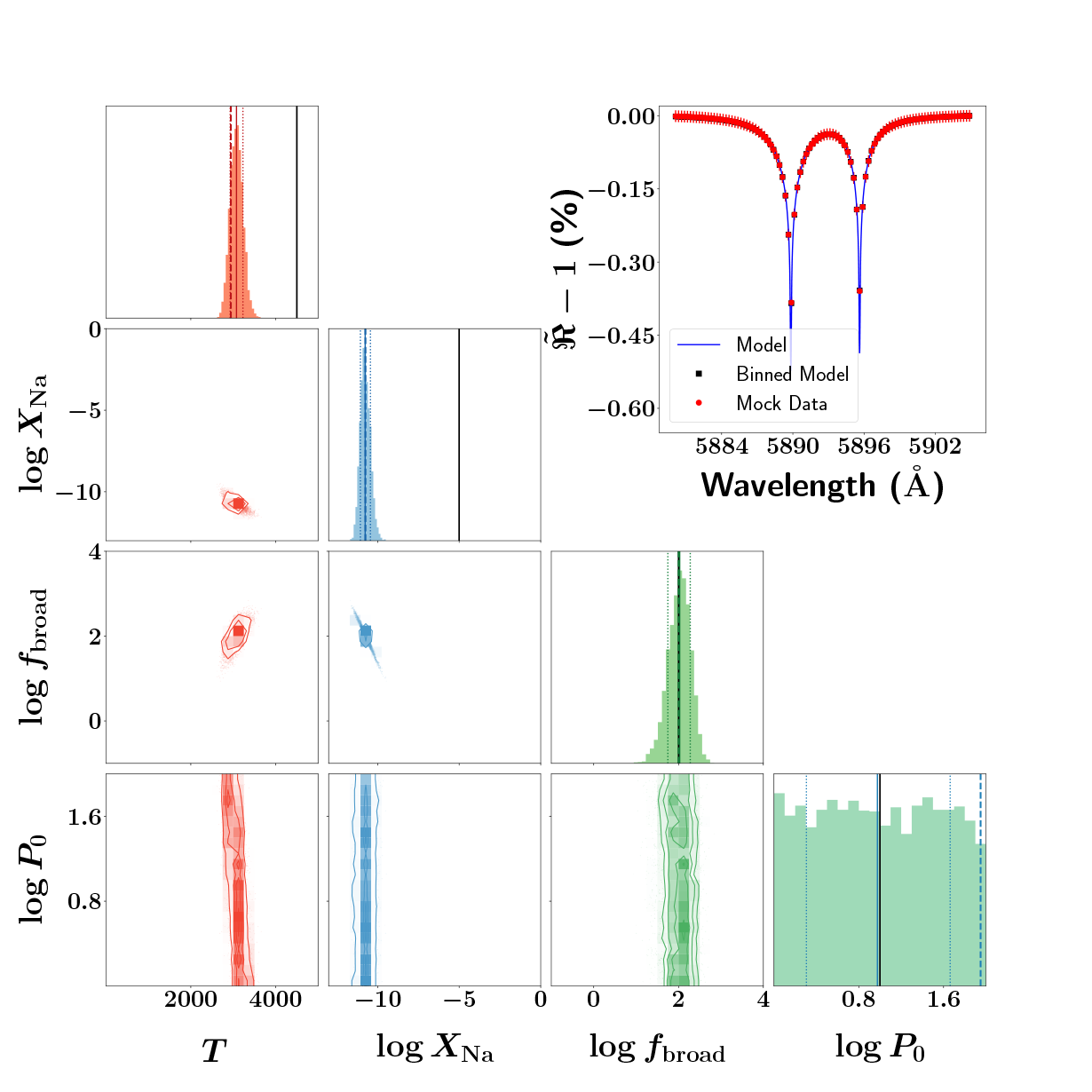}
\includegraphics[width=0.9\columnwidth]{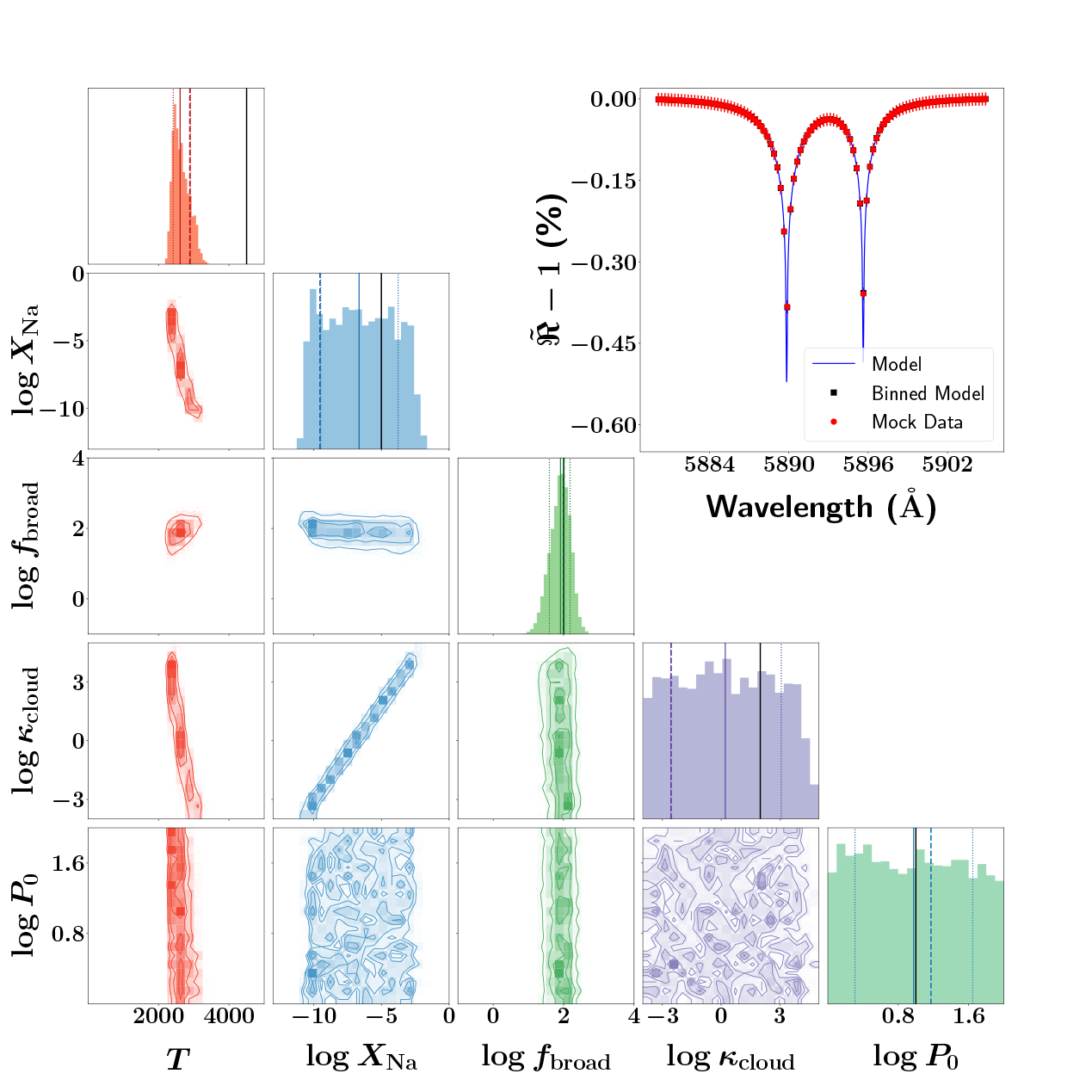}
\includegraphics[width=0.9\columnwidth]{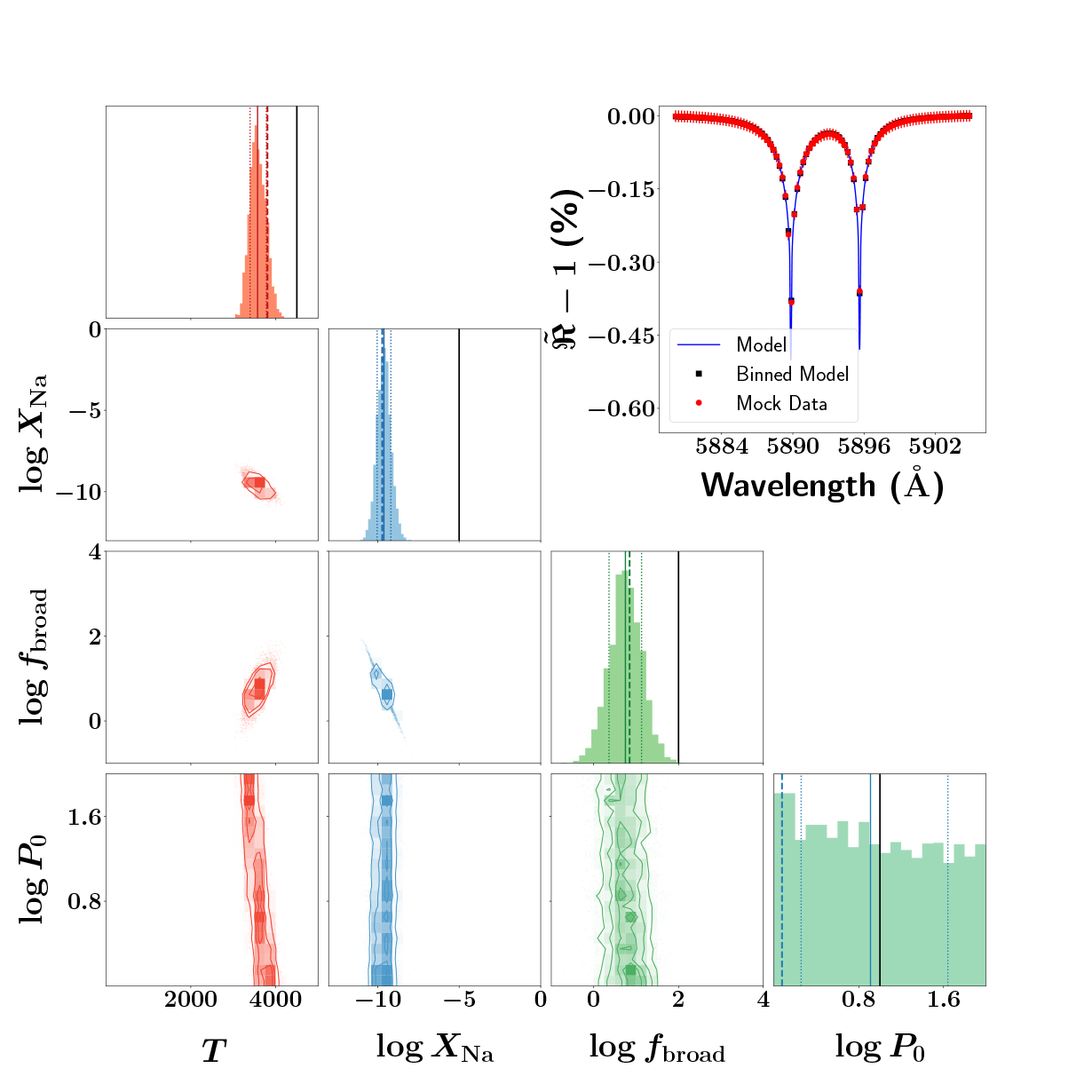}
\includegraphics[width=0.9\columnwidth]{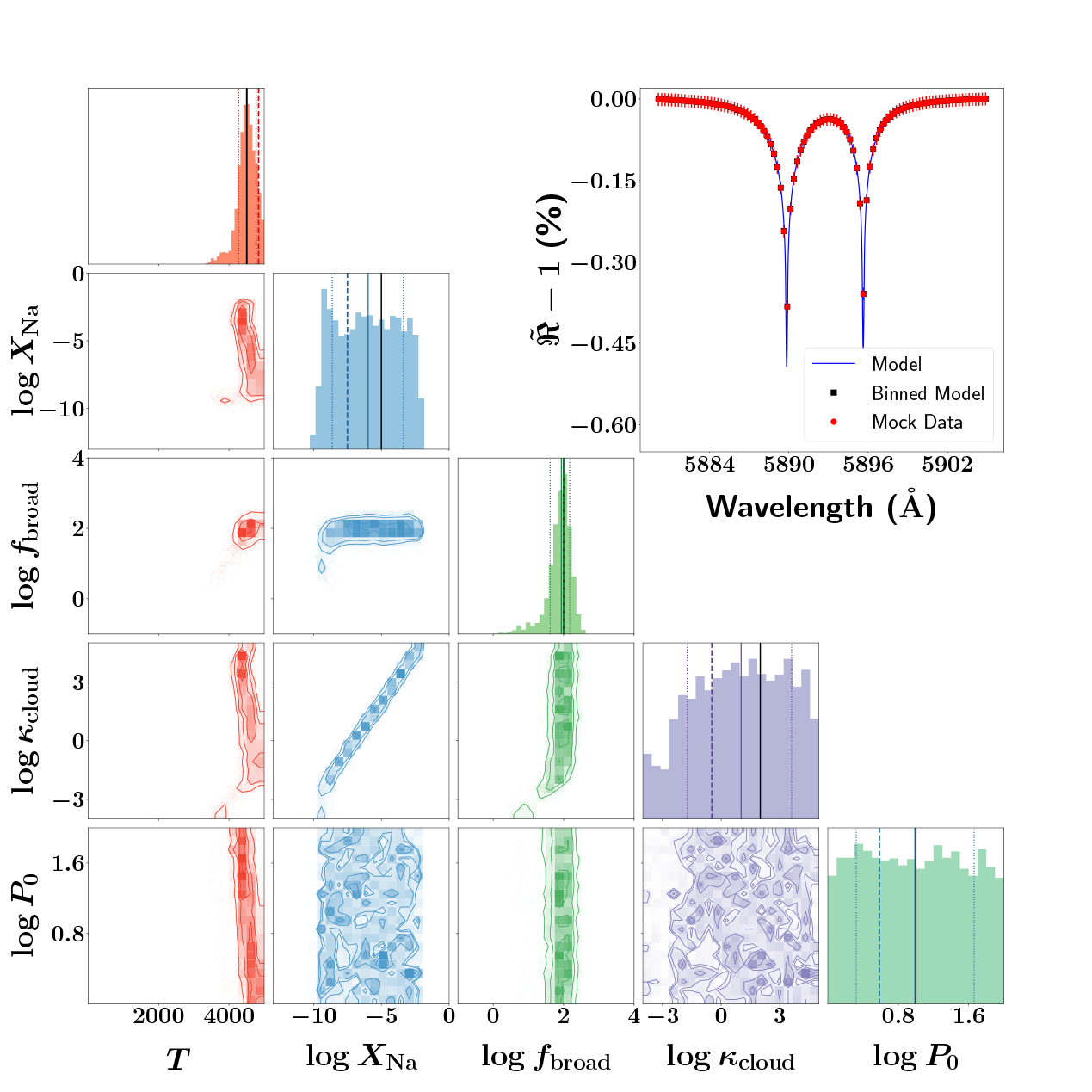}
\caption{Suite of retrievals on a mock transmission spectrum, with HARPS-like spectral resolution, constructed using a cloudy NLTE model of the sodium doublet.  Retrievals are performed assuming a cloudfree LTE model (top left panel), a cloudy LTE model (top right panel), a cloudfree NLTE model (bottom left panel) or a cloudy NLTE model (bottom right panel). The solid and dotted vertical lines show the median and 1-sigma limits for the posteriors, respectively. The thick dashed lines show the best-fit values. The solid black lines show the truth values for the mock data.}
\label{fig:mock_retrievals}
\end{figure*}

\begin{figure}
\centering
\includegraphics[width=\columnwidth]{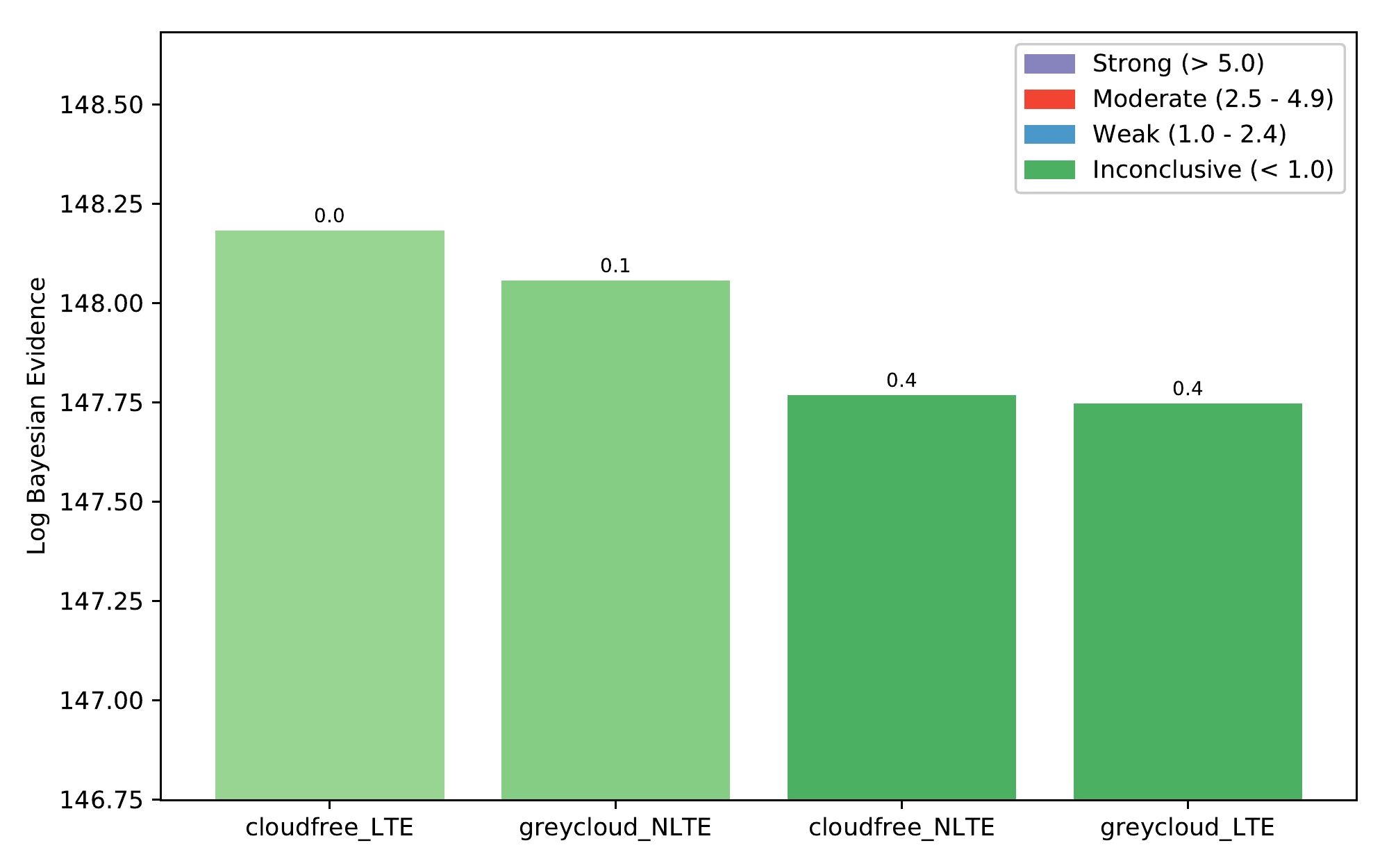}
\caption{Logarithm of the Bayesian evidence and corresponding Bayes factor between each model compared to the best model (as indicated by the number on top of each bar) for the suite of mock retrievals in Figure \ref{fig:mock_retrievals}.  The entry marked by ``0" is the best model, i.e., the model with the highest Bayesian evidence. The legend lists the correspondence between the Bayes factor and the strength or weakness of the evidence in favour of a given model (compared to the best model).}
\label{fig:mock_bayesian_evidence}
\end{figure}

\subsection{Does the normalization degeneracy affect our interpretation of the sodium doublet?}

Before analyzing measured transmission spectra of the sodium doublet, we study a suite of mock retrievals to elucidate the degeneracies between the various model parameters.  We also wish to understand the effect of the normalization degeneracy on spectra with a narrow wavelength range centered on the sodium doublet.  To generate mock spectra, we assume 100 wavelength bins between 5880 and 5905 ~\AA.  Each data point is assumed to have an uncertainty of 0.005 ~\AA~ on the transit radius. These numbers are guided by the measured HARPS transmission spectrum of WASP-49b \citep{w17}.  We use the measured gravity of WASP-49b of 689 cm$^2$ s$^{-1}$ \citep{w17}.  For illustration, we assume $T=4500$ K, $X_{\rm Na} = 10^{-5}$, $f_{\rm broad} = 100$ and $\kappa_{\rm cloud} = 100$ cm$^2$ g$^{-1}$ in our NLTE model of a mock transmission spectrum.

We study two types of retrievals of mock spectra.  Un-normalized spectra are presented in the ${\tilde{\Re}}-1$ format previously described in \S\ref{subsect:data_format}.  Normalized spectra are presented in transit radii ($R$), where we match the continuum of the mock spectrum to the measured white-light radius of $1.198 ~R_{\rm J}$ for WASP-49b \citep{w17}.  For the choice of $P_0=10$ bar, this requires setting $R_0=0.752 ~R_{\rm J}$.

The top panel of Figure \ref{fig:degeneracies} shows pairs of retrievals where we fix $P_0=10$ bar and $R_0=0.752 ~R_{\rm J}$ versus if we fix only $R_0=0.752 ~R_{\rm J}$ and allow $P_0$ to be part of the retrieval.  Several striking trends are revealed even by examining the un-normalized spectrum (top left panel).  First, the behavior of the temperature and broadening parameter are distinct enough from the other parameters that they may be accurately retrieved, regardless of whether $P_0$ is known.  It demonstrates that our parametrisation of line broadening using $f_{\rm broad}$ effectively isolates it from other effects, including temperature variations. 

Second, there is value added to the retrieval when the transmission spectrum is normalized (top right panel).  The posterior distribution of sodium abundance spans about 5 orders of magnitude (for the full width at half-maximum or FWHM) for the un-normalized spectrum, but shrinks to about 2--3 orders of magnitude simply by matching the spectral continuum to the HST white-light radius.  Our ignorance of $P_0$ for the normalized spectrum results in the cloud opacity being loosely unconstrained, but this does not affect our ability to accurately retrieve the temperature and broadening parameter.

Third, for the normalized spectrum (top right panel of Figure \ref{fig:degeneracies}), both the degeneracy with cloudiness (i.e., any increase in the sodium abundance may be compensated by an increase in the gray cloud opacity) and the normalization degeneracy (i.e., any increase in the sodium abundance may be compensated by a decrease in $P_0$) are clearly present in the posterior distributions.  These degeneracies clearly drive the somewhat large widths of the posterior distributions of both the gray cloud opacity and sodium abundance.  When the spectrum is un-normalized (top left panel of Figure \ref{fig:degeneracies}), the normalization degeneracy vanishes.  Essentially, over the width of the posterior distributions of both $X_{\rm Na}$ and $\kappa_{\rm cloud}$ as determined by the cloudiness degeneracy, the reference pressure is unconstrained.  Figure \ref{fig:mock_cloudiness_index} shows the corresponding posterior distributions for the cloudiness index, which is robust to whether $P_0$ is treated as a fitting parameter and whether the retrieval is performed on un-normalized or normalized spectra.

In a real situation, the value of $R_0$ is unknown.  As demonstrated by \cite{fh18}, it suffices to make a reasonable guess for $R_0$ and then retrieve for $P_0$.  The bottom panels of Figure \ref{fig:degeneracies} elucidates the effect of good ($R_0 = 0.85 R_{\rm J}$), not-so-good ($R_0 = 0.95 R_{\rm J}$) and bad ($R_0 = 1.05 R_{\rm J}$) guesses. For the \textbf{normalized} spectrum, a bad guess results in a posterior for the cloud opacity that bumps up against its prior value.  More insidiously, it also gives a posterior distribution of the sodium abundance that is \textit{narrower} than the one for the good guess.  A not-so-good guess also produces misleadingly narrow posterior distributions of the sodium abundance and cloud opacity.

The bottom left panel of Figure \ref{fig:degeneracies} shows that these concerns about guessing $R_0$ are mostly irrelevant for the un-normalized spectrum, because the posterior distributions of parameters are nearly identical for all quantities.  The exception is the temperature, which is somewhat different for the bad guess.  In other words, the retrieval of un-normalized spectra is unaffected by the normalization degeneracy, but at the expense of more uncertain posterior distributions compared to normalized spectra.

\subsection{Can we distinguish between LTE and NLTE scenarios?}

We now use mock retrievals to study a different question, which is whether a suite of nested-sampling retrievals are capable of distinguishing between LTE versus NLTE and cloudfree versus cloudy models?  Using the same cloudy NLTE mock spectrum, we perform four retrievals using the cloudfree LTE model, cloudy LTE model, cloudfree NLTE model and cloudy NLTE model.  Figures \ref{fig:mock_retrievals} and \ref{fig:mock_bayesian_evidence} show the outcomes of these retrievals.  The first major outcome is that the shape of the sodium doublet, in un-normalized transmission spectra, does not encode enough information to distinguish between cloudfree and cloudy models based on the Bayesian evidence.  It has the implication that the temperature is under-estimated if one uses a cloudfree model to interpret a cloudy transit chord.

A comparison of the Bayesian evidence also does not allow us to distinguish between NLTE and LTE models, which in our formulation have exactly the same number of parameters.  A robust outcome of the retrievals is that\textbf{ the value of the broadening parameter is accurately retrieved regardless of whether cloudy NLTE or LTE models are used}.  However, when a LTE model is used to perform retrieval on the NLTE mock spectrum, the retrieved temperature is substantially under-estimated.  In the example shown, we retrieve a temperature of about 2600 K compared to the true input value of 4500 K. This suite of mock retrieval suggest that retrievals performed on un-normalized HARPS transmission spectra of the sodium doublet are unable to discern between NLTE and LTE models.

In both the cloudy LTE and NLTE models, the width of the posterior distribution of the sodium abundance is determined by the degeneracy between the gray cloud opacity and the sodium abundance.

\subsection{Retrieval analysis of HARPS data of WASP-49b}

\begin{figure*}
\centering
\includegraphics[width=2\columnwidth]{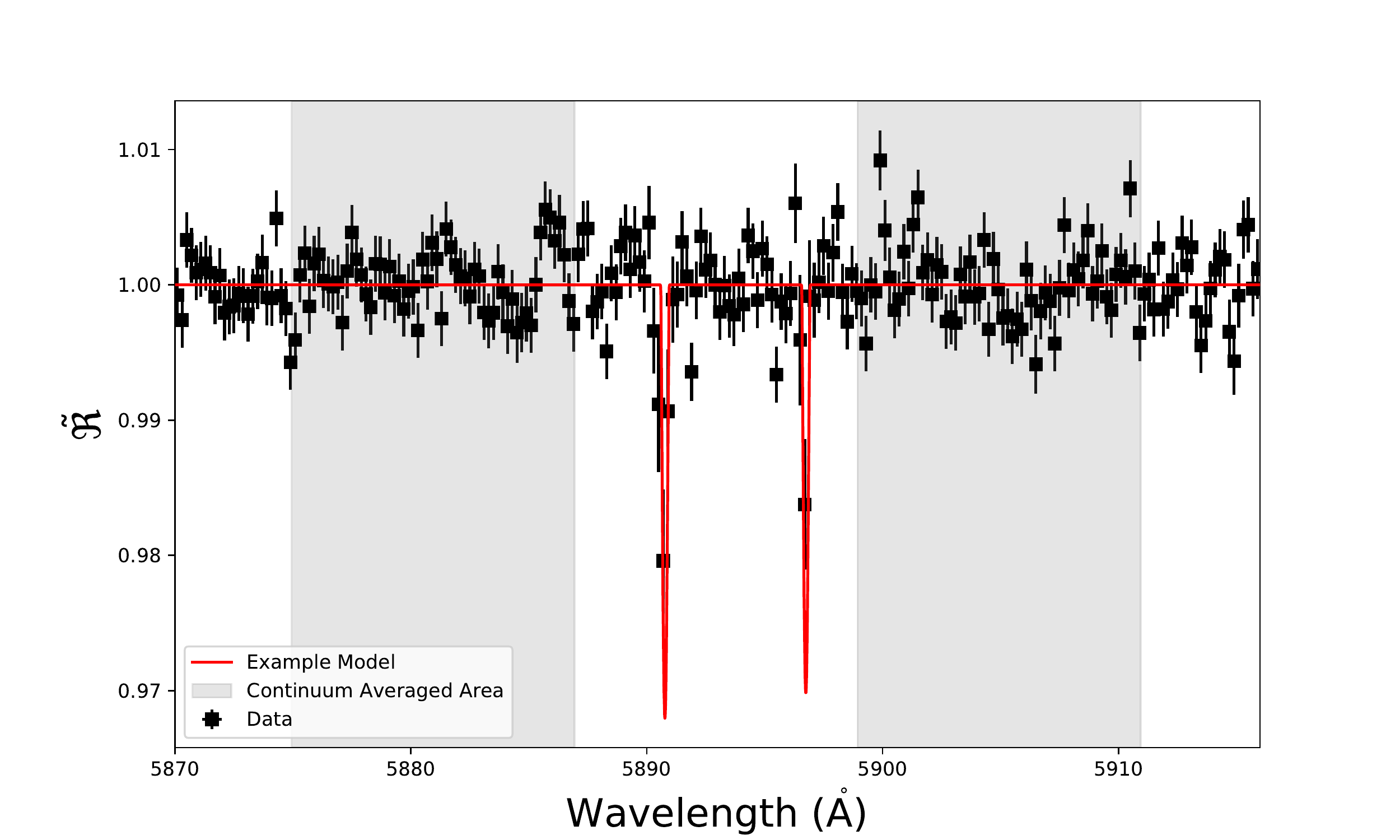}
\caption{Illustration of how the averaged continuum is measured in blue and red bands (shaded areas) and used to set the reference transit depth, $D_{\rm min}$.  The line peaks are measured in narrow bands with widths of 0.4 ~\AA.  As an illustration, we have used a model with $T=7208$ K, $X_{\rm Na}=10^{-5.64}$, $\kappa_{\rm cloud}=10^{-0.18}$ cm$^2$ g$^{-1}$ and $f_{\rm broad}=10^{-5.70}$.}
\label{fig:wasp49_continuum}
\end{figure*}

\begin{figure*}
\centering
\includegraphics[width=0.9\columnwidth]{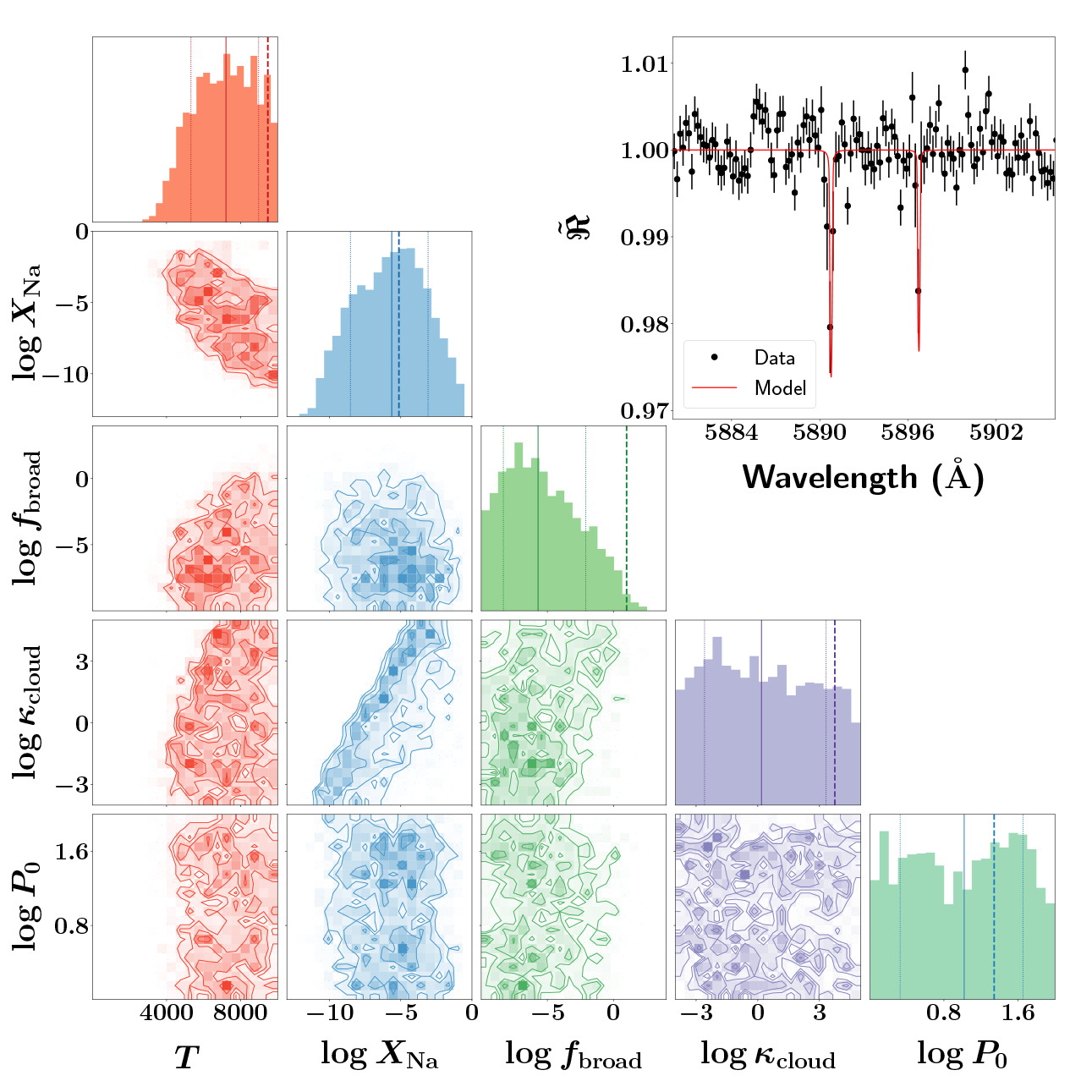}
\includegraphics[width=0.9\columnwidth]{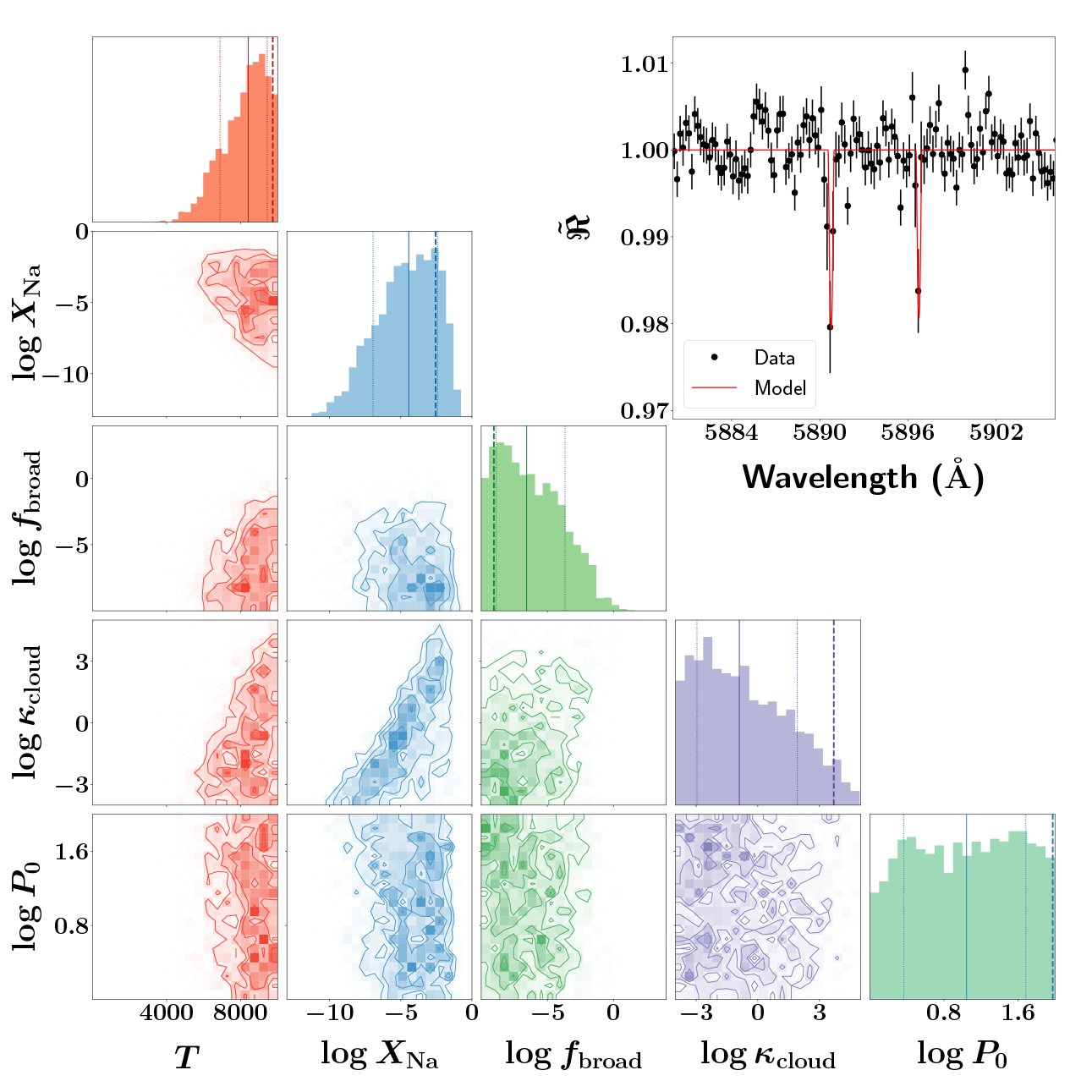}
\includegraphics[width=0.9\columnwidth]{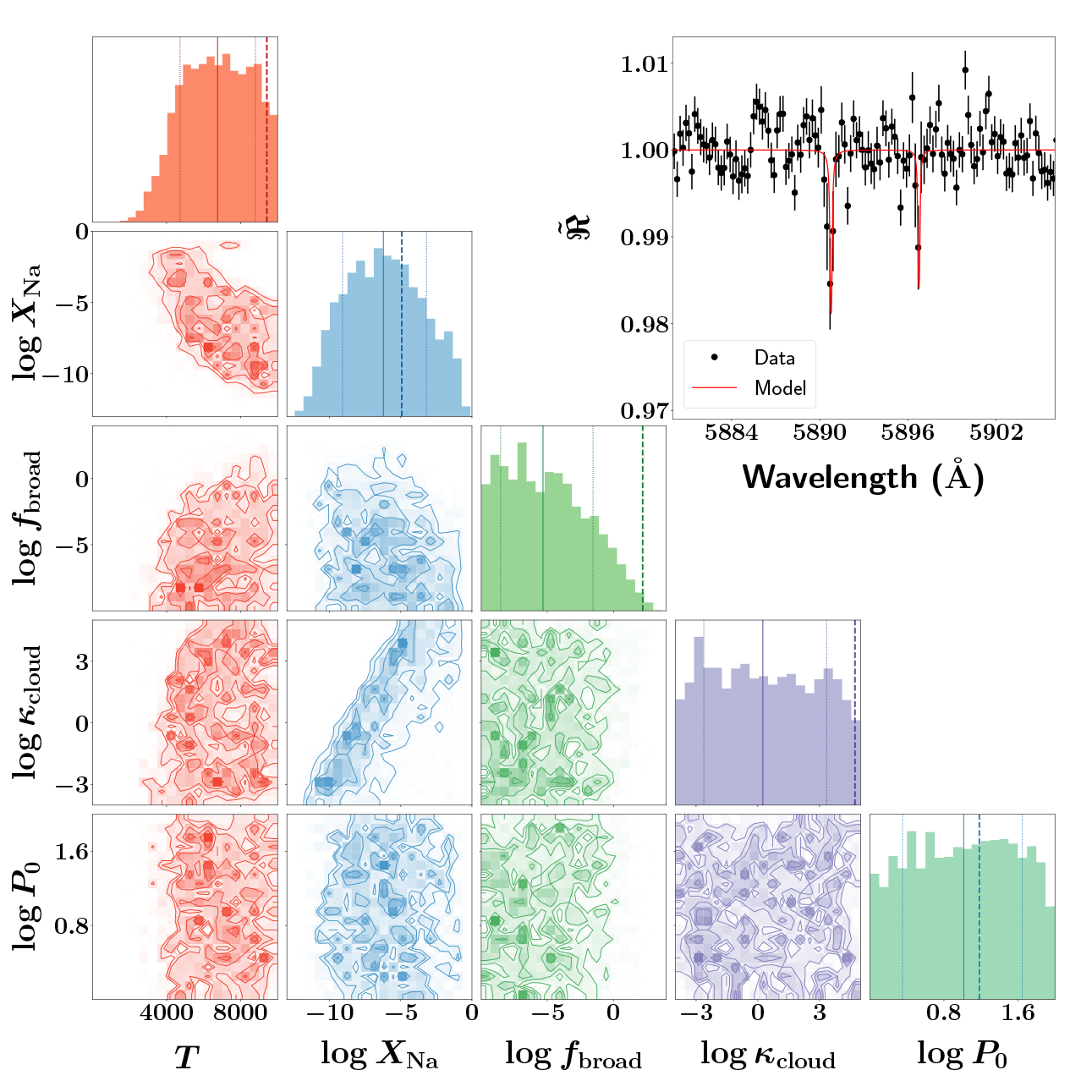}
\includegraphics[width=0.9\columnwidth]{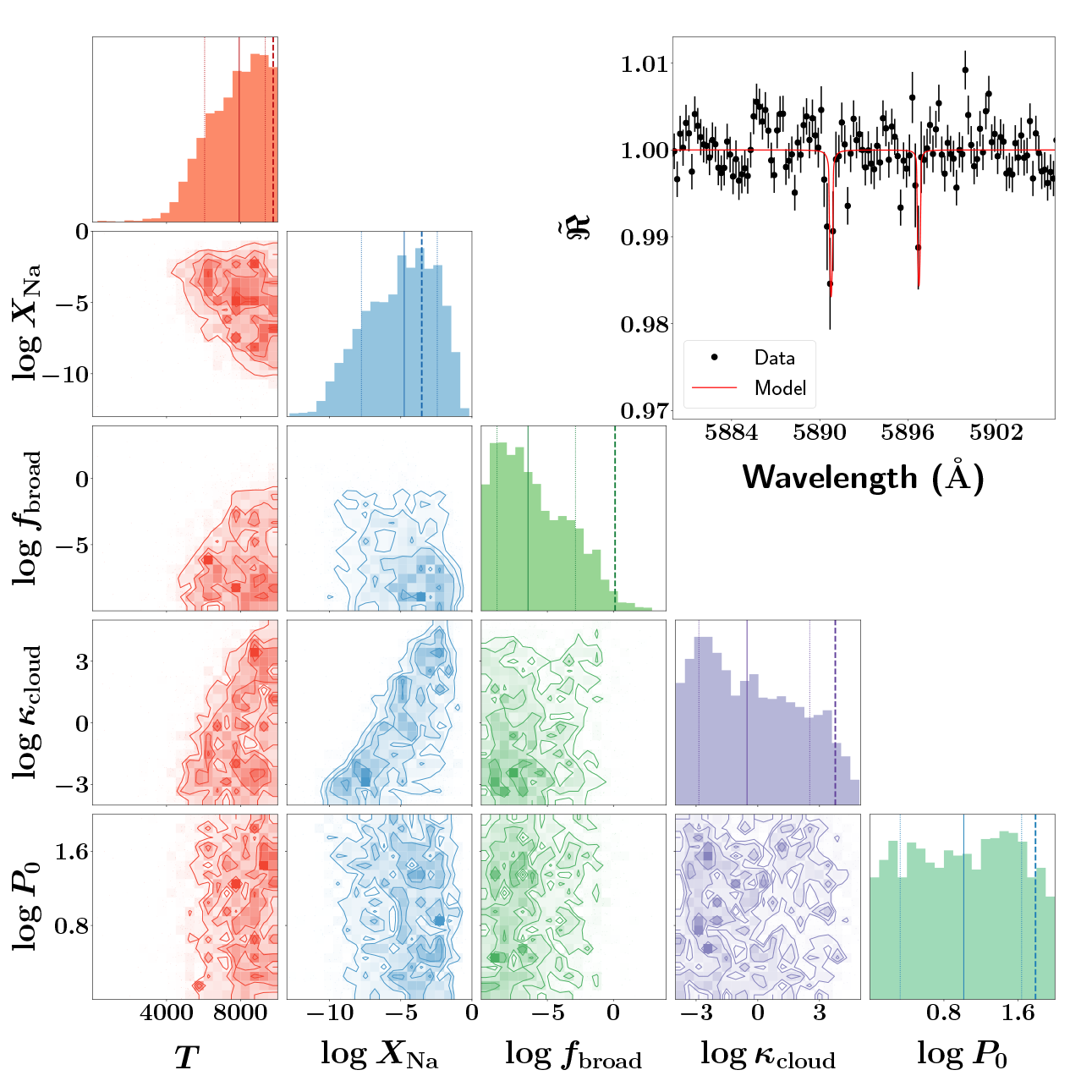}
\caption{Retrievals on the real HARPS dataset for WASP-49b using the LTE (left column) and non-LTE (right column) models.  The top row shows the retrievals performed on the original data.  For the bottom row, we have artificially increased the two data points corresponding to the line peaks by 0.005 in order to test the sensitivity of the retrieval outcomes to fluctuations in the spectral continuum.}
\label{fig:wasp49_retrievals}
\end{figure*}

\textbf{The mock retrievals of Figure \ref{fig:mock_retrievals} assume idealized conditions, meaning that sources of contamination due to stellar activity, telluric lines, imperfect co-addition of multiple exposures, etc, are not considered.  In analyzing the WASP-49b HARPS transmission spectrum of \cite{w17}, we assume that these issues have been addressed by the authors. }

\cite{w17} recorded the peak of each sodium line within a narrow band with a width of 0.4 ~\AA~(see their Section 4.2). \textbf{As discussed in Section \ref{subsect:data_format}, the reference transit depth, $D_{\rm min}$, is theoretically taken as the minimum transit depth across the spectrum. However, due to the fluctuations in the continuum points in the data, \cite{w17} set this reference value to the average measured continuum within two reference bands: blue (5874.94--5886.94 ~\AA) and red (5898.94--5910.94 ~\AA).  We apply the same procedure to our models: within the same pair of reference bands, we compute the average value of the continuum.  We then take $D_{\rm min}$ as this value when calculating ${\tilde{\Re}}$, before shifting it to ${\tilde{\Re}}=1$. Our fit to the data using this procedure is shown in Figure \ref{fig:wasp49_continuum}.}   

The top row of Figure \ref{fig:wasp49_retrievals} shows our retrieval analysis of the measured high-resolution, un-normalized transmission spectrum of WASP-49b \citep{w17}. In our initial retrievals (not shown), we struggled to fit the deep line peaks with our model. This motivated us to extend our prior for $f_{\rm broad}$ to lower values, compared with those stated in Table \ref{tab:priors}. For this data, we use the range $[10^{-10}, 10^4]$. As expected, we are unable to distinguish between the LTE versus NLTE interpretation from the computed Bayesian evidence (448.8 for LTE versus 448.4 for NLTE).  On physical grounds alone, we favor the NLTE interpretation.  The temperature is $7209^{+1763}_{-1892}$ K for the LTE interpretation versus $8415^{+1020}_{-1526}$ K for the NLTE interpretation, consistent with the lesson learned from our mock retrievals that the LTE model tends to predict a lower temperature.  These retrieved temperatures are discrepant from the $2950^{+400}_{-500}$ K value reported by \cite{w17}, but we note that \cite{w17} fitted Gaussians rather than Voigt profiles to the measured sodium doublets.  The use of Gaussians is equivalent to fitting with a Doppler profile in the absence of Lorentzian wings.  Our retrieved temperatures remain consistent with the expectation that radiative cooling by collisionally excited atomic hydrogen thermostats the temperature to $\sim 10^4$ K \citep{mc09}.  The retrieved sodium volume mixing ratios are loosely constrained and somewhat insensitive to whether the LTE or NLTE interpretation is assumed ($\log{X_{\rm Na}} = -5.64^{+2.55}_{-2.90}$ versus $-4.44^{+2.02}_{-2.51}$), as is the retrieved cloud opacity ($\log{\kappa_{\rm cloud}} = -0.18^{+3.14}_{-2.77}$ versus $-0.90^{+2.82}_{-2.06}$).  The broadening parameter is essentially unconstrained, but takes on $f_{\rm broad} < 1$ values as the narrow wavelength range precludes full coverage of the sodium line wings. 

The continuum of the measured transmission spectrum appears to possess a scatter with a half-width of about 0.005.  To test the robustness of our retrieval results, we artificially increased the two line peaks by 0.005 each and re-ran our retrievals.  The retrieval outcomes are shown in the bottom row of Figure \ref{fig:wasp49_retrievals}.  For the LTE retrieval, we obtained a reduced temperature of $T=6756^{+2048}_{-2030}$ K and a similar sodium abundance ($\log{X_{\rm Na}} = -6.22^{+3.03}_{-2.86}$).  For the NLTE retrieval, the temperature becomes $T=7925^{+1408}_{-1869}$ K and the sodium abundance becomes $\log{X_{\rm Na}} =-4.76^{+2.32}_{-3.01}$.  This pair of tests suggests that the retrieval outcomes are somewhat robust to the uncertainty in the continuum.

\subsection{Retrieval analysis of low-resolution transmission spectra}

\begin{figure*}
\centering
\includegraphics[width=0.9\columnwidth]{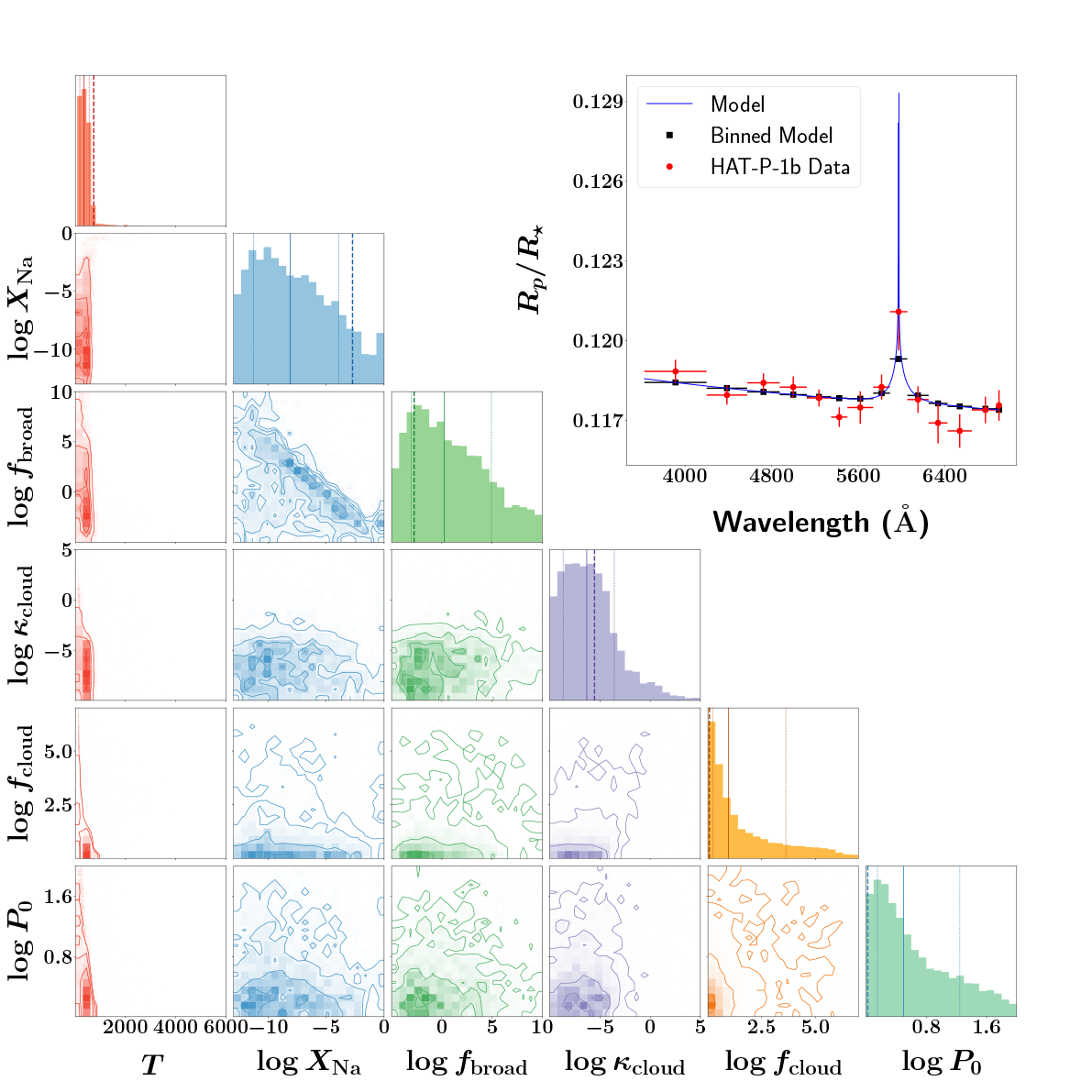}
\includegraphics[width=0.9\columnwidth]{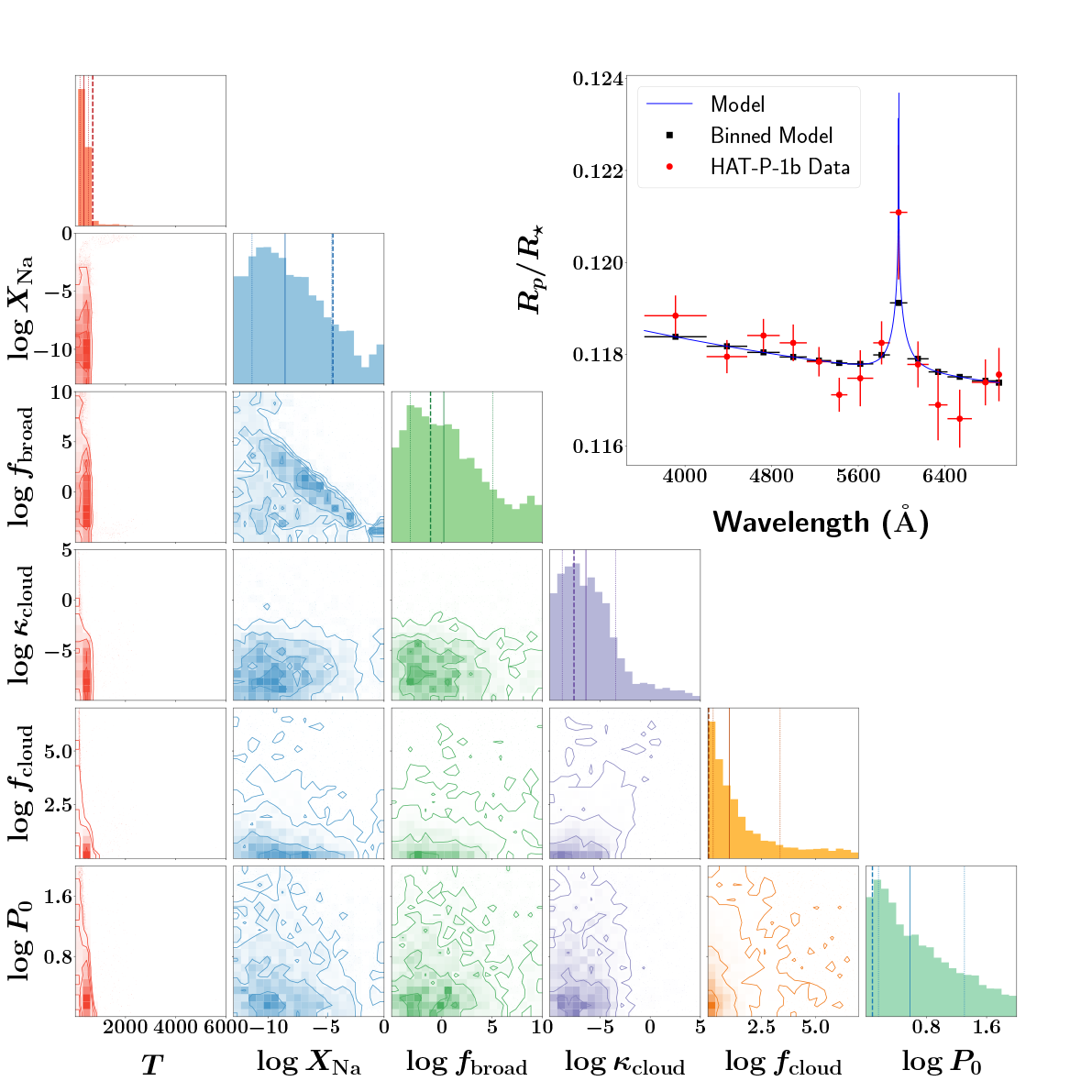}
\includegraphics[width=0.9\columnwidth]{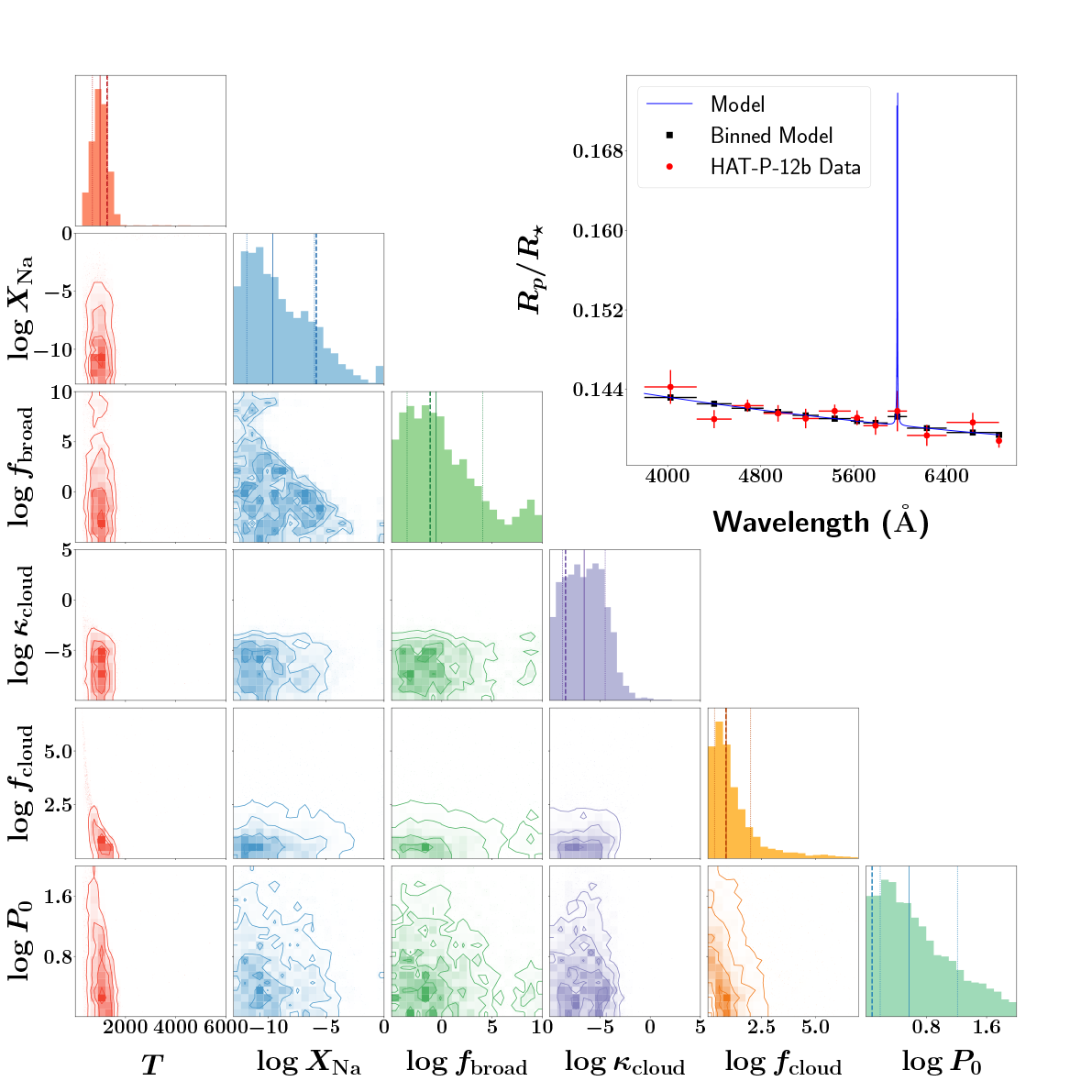}
\includegraphics[width=0.9\columnwidth]{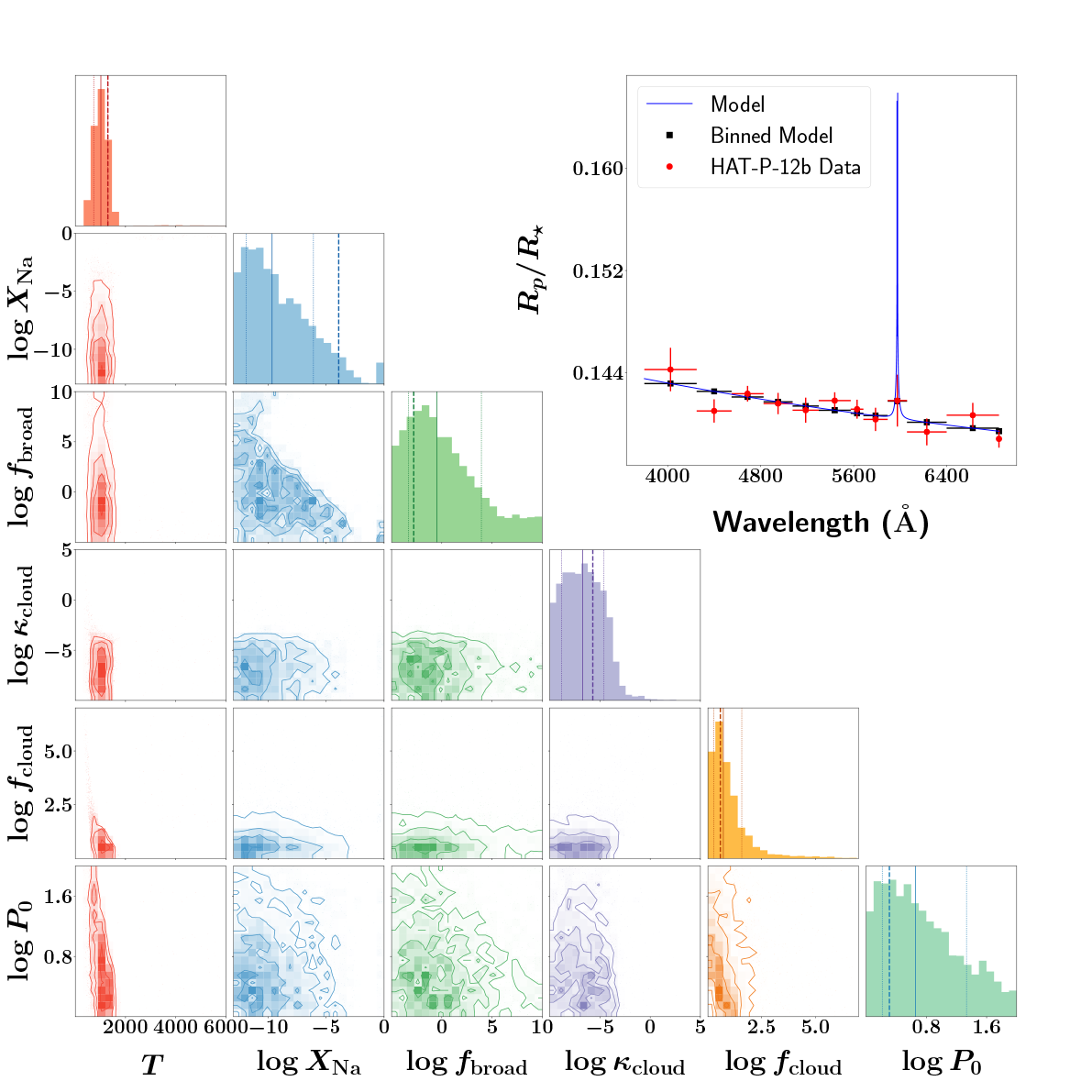}
\includegraphics[width=0.9\columnwidth]{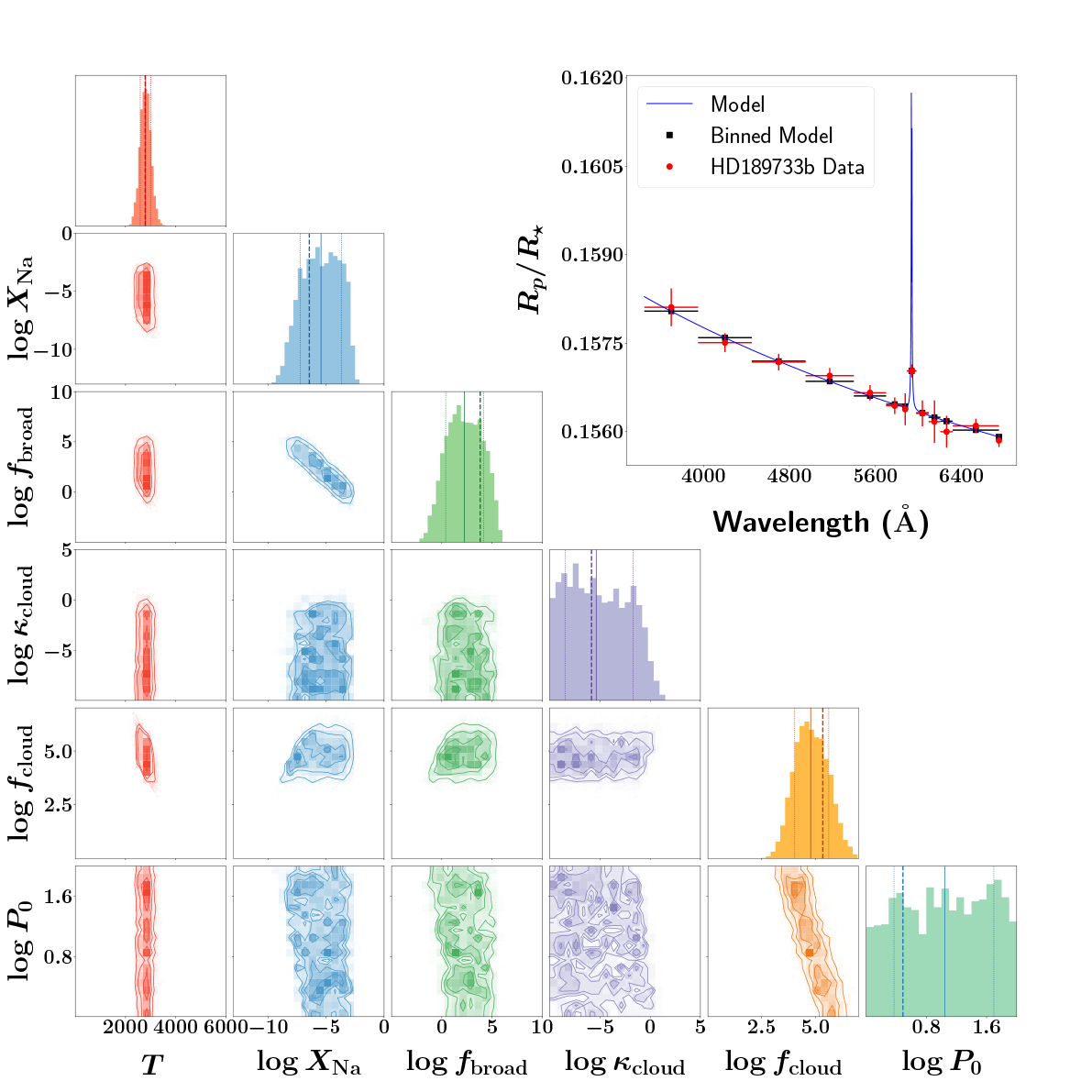}
\includegraphics[width=0.9\columnwidth]{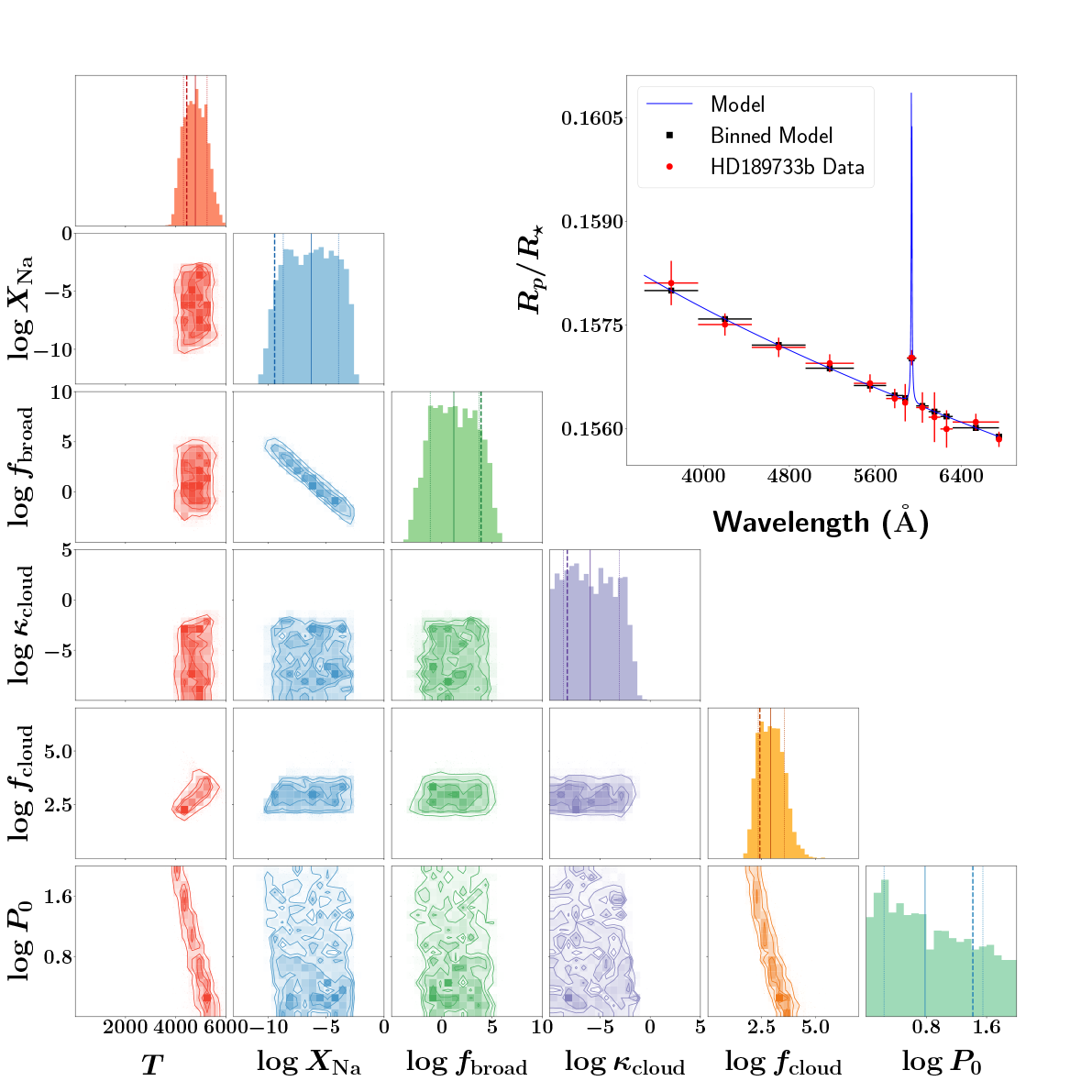}
\caption{Retrieval analysis of low-resolution transmission spectra.  The left and right columns are for the LTE and NLTE interpretations, respectively.  The top, middle and bottom rows are for HAT-P-1b, HAT-P-12b and HD 189733b, respectively.}
\label{fig:low_res}
\end{figure*}

\begin{figure*}
\centering
\includegraphics[width=0.9\columnwidth]{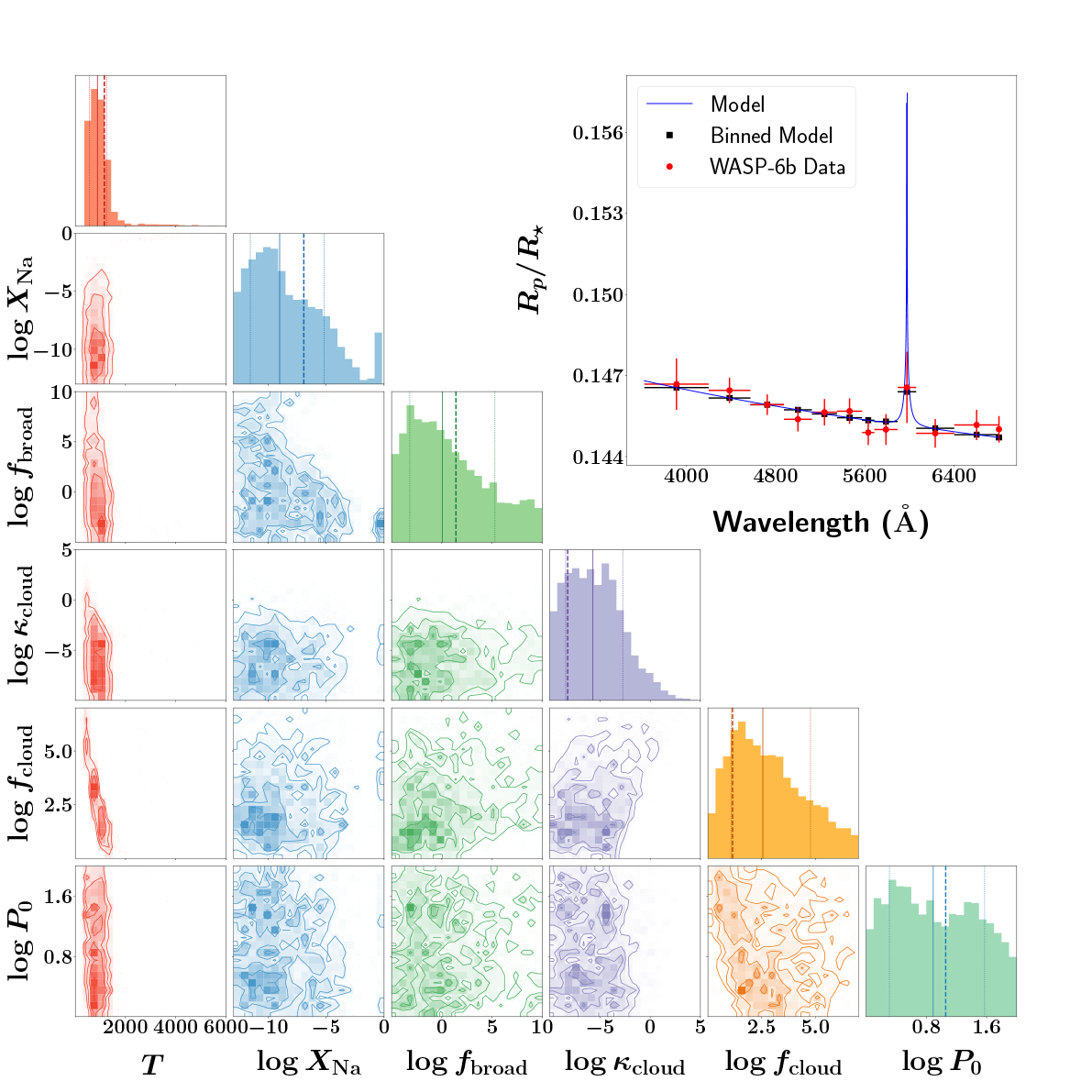}
\includegraphics[width=0.9\columnwidth]{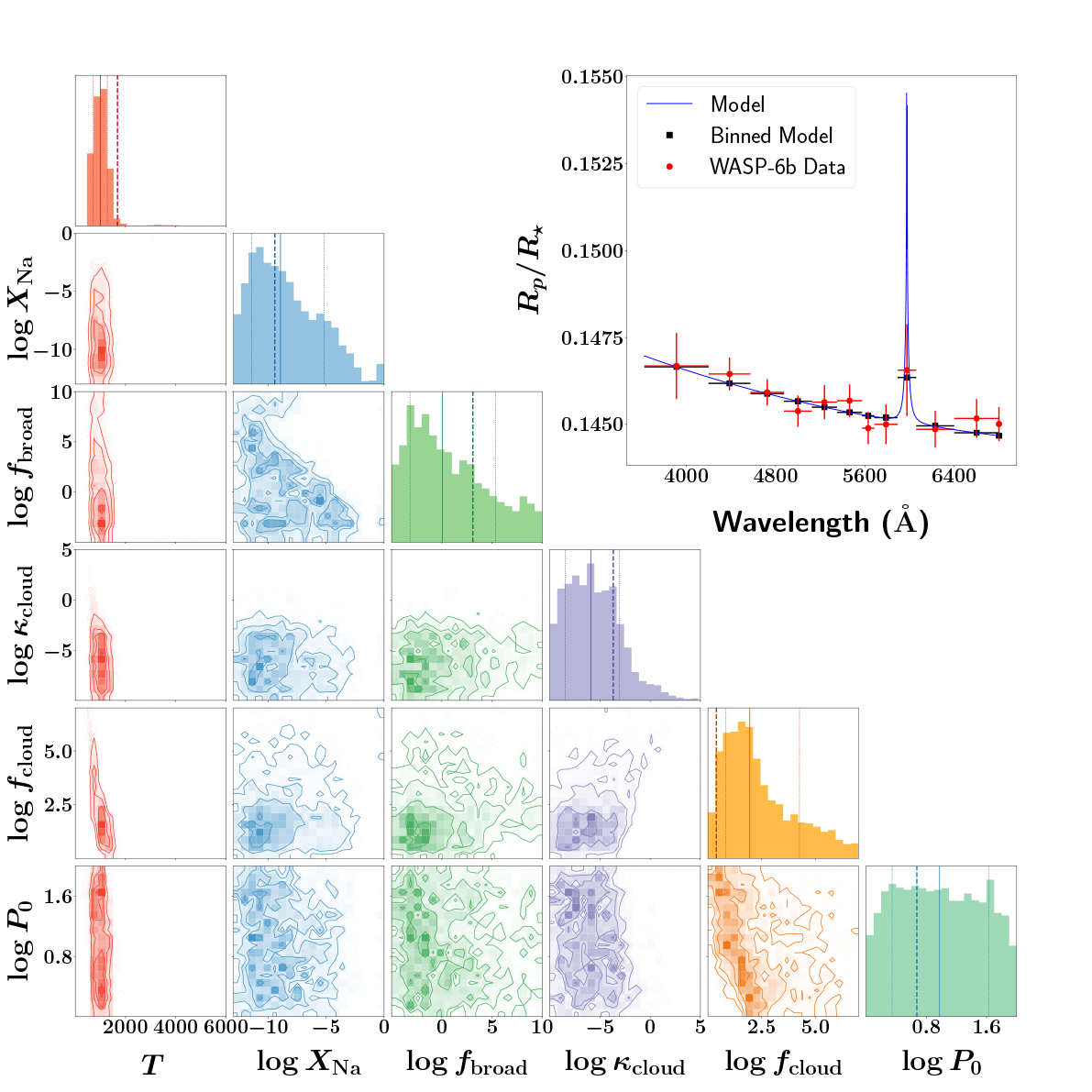}
\includegraphics[width=0.9\columnwidth]{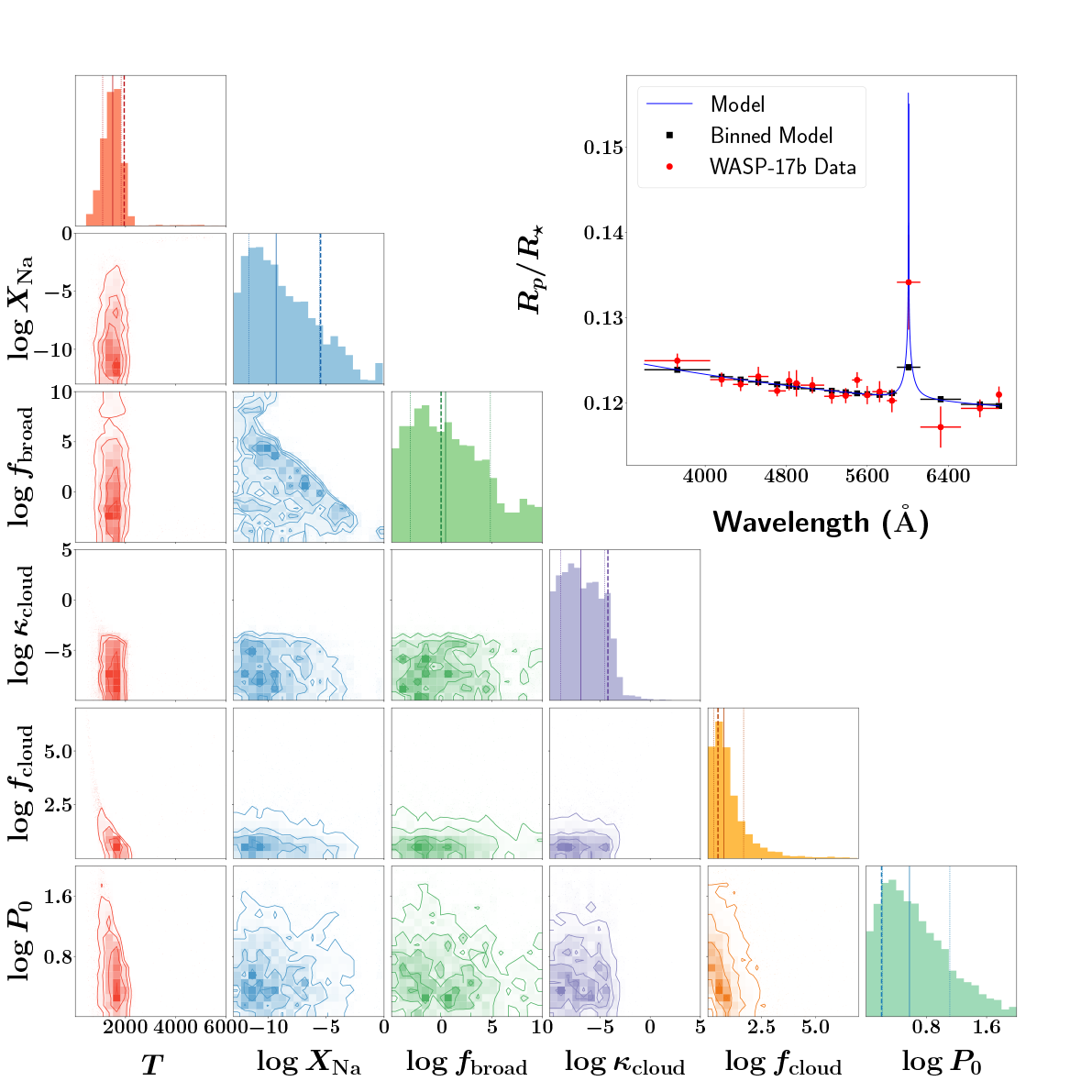}
\includegraphics[width=0.9\columnwidth]{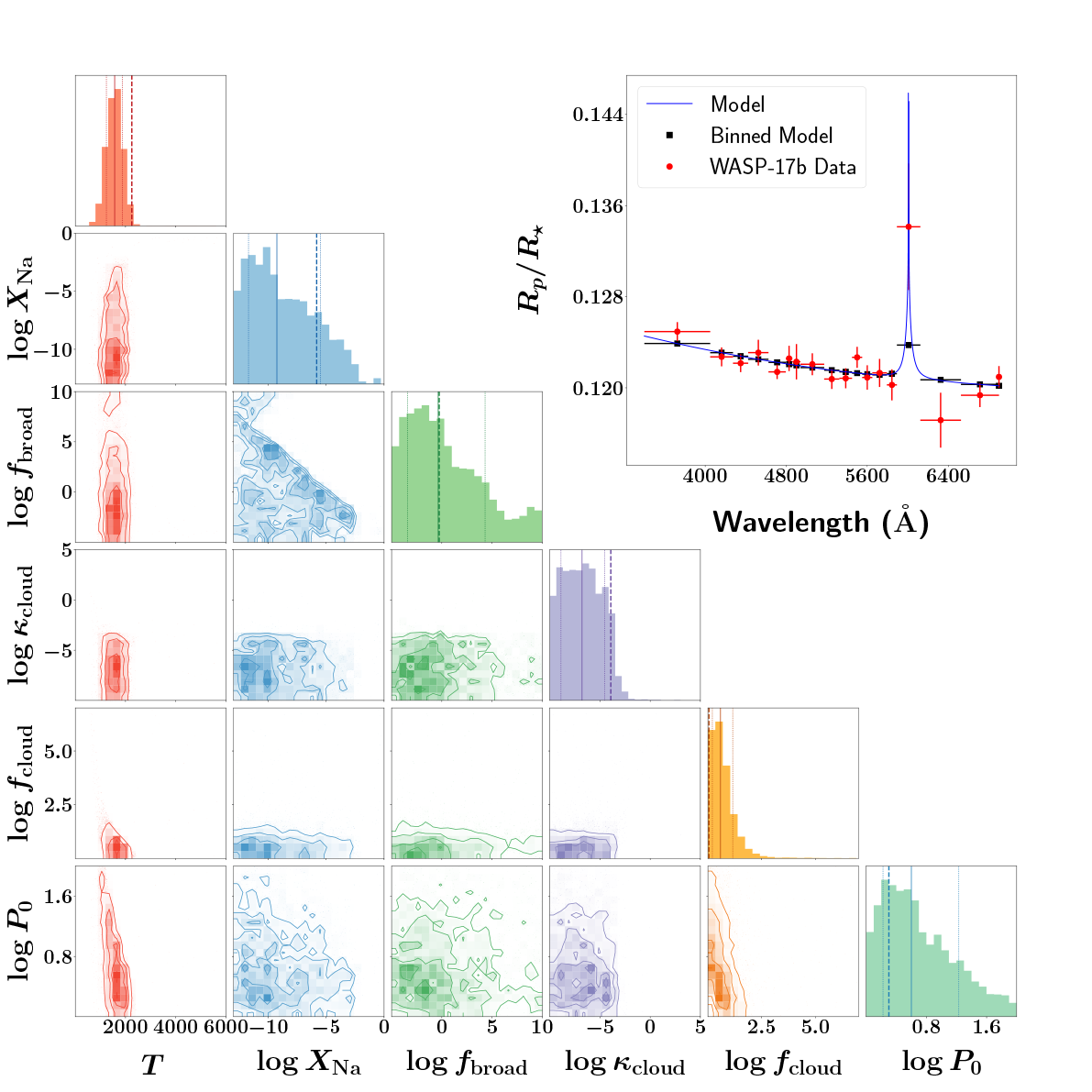}
\includegraphics[width=0.9\columnwidth]{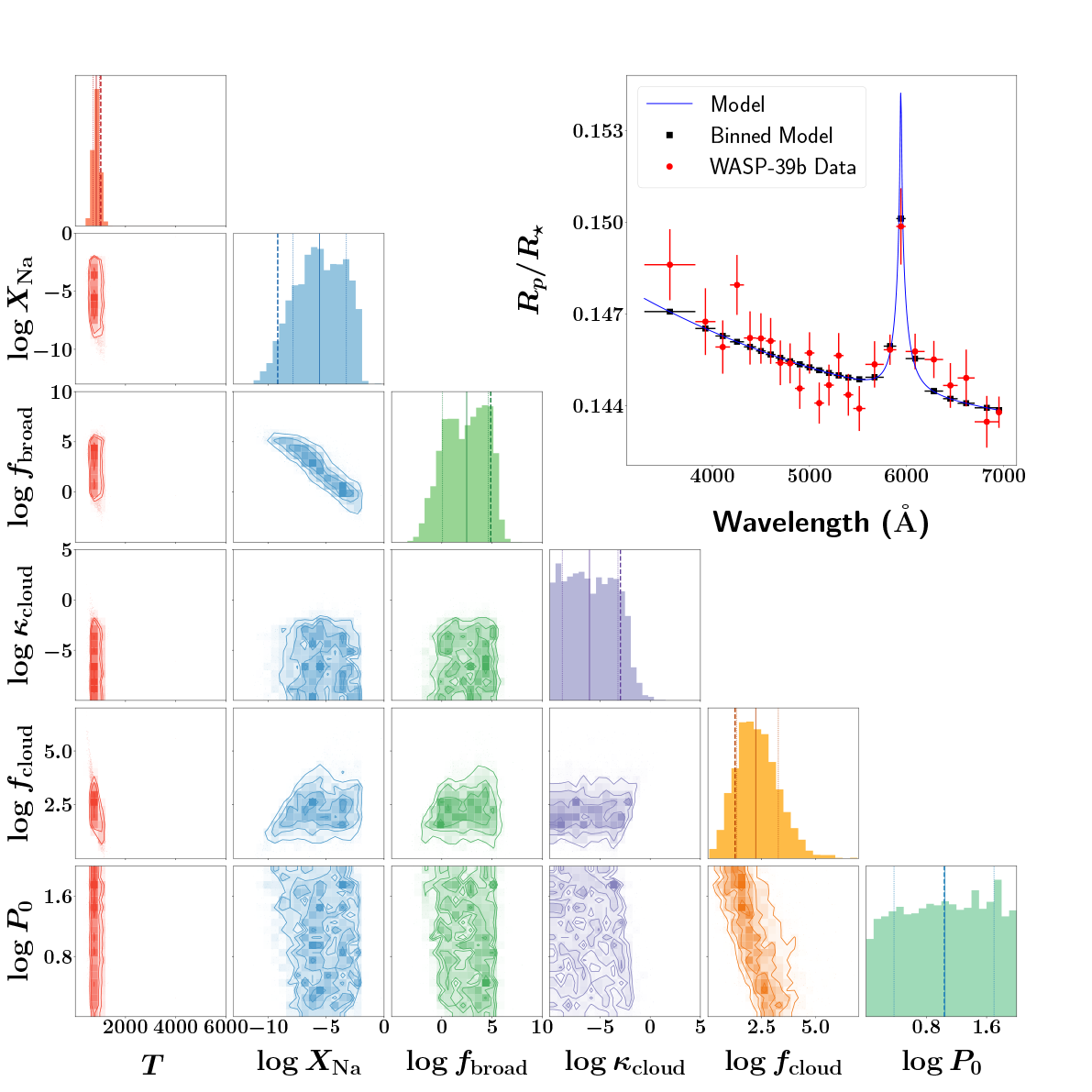}
\includegraphics[width=0.9\columnwidth]{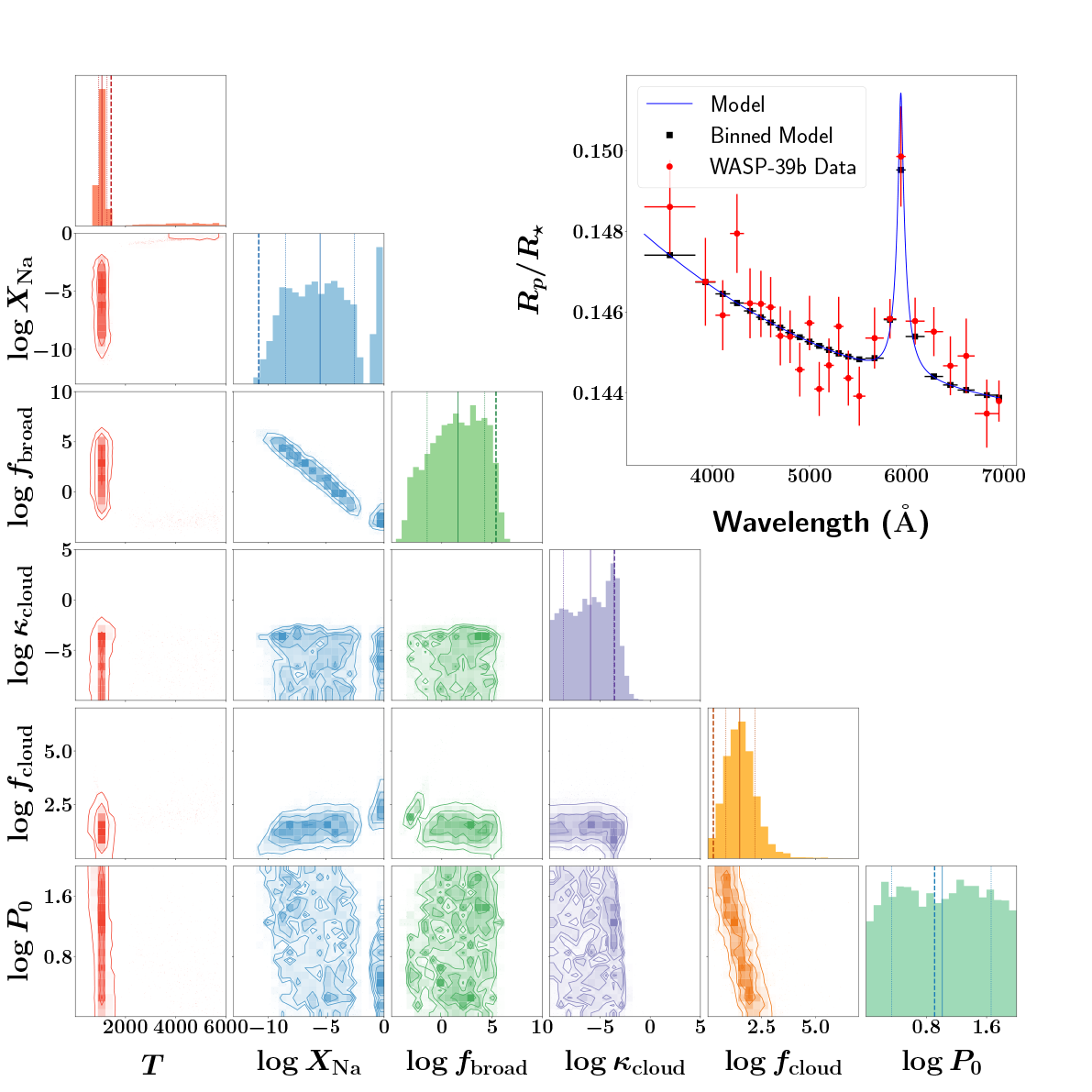}
\caption{Same as Figure \ref{fig:low_res}, but for WASP-6b (top row), WASP-17b (middle row) and WASP-39b (bottom row).}
\label{fig:low_res_2}
\end{figure*}

\begin{figure*}
\centering
\includegraphics[width=0.9\columnwidth]{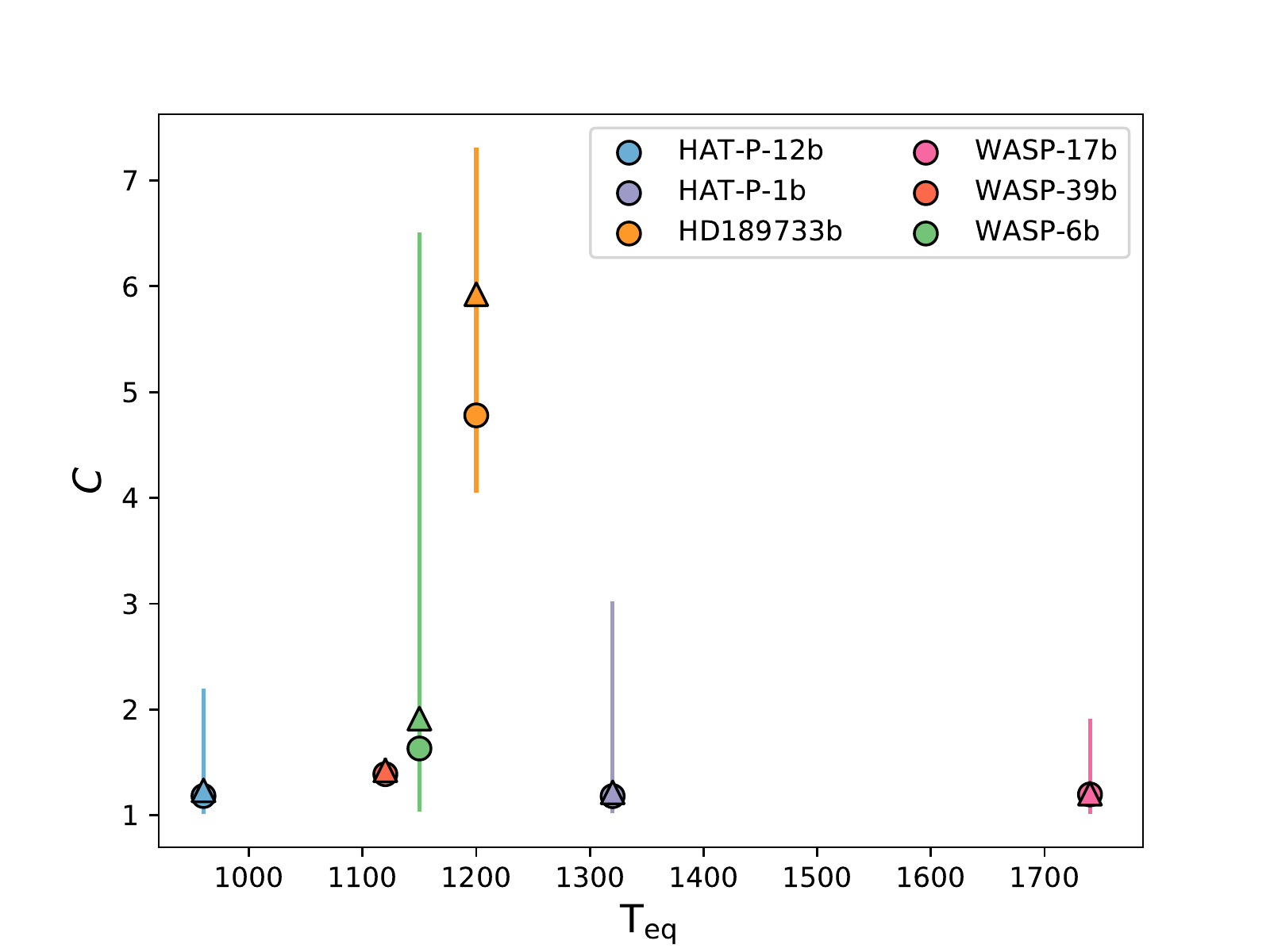}
\includegraphics[width=0.9\columnwidth]{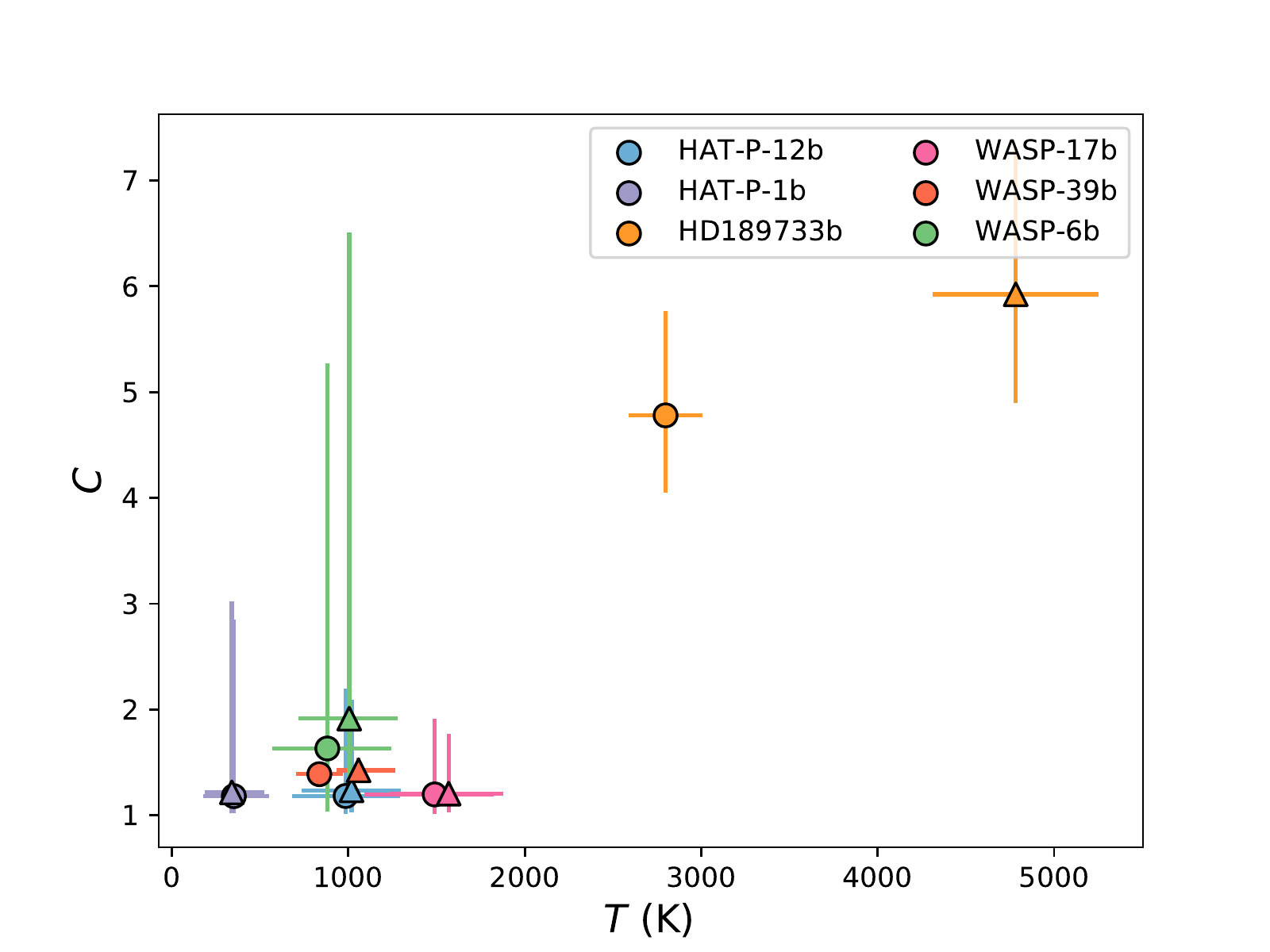}
\includegraphics[width=0.9\columnwidth]{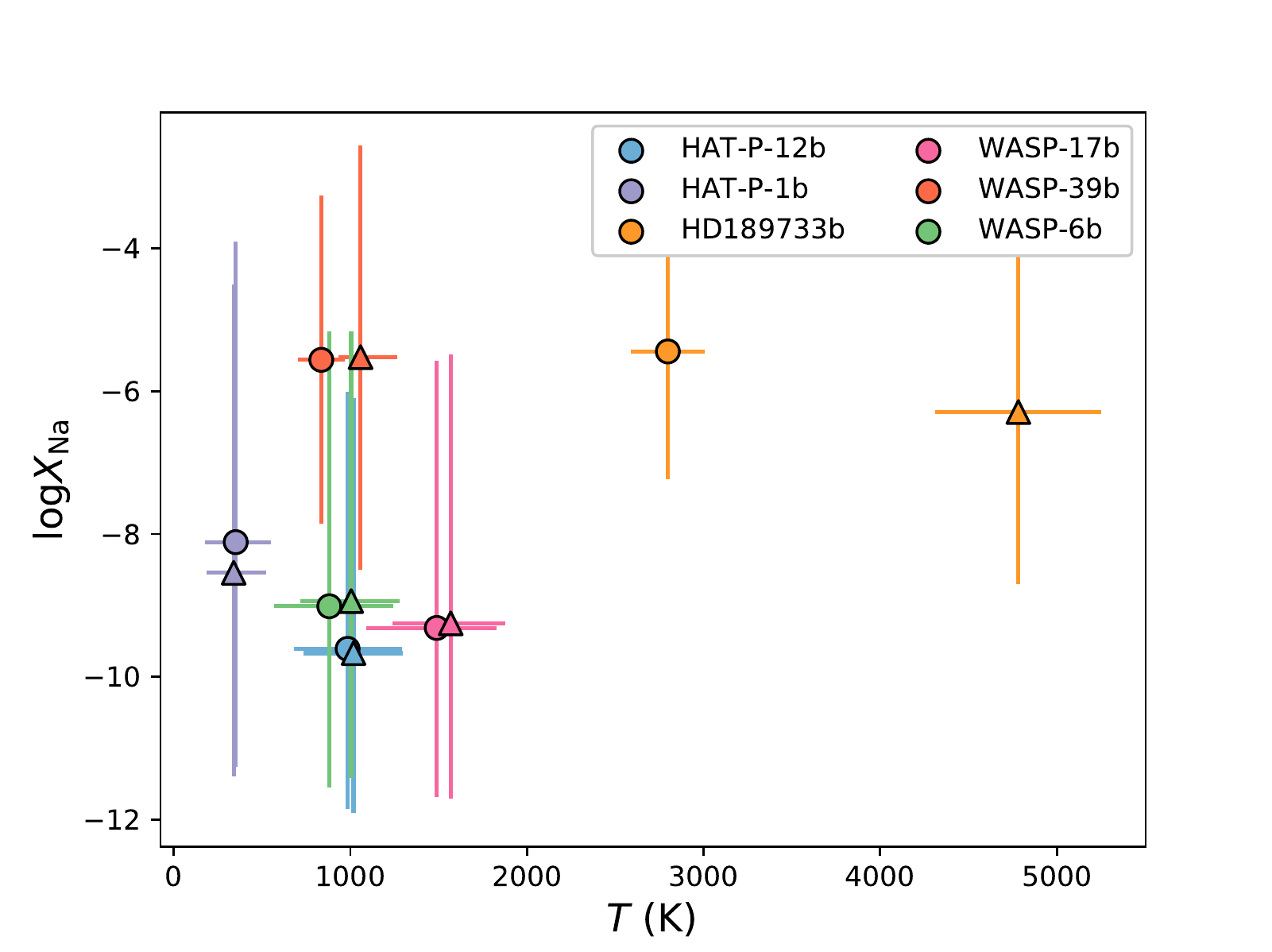}
\includegraphics[width=0.9\columnwidth]{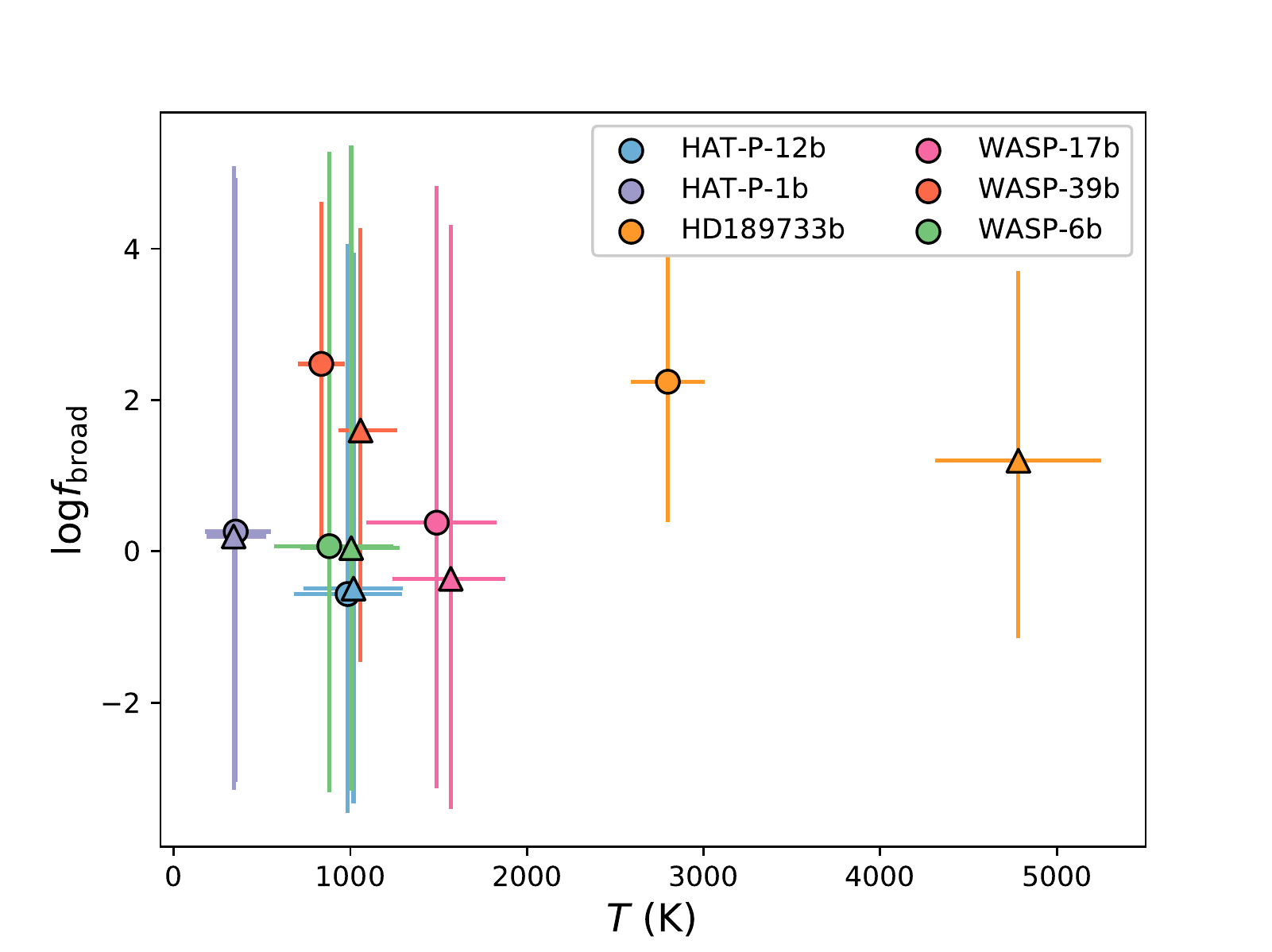}
\includegraphics[width=0.9\columnwidth]{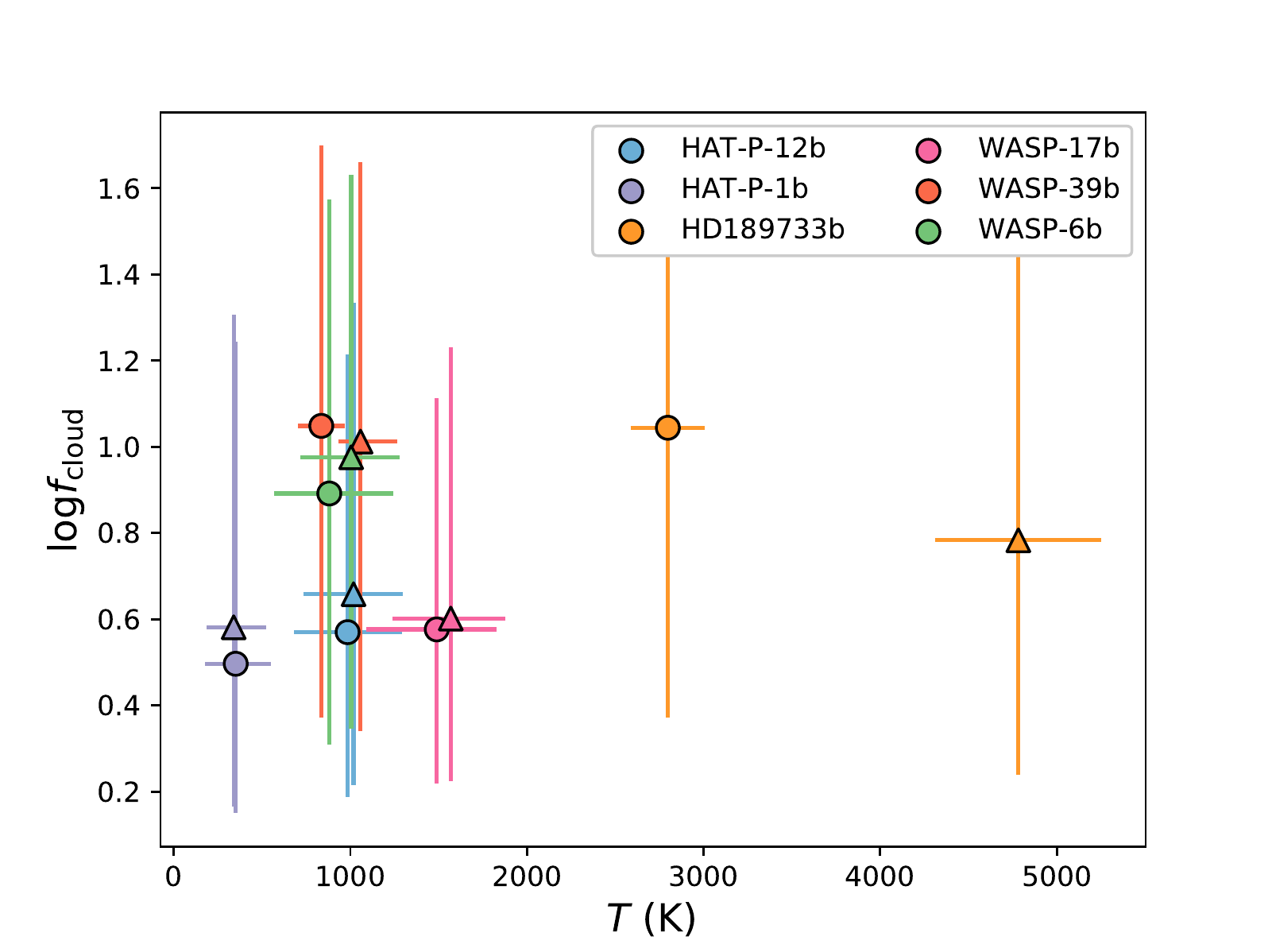}
\includegraphics[width=0.9\columnwidth]{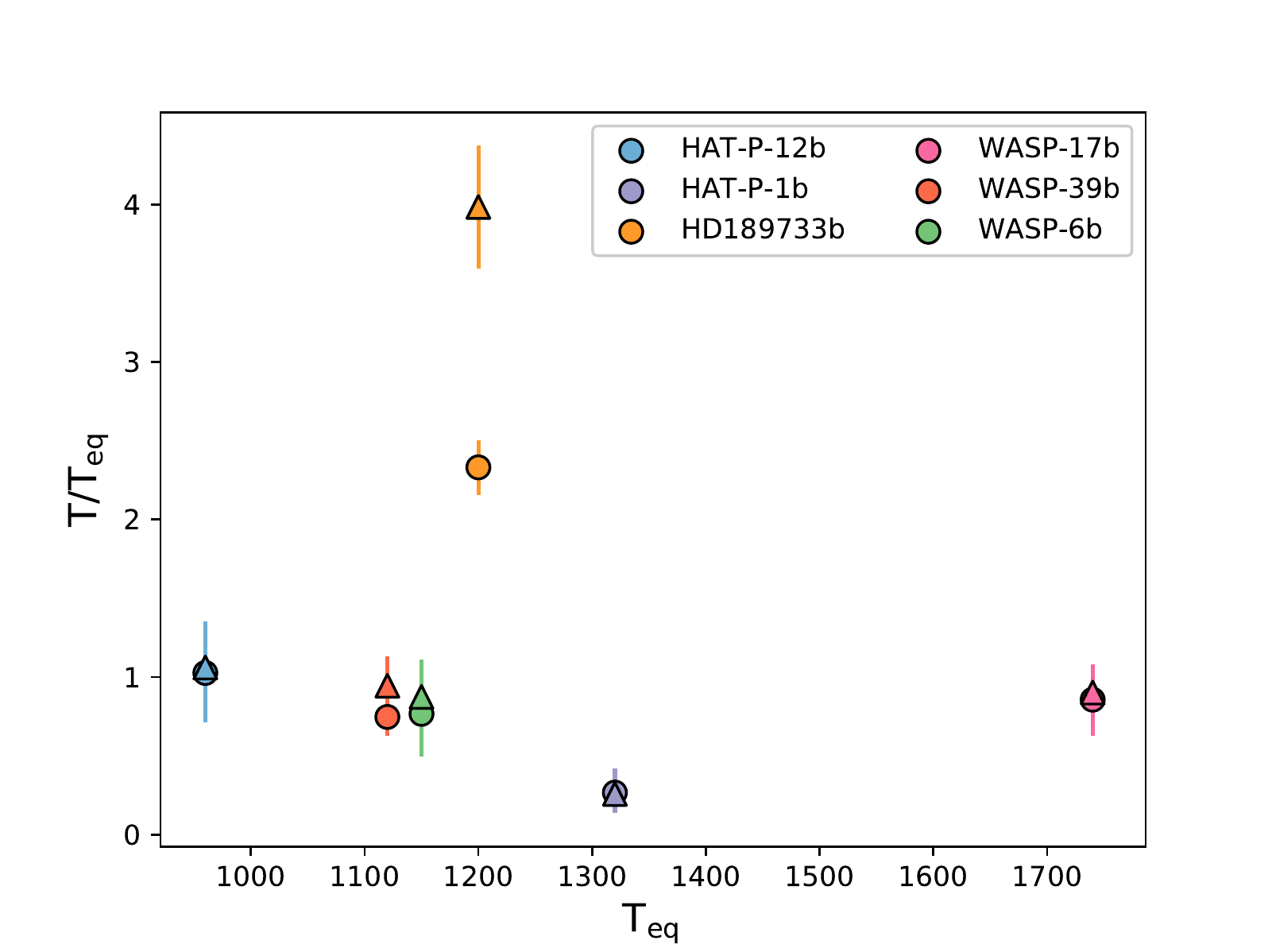}
\caption{Retrieved properties of the six exoplanets based on the retrieval analysis of their low-resolution transmission spectra.  Shown are the cloudiness index versus equilibrium temperature (top left panel), cloudiness index versus retrieved temperature (top right panel), sodium abundance versus temperature (middle left panel), broadening parameter versus temperature (middle right panel), cloud factor versus temperature (bottom left panel), and ratio of retrieved to equilibrium temperature versus equilibrium temperature (bottom right panel). The LTE models are labeled by circles, while the NLTE models are labeled by triangles.}
\label{fig:trends2}
\end{figure*}

We examine a sample of 6 low-resolution transmission spectra: HAT-P-1b, HAT-P-12b, HD 189733b, WASP-6b, WASP-17b and WASP-39b \citep{sing16}.  Given the sparseness of the data, we assume a simple model that includes the sodium doublet and a spectral continuum sourced by clouds with both small and large particles of unspecified composition.  Large cloud particles are represented by a constant opacity ($\kappa_{\rm cloud}$).  Following \cite{sing16}, small cloud particles are assumed to contribute Rayleigh scattering that has a behavior, across wavelength, which is identical to that of molecular hydrogen, but with a magnitude that is offset by some constant, dimensionless factor $f_{\rm cloud} \ge 1$.  The cross section for Rayleigh scattering by molecular hydrogen is \citep{su05},
\begin{equation}
\sigma_{\rm H_2} = \frac{24 \pi^3}{n_{\rm ref}^2 \lambda^4} \left( \frac{n_r^2 - 1}{n_r^2+2} \right)^2,
\end{equation}
where $n_{\rm ref} = 2.68678 \times 10^{19}$ cm$^{-3}$ and the real part of the index of refraction is \citep{cox}
\begin{equation}
n_r = 1.358 \times 10^{-4} \left[ 1 + 7.52 \times 10^{-3} ~\left(\frac{\lambda}{1 ~\mu\mbox{m}} \right)^{-2} \right] + 1.
\end{equation}
Collectively, the opacity function is
\begin{equation}
\begin{split}
\kappa =& \left( \frac{X_{\rm Na} \pi e^2 f_{12}}{m m_e c} \Phi + \frac{X_{\rm H_2} f_{\rm cloud} \sigma_{\rm H_2}}{m} + \kappa_{\rm cloud} \right) \\
&\times \left( 1 + \frac{n_{\rm crit} k_{\rm B} T}{P} \right)^{-1},
\end{split}
\end{equation}
where $m_{\rm H_2}=2 m_{\rm amu}$ is the mass of the hydrogen molecule.

The computed model spectrum is degraded to the measured one by taking the average value of the transit radius in each wavelength bin.  Figures \ref{fig:low_res} and \ref{fig:low_res_2} show the outcomes of these 12 retrievals.  The reference pressure is either unconstrained or only loosely constrained and its posterior distribution is largely prior-dominated.  This is unsurprising, because the pressure probed by the transit chord is
\begin{equation}
P \sim \frac{m g}{f_{\rm cloud} X_{\rm H_2} \sigma_{\rm H_2} } \sqrt{\frac{H}{R}}.
\end{equation}
If the atmosphere is cloudfree ($f_{\rm cloud}=1$) and hydrogen-dominated ($X_{\rm H_2} \approx 0.91$), then the pressure can be uniquely determined.  The transit-chord pressure and reference pressure are related by hydrostatic equilibrium.  However, if the value of $f_{\rm cloud}$ is a priori unknown, then $P$ and hence $P_0$ cannot be uniquely determined.

At low spectral resolutions, the trend of decreasing $f_{\rm broad}$ producing strong lines (relative to the continuum) is negated by a higher abundance of sodium, and it is this degeneracy that dominates the width of the posterior distributions of $X_{\rm Na}$.  In all 12 retrievals, the temperature is tightly constrained as it functions like an independent ``stretch mode" in the retrieval (\citealt{fh18}; see also Figure \ref{fig:trends}).

HD 189733b is singled out for discussion due to its historical importance in the literature \citep{pont08,pont13,sing11}.  It is the only object among the sample of 6 objects that has a clearly cloudy STIS transit chord ($f_{\rm cloud} \gg 1$), but it is important to note that the range of $f_{\rm cloud}$ values retrieved is prior-dominated by our choice of $1 \le P_0 \le 100$ bar.  The temperature corresponding to the STIS transit chord is about 2800 K (assuming LTE) or about 4800 K (for NLTE), but these values are not inconsistent with those reported by \cite{w15} and \cite{heng15}.  In all 12 retrievals, the values of $\kappa_{\rm cloud}$ retrieved are small, consistent with the continuum being non-gray.

Figure \ref{fig:trends2} shows the empirical trends (or lack thereof) between the various retrieved quantities.  For the purpose of computing cloudiness index, the cloudfree scenario is obtained by setting $f_{\rm cloud}=1$ and $\kappa_{\rm cloud}=0$.  No clear trend between the cloudiness index ($C$) and the equilibrium temperature, which is a proxy for the insolation or stellar irradiation, is seen, contrary to the tentative trend reported by \cite{heng16}.  The retrieved $C$ values range from being consistent with unity to larger than unity, in broad agreement with the study of \cite{barstow17}.  NLTE interpretations yield a higher degree of cloudiness, because of the diminished strength of the sodium lines relative to LTE.

The uncertainties on the retrieved sodium abundances are large, but the general trend is that $X_{\rm Na}$ is consistent with being solar ($\sim 10^{-6}$) or subsolar.  The uncertainties on the retrieved broadening parameters are also large, but the retrieved $f_{\rm broad}$ values are broadly consistent with unity, suggesting that sub-Lorentzian wings are not needed to fit the data.  Generally, $T/T_{\rm eq} \sim 1$, where $T$ is the retrieved temperature of the transit chord, with the exception of HD 189733b where $T/T_{\rm eq} > 2$.

\section{Discussion}
\label{sect:discussion}

Our retrieval outcomes in Figure \ref{fig:wasp49_retrievals} clearly demonstrate that ground-based, high-resolution spectra of the sodium doublet alone do not encode enough information to infer the pressure level being probed by the lines---not even at the order-of-magnitude level.  This finding is consistent with the lessons learned from our mock retrievals in Figure \ref{fig:degeneracies}, where we demonstrated that the normalization degeneracy \textit{and} the lack of an empirical normalization prevents $P_0$ from being meaningfully constrained.  The finding that ground-based high-resolution spectroscopy is incapable of accurately retrieving the pressures probed appears to contradict the work of \cite{pino18} for HD 189733b, but we note that this work did not address the normalization degeneracy and instead assumed a fixed value for $R_0$ that corresponds to $P_0=10$ bar.  It is this assumption of fixing the reference radius that allows pressure levels to be inferred.  The broader implication of this finding is that one cannot easily infer which part of the atmosphere one is probing (i.e., thermosphere, exosphere) if the sodium lines are analyzed in isolation.  Future work should elucidate if ground-based, high-resolution spectra alone (with their lack of an empirical normalization), even across an extended wavelength range, encode enough information \textbf{(e.g., in the sodium line wings)} to provide precise information on the pressures probed by the transit chord.

In the current study, we have chosen to focus only on the optical part of the spectrum that contains the sodium lines in order to elucidate the limitations associated with such a restricted analysis.  Future work should combine spectra from the optical and infrared, and elucidate if the near-infrared water lines probed by WFC3 also require a NLTE treatment as has been presented in the current study.

While the retrieved outcomes in the current study are consistent with Voigt profiles for the sodium line shapes, future work should elucidate a theory of $f_{\rm broad} = f_{\rm broad}(P)$ that reconciles and unifies the work of \cite{burrows00} and \cite{allard12}.

\begin{acknowledgments}
We are grateful to Chris Hirata for guidance on the solution to the Saha equation for the sodium atom and to Jens Hoeijmakers and David Ehrenreich for useful discussions on the relative transit depth.  We acknowledge financial support from the Swiss National Science Foundation, the European Research Council (via a Consolidator Grant to KH; grant number 771620), the PlanetS National Center of Competence in Research (NCCR), the Center for Space and Habitability (CSH) and the Swiss-based MERAC Foundation.
\end{acknowledgments}

\begin{table*}
\begin{center}
\caption{Summary of retrieval outcomes (both LTE and NLTE models)}
\label{tab:retrievals}
\begin{tabular}{lccccccccc}
\hline
Object & $T_{\rm eq}$ (K) & $T$ (K) & $\log{X_{\rm Na}}$ & $\log{f_{\rm broad}}$ & $\log{\kappa_{\rm cloud}}$ (cm$^2$ g$^{-1}$) & $\log{f_{\rm cloud}}$ & $C$ & $\log{P_0}$ (bar) & $R_0$ ($R_{\rm J}$)\\
\hline
HAT-P-1b$\dagger$ & 1320 & 351$^{+200}_{-172}$ & -8.11$^{+4.2}_{-3.15}$ & 0.26$^{+4.67}_{-3.31}$ & -6.32$^{+2.75}_{-2.34}$ & 0.95$^{+2.67}_{-0.73}$ & 1.18$^{+1.67}_{-0.16}$ & 0.5$^{+0.75}_{-0.35}$ & 1.26 \\
\multicolumn{1}{c}{NLTE} &  & 339$^{+185}_{-152}$ & -8.54$^{+4.03}_{-2.85}$ & 0.2$^{+4.89}_{-3.35}$ & -6.4$^{+2.96}_{-2.35}$ & 1.0$^{+2.35}_{-0.76}$ & 1.22$^{+1.81}_{-0.2}$ & 0.5$^{+0.75}_{-0.35}$ & 1.26 \\
HAT-P-12b$\dagger$ & 960 & 984$^{+308}_{-302}$ & -9.61$^{+3.59}_{-2.24}$ & -0.56$^{+4.63}_{-2.9}$ & -6.57$^{+2.1}_{-2.15}$ & 0.85$^{+1.13}_{-0.55}$ & 1.18$^{+1.01}_{-0.17}$ & 0.57$^{+0.64}_{-0.38}$ & 0.9 \\
\multicolumn{1}{c}{NLTE} &  & 1018$^{+278}_{-283}$ & -9.67$^{+3.57}_{-2.24}$ & -0.49$^{+4.43}_{-2.84}$ & -6.74$^{+2.13}_{-2.09}$ & 0.72$^{+0.86}_{-0.44}$ & 1.24$^{+0.86}_{-0.21}$ & 0.57$^{+0.64}_{-0.38}$ & 0.9 \\
HD189733b$\dagger$ & 1200 & 2797$^{+208}_{-211}$ & -5.44$^{+1.77}_{-1.79}$ & 2.24$^{+1.94}_{-1.86}$ & -5.37$^{+3.65}_{-3.11}$ & 4.78$^{+0.84}_{-0.75}$ & 4.78$^{+0.99}_{-0.73}$ & 1.04$^{+0.65}_{-0.67}$ & 1.12 \\
\multicolumn{1}{c}{NLTE} &  & 4782$^{+467}_{-472}$ & -6.29$^{+2.36}_{-2.41}$ & 1.2$^{+2.51}_{-2.35}$ & -5.97$^{+2.9}_{-2.67}$ & 2.92$^{+0.63}_{-0.58}$ & 5.92$^{+1.39}_{-1.03}$ & 1.04$^{+0.65}_{-0.67}$ & 1.12 \\
WASP-6b$\dagger$ & 1150 & 880$^{+362}_{-312}$ & -9.01$^{+3.84}_{-2.53}$ & 0.07$^{+5.21}_{-3.26}$ & -5.7$^{+2.97}_{-2.72}$ & 2.55$^{+2.22}_{-1.48}$ & 1.63$^{+3.64}_{-0.6}$ & 0.89$^{+0.68}_{-0.58}$ & 1.18 \\
\multicolumn{1}{c}{NLTE} &  & 1004$^{+272}_{-287}$ & -8.94$^{+3.77}_{-2.48}$ & 0.04$^{+5.32}_{-3.21}$ & -5.89$^{+2.8}_{-2.58}$ & 1.95$^{+2.31}_{-1.11}$ & 1.91$^{+4.6}_{-0.82}$ & 0.89$^{+0.68}_{-0.58}$ & 1.18 \\
WASP-17b$\dagger$ & 1740 & 1489$^{+336}_{-399}$ & -9.32$^{+3.74}_{-2.37}$ & 0.38$^{+4.45}_{-3.51}$ & -6.91$^{+2.39}_{-2.01}$ & 0.75$^{+0.91}_{-0.47}$ & 1.2$^{+0.71}_{-0.18}$ & 0.58$^{+0.54}_{-0.36}$ & 1.73 \\
\multicolumn{1}{c}{NLTE} &  & 1568$^{+307}_{-329}$ & -9.25$^{+3.77}_{-2.45}$ & -0.36$^{+4.67}_{-3.04}$ & -6.78$^{+2.24}_{-2.1}$ & 0.58$^{+0.57}_{-0.38}$ & 1.2$^{+0.56}_{-0.18}$ & 0.58$^{+0.54}_{-0.36}$ & 1.73 \\
WASP-39b$\dagger$ & 1120 & 835$^{+131}_{-132}$ & -5.56$^{+2.3}_{-2.29}$ & 2.48$^{+2.14}_{-2.4}$ & -6.04$^{+2.84}_{-2.71}$ & 2.22$^{+1.05}_{-0.89}$ & 1.39$^{+0.1}_{-0.08}$ & 1.05$^{+0.65}_{-0.68}$ & 1.18 \\
\multicolumn{1}{c}{NLTE} &  & 1057$^{+205}_{-125}$ & -5.52$^{+2.96}_{-2.98}$ & 1.6$^{+2.68}_{-3.06}$ & -5.94$^{+2.31}_{-2.7}$ & 1.48$^{+0.7}_{-0.64}$ & 1.42$^{+0.12}_{-0.1}$ & 1.05$^{+0.65}_{-0.68}$ & 1.18 \\
WASP-49b$^\heartsuit$ & 1400 & $7209^{+1763}_{-1892}$ & $-5.64^{+2.55}_{-2.90}$ & $-5.70^{+3.61}_{-2.62}$ & $-0.18^{+3.14}_{-2.77}$ & $-$ & $1.52^{+0.84}_{-0.37}$ & $1.02^{+0.63}_{-0.69}$ & $1.198$ \\
\multicolumn{1}{c}{NLTE} & & $8415^{+1020}_{-1526}$ & $-4.44^{+2.02}_{-2.51}$ & $-6.56^{+2.91}_{-2.30}$ & $-0.90^{+2.82}_{-2.06}$ & $-$ & $1.36^{+0.42}_{-0.24}$ & $1.04^{+0.64}_{-0.68}$ & $1.198$ \\
\hline
\hline
\end{tabular}\\
%\vspace{0.05in}
$\dagger$: HST STIS data from \cite{sing16}. $\heartsuit$: HARPS data from \cite{w17}.
\end{center}
\end{table*}

\appendix

\section{Electron density via solution of Saha equation}
\label{append:saha}

For completeness, we provide the expression for $n_e$ if the electrons were entirely sourced by the collisional ionization of the sodium atom, 
\begin{equation}
\mbox{Na} \longleftrightarrow \mbox{Na}^+ + e^-.
\end{equation}
To render the problem tractable, we retain the assumption that $n_e$ is a solution of the Saha equation, i.e., the ionization states are in LTE, which reads (e.g., Section 3.4 of \citealt{draine11})
\begin{equation}
\frac{n_e n_{{\rm Na}^+}}{n_{\rm Na}} = \frac{2 \left( 2 \pi m_e k_{\rm B} T \right)^{3/2}}{h^3} \frac{g_{{\rm Na}^+}}{g_{\rm Na}} ~e^{-E_{\rm ion}/k_{\rm B} T},
\end{equation}
where $E_{\rm ion}$ is the ionization energy.  $g_{{\rm Na}^+}$ and $g_{\rm Na}$ are the quantum degeneracies associated with Na$^+$ and Na, respectively.  If we assume that the electrons are solely provided by the ionization of the sodium atom, then we have $n_e = n_{{\rm Na}^+}$ and
\begin{equation}
n_e = \frac{\left( 2 \pi m_e k_{\rm B} T \right)^{3/4}}{h^{3/2}} \sqrt{\frac{g_{{\rm Na}^+}}{g_{\rm Na}}} ~\sqrt{2 X_{\rm Na} n_{\rm total}} ~e^{-E_{\rm ion}/2k_{\rm B} T},
\label{eq:ne}
\end{equation}
where $X_{\rm Na} \equiv n_{\rm Na}/n_{\rm total}$ is the volume mixing ratio of sodium.  Previously, the preceding expression was used to describe the number density of electrons in protoplanetary disks for the purpose of studying the magneto-rotational instability, where it was assumed that all of the electrons were sourced by potassium, which has an ionization energy of $E_{\rm ion}=4.3407$ eV \citep{bh00}.  It was also used to study Ohmic dissipation in hot Jupiters \citep{perna10}.  In the current situation, we assume that the electrons are sourced only by sodium, which has $E_{\rm ion}=5.1391$ eV.  We have $g_{{\rm Na}^+}/g_{\rm Na}=1/2$ because the singly-charged ion possesses a closed shell of electrons, but the neutral atom has a single valence electron in the $3s$ orbital.

\label{lastpage}

\end{document}